\documentclass[usenatbib]{mn2e}
\usepackage{apjfonts,amsfonts,amsmath,amssymb,bm,ctable,verbatim}

\usepackage[colorinlistoftodos]{todonotes}

\usepackage[percent]{overpic}

\newcommand{\paperone}{Paper {\small I}}

\newcommand{\movieurl}{\href{http://www.tapir.caltech.edu/~phopkins/Site/animations/dust-and-gas-in-astrophysic}{\url{http://www.tapir.caltech.edu/~phopkins/Site/animations/dust-and-gas-in-astrophysic}}}

\newcommand{\acknowledgments}[1]{\begin{small}\section*{Acknowledgments}\end{small}{\noindent #1}\vspace{5pt}}
\newcommand{\fref}[1]{Fig.~\ref{#1}}
\newcommand{\sref}[1]{\S~\ref{#1}}

\newcommand{\tref}[1]{Table~\ref{#1}}
\newcommand{\figthreeS}[1]{\includegraphics[width=0.32\columnwidth]{figures/#1}}
\newcommand{\figthreeSlab}[2]{\begin{overpic}[width=0.32\columnwidth]{figures/#1}\put (54,88) {\tiny #2}\end{overpic}}
\newcommand{\bpanelfigS}{\begin{figure}\begin{center}}
\newcommand{\epanelfigS}{\end{center}\end{figure}}
\newcommand{\figthreeL}[1]{\includegraphics[width=0.33\textwidth]{figures/#1}}
\newcommand{\figthreeLlab}[2]{\begin{overpic}[width=0.33\textwidth]{figures/#1}\put (62,85) {#2}\end{overpic}}
\newcommand{\bpanelfigL}{\begin{figure*}\begin{center}}
\newcommand{\epanelfigL}{\end{center}\end{figure*}}

\newcommand{\figthree}[1]{\figthreeS{#1}}
\newcommand{\bpanelfig}{\bpanelfigS}
\newcommand{\epanelfig}{\epanelfigS}

\newcommand{\figthreeX}[3]{\figthree{#1}\figthree{#2}\figthree{#3}\\}

\newcommand{\figthreeXSlab}[6]{\figthreeSlab{#1}{#2}\figthreeSlab{#3}{#4}\figthreeSlab{#5}{#6}\\}
\newcommand{\figthreeXL}[3]{\figthreeL{#1}\figthreeL{#2}\figthreeL{#3}\\}
\newcommand{\figthreeXLlab}[6]{\figthreeLlab{#1}{#2}\figthreeLlab{#3}{#4}\figthreeLlab{#5}{#6}\\}
\newcommand{\Dt}[1]{\frac{\mathrm{d} #1}{\mathrm{dt}}}
\newcommand{\initvalupper}[1]{#1^{0}}
\newcommand{\initvallower}[1]{#1_{0}}
\newcommand{\driftvel}{{\bf w}_{s}}
\newcommand{\driftvelmag}{w_{s}}
\newcommand{\driftvelhat}{\hat{{\bf w}}_{s}}
\newcommand{\dustvel}{{\bf v}_{d}}
\newcommand{\gasvel}{{\bf u}_{g}}
\newcommand{\gasden}{\rho_{g}}
\newcommand{\gaspressure}{P}
\newcommand{\dustden}{\rho_{d}}
\newcommand{\resolution}{\Delta x_{0}}
\newcommand{\ts}{t_{s}}
\newcommand{\cs}{c_{s}}
\newcommand{\vA}{v_{A}}
\newcommand{\internaldensity}{\bar{\rho}_{\rm grain}^{\,i}}
\newcommand{\grainsize}{\epsilon_{\rm grain}}
\newcommand{\grainmass}{m_{\rm grain}}
\newcommand{\graincharge}{q_{\rm grain}}
\newcommand{\grainchargeZ}{Z_{\rm grain}}
\newcommand{\tL}{t_{L}}

\newcommand{\B}{{\bf B}}
\newcommand{\Bhat}{\hat{\bf B}}
\newcommand{\acc}{{\bf a}}
\newcommand{\Lbox}{L_{\rm box}}
\newcommand{\Alf}{{Alfv\'en}}
\newcommand{\BV}{Brunt-V\"ais\"al\"a}

\title[Simulating MHD Drag Instabilities]{Simulating Diverse Instabilities of Dust in Magnetized Gas}

\author[Hopkins et al.]{
\parbox[t]{\textwidth}{
Philip F.~Hopkins$^1$, Jonathan Squire$^{2}$, Darryl Seligman$^3$} \vspace*{4pt} \\
$^1$ TAPIR, Mailcode 350-17, California Institute of Technology, Pasadena, CA 91125, USA. E-mail:phopkins@caltech.edu \\
$^2$ Physics Department, University of Otago, 730 Cumberland St., Dunedin 9016, New Zealand \\
$^3$ Department of Astronomy, Yale University, 52 Hillhouse Ave., New Haven CT 06511, USA \\
}

\date{}
\hypersetup{draft}
\begin{document}
\maketitle
\mathchardef\mhyphen="2D 

\begin{abstract}
Recently \citet{squire.hopkins:RDI} showed that charged dust grains moving through magnetized gas under the influence of a uniform external force (such as radiation pressure or gravity) are subject to a spectrum of instabilities. Qualitatively distinct instability families are associated with different \Alf\ or magnetosonic waves and drift or gyro motion. We present a suite of simulations exploring these instabilities, for grains in a homogeneous medium subject to an external acceleration. We vary parameters such as the ratio of Lorentz-to-drag forces on dust, plasma $\beta$, size scale, and acceleration. All regimes studied drive turbulent motions and dust-to-gas fluctuations in the saturated state, rapidly amplify magnetic fields into equipartition with velocity fluctuations, and produce instabilities that persist indefinitely (despite random grain motions). Different parameters produce diverse morphologies and qualitatively different features in dust, but the saturated gas state can be broadly characterized as anisotropic magnetosonic or \Alf{ic} turbulence. Quasi-linear theory can qualitatively predict the gas turbulent properties. Turbulence grows from small to large scales, and larger-scale modes usually drive more vigorous gas turbulence, but dust velocity and density fluctuations are more complicated. In many regimes, dust forms structures (clumps, filaments, sheets) that reach extreme over-densities (up to $\gg 10^{9}$ times mean), and exhibit substantial sub-structure even in nearly-incompressible gas. These can be even more prominent at lower dust-to-gas ratios. In other regimes, dust self-excites scattering via magnetic fluctuations that isotropize and amplify dust velocities, producing fast, diffusive dust motions.
\end{abstract}

\begin{keywords}
instabilities --- turbulence --- ISM: kinematics and dynamics --- star formation: general --- 
galaxies: formation --- cosmology: theory --- planets and satellites: formation --- accretion, accretion disks
\end{keywords}

\section{Introduction}

%
%
\begin{figure}
\begin{center}
\includegraphics[width=\columnwidth]{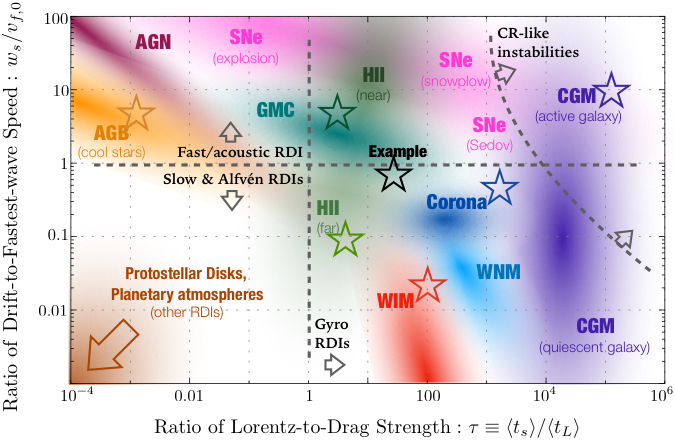}\vspace{-0.35cm}
\caption{The simulations studied in this work are shown with stars  (see \tref{table:sims}), in an illustration of two important parameters of the MHD RDI (adapted from \paperone). Axes show dust drift speed normalized by the fastest wave speed ($\driftvelmag / v_{f,0} \equiv |\driftvel| / (\cs^{2} + \vA^{2})^{1/2}$), and the ratio of Lorentz force to drag force (or drag/stopping time to gyro time, $\tau\equiv \langle t_s \rangle/\langle t_L \rangle$). Different shaded regions illustrate different astrophysical environments (see \paperone, \S~9), including the warm ionized and warm neutral medium (WIM/WNM), giant molecular clouds (GMCs) and near/far vicinity of O-stars in HII regions (HII), supernovae in various phases of evolution (SNe), stellar coronal dust (Corona), cool/giant/AGB star photospheres and outflows (AGB), dusty ``torii'' around active galactic nuclei (AGN), the circum and/or inter-galactic medium around AGN/starburst systems or quiescent galaxies (CGM), and proto-stellar/planetary disks and planetary atmospheres (which extend off the plotted range). Lines/arrows illustrate where different forms of the RDIs should appear: the fast (acoustic) RDI is unstable for $\driftvelmag/v_{f,0}\gtrsim 1$, gyro-resonant RDIs can be dominant at $\tau \gtrsim 1$, and cosmic ray-like RDIs can dominate at very large $\tau$.
\label{fig:parameter.space}}
\end{center}
\end{figure}
%
%
\begin{table*}
\begin{center}
 \begin{tabular}{| l c c c c c r |}
 \hline
Name & $\frac{|\initvalupper{ \driftvel }|}{\initvalupper{ \cs }}$ ($\tilde{a}$) & $\tau$ ($\tilde{\phi}$) & $|\initvallower{\Bhat}\cdot\hat{\acc}|$ & $\frac{\Lbox}{\initvalupper{ \cs}\, \initvalupper{\ts }}$ ($\tilde{\alpha}$) & $\beta$ & Notes \\
 \hline\hline
{\bf Example} & 0.89 (25) & 29 (50) & $0.05$ & 0.34 (5) & 2 & Case study in \citet{seligman:2018.mhd.rdi.sims}  \\ 
\hline
 {\bf AGB} & 3 (8.4) & 1e-3 (2.8e-3) & $1/\sqrt{2}$ & 0.01, 3.1, 920, 3e5 & 2 & AGB wind, $r\sim100\,R_{\odot}$, $\dot{M}\sim 10^{-6}\,M_{\odot}\,{\rm yr}^{-1}$ \\ 
 (S/M/L/XL) &   &   &   & (270, 0.9, 0.003, 1e-5) &   & $\Lbox \sim $ (5\,km, 2000\,km, 1\,$R_{\odot}$, 300\,$R_{\odot}$)  \\ 
\hline
{\bf HII-near} & 4 (20) & 2.3 (24) & $1/\sqrt{2}$ & 0.0088, 2.6, 1000 & 20 & HII region, $r=0.1\,{\rm pc}$, $n=10^{4}\,{\rm cm^{-3}}$ \\
(S/M/L)  &   &   &   & (390, 1.3, 0.0034) &   & $\Lbox \sim $ (0.2 au, 50 au, 0.1 pc)  \\ 
\hline
 {\bf HII-far} & 0.15 (0.45) & 3.5 (16) & $1/\sqrt{2}$ & 0.0032, 1.0, 330 & 20 & HII region, $r=1\,{\rm pc}$, $n=10^{2}\,{\rm cm^{-3}}$  \\
 (S/M/L) &   &   &   & (500, 1.7, 0.005) &   & $\Lbox \sim $ (0.7 au, 200 au, 0.3 pc)  \\ 
\hline
 {\bf WIM} & 0.05 (0.12) & 100 (160) & $1/\sqrt{2}$ & 4.8e-4, 0.21, 100 & 2 & WIM, $n\sim 1\,{\rm cm^{-3}}$, $T\sim 10^{4}\,$K \\ 
(S/M/L)  &   &   &   & (3400, 7.6, 0.017) &   & $\Lbox \sim $ (1 au, 500 au, 1 pc)  \\ 
\hline
 {\bf Corona} & 20 (480) & 3200 (1700) & $1/\sqrt{2}$ & 3.3e-4, 0.067, 20 & 0.002 & Solar corona, $r\sim R_{\odot}$ \\ 
 (S/M/L) &   &   &   & (4.8e4, 240, 0.8) &   & $\Lbox \sim $ (10 km, 2000 km, 1 $R_{\odot}$)  \\ 
\hline
 {\bf CGM} & 9.5 (100) & 1.4e5 (3e7) & $1/\sqrt{2}$ & 0.29 (25) &  2000 & CGM at $100\,$kpc from QSO; $\Lbox\sim1\,$kpc \\ 
\hline
\end{tabular}
\end{center}\vspace{-0.25cm}
\caption{The default initial conditions for the simulations studied in this paper. Each simulation (by default) adopts an isothermal gas equation-of-state, Epstein drag, constant grain charge, and follows a single population of grains, with resolution $2\times128^{3}$ elements (equal number gas and dust), and total dust-to-gas mass ratio  $\mu \equiv \initvalupper{\dustden}/\initvalupper{\gasden} = 0.01$. These choices are varied below. Columns show: {\bf (1)} Simulation name (used throughout). {\bf (2)} ${|\initvalupper{ \driftvel }|}/{\initvalupper{ \cs }}$: The initial equilibrium drift velocity, in units of the sound speed. {\bf (2)} $\tau \equiv \initvalupper{\ts}/\initvalupper{\tL}$: Ratio of Lorentz to drag forces (stopping time to gyro time). {\bf (3)} $|\initvallower{\Bhat}\cdot\hat{\acc}|$: Angle between initial magnetic field and direction of differential acceleration/force between dust and gas, $\acc \equiv \acc_{\rm ext,\,dust}-\acc_{\rm ext,\,gas}$. {\bf (4)} ${\Lbox}/{\initvalupper{ \cs}\, \initvalupper{\ts }}$: Box size in dimensionless units.  {\bf (5)} $\beta \equiv \initvalupper{P_{\rm gas}} / \initvalupper{P_{\rm B}}  = 2\,(c_{s}^{0}/v_{A}^{0})^{2}$: Initial ratio of gas thermal to magnetic pressure. {\bf (6)} Notes: we provide an example physical regime where these parameters are plausible for typical interstellar grains (from \paperone). 
We also quote an equivalent set of dimensionless parameters: the ``acceleration parameter'' $\tilde{a} \equiv |\acc|\,\internaldensity\,\grainsize /((\initvalupper{\cs})^{2}\,\initvalupper{ \gasden })$,  ``charge parameter'' $\tilde{\phi} \equiv 3\,\initvalupper{ \grainchargeZ } \,e / (4\pi\,c\,\grainsize^{2}\,(\initvalupper{ \gasden })^{1/2})$, and ``size parameter'' $\tilde{\alpha} \equiv \internaldensity\,\grainsize / \initvalupper{ \gasden }\, \Lbox$.
\label{table:sims}\vspace{-0.25cm}}
\end{table*}

%
%
\bpanelfigL
\figthreeL{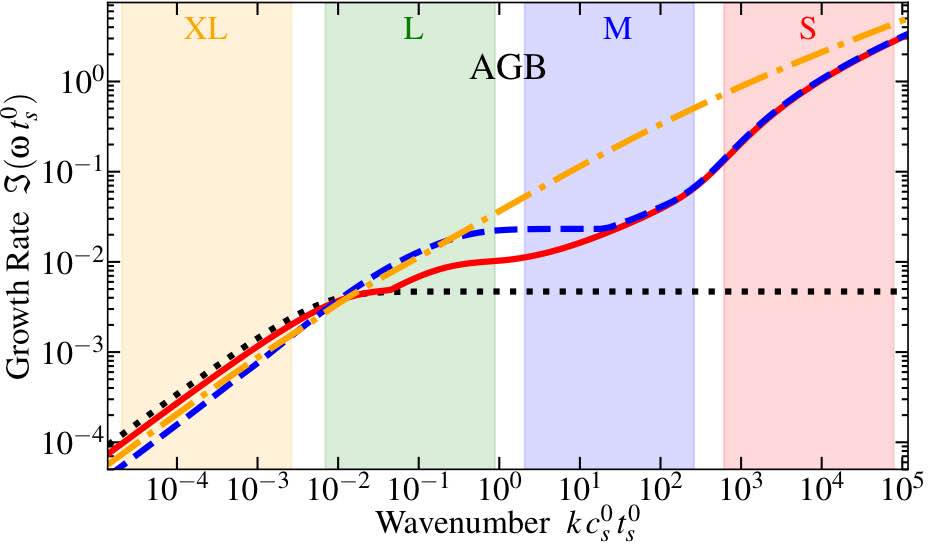}
\figthreeL{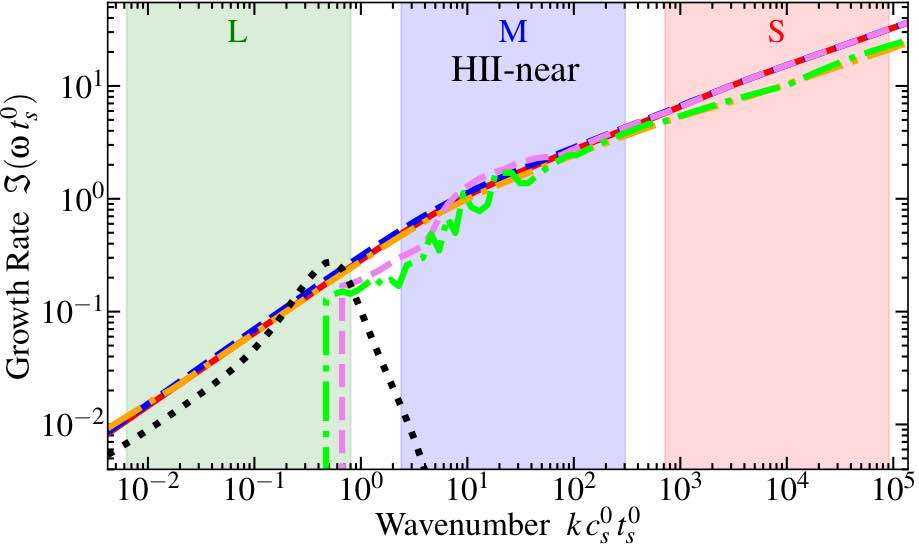}
\figthreeL{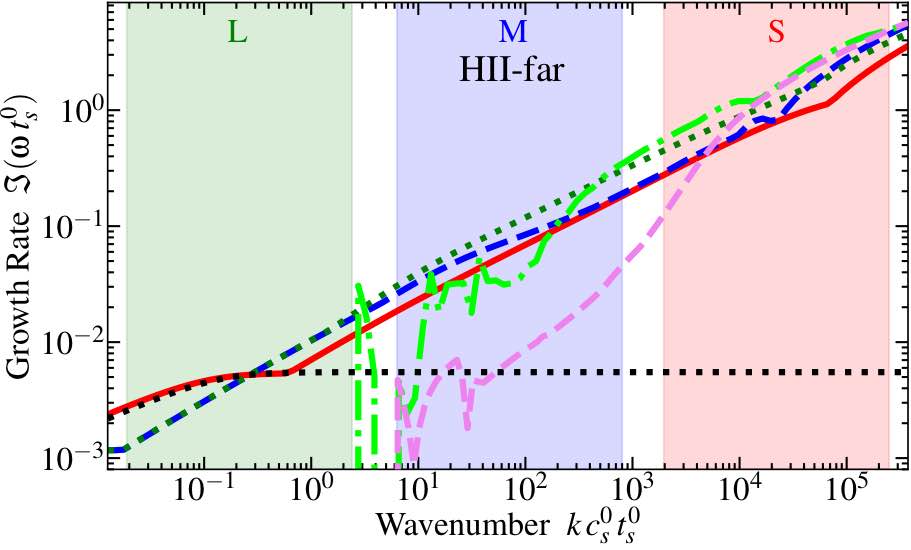}\\
\figthreeL{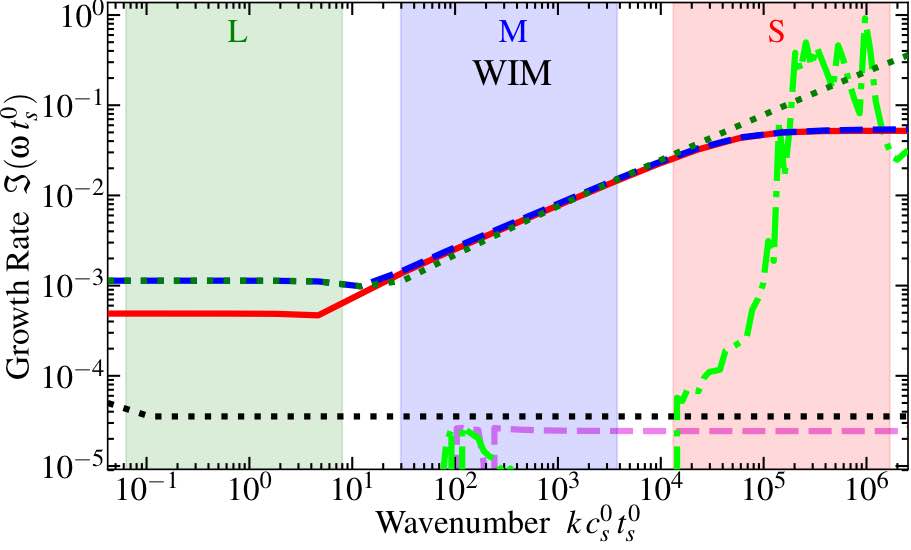}
\figthreeL{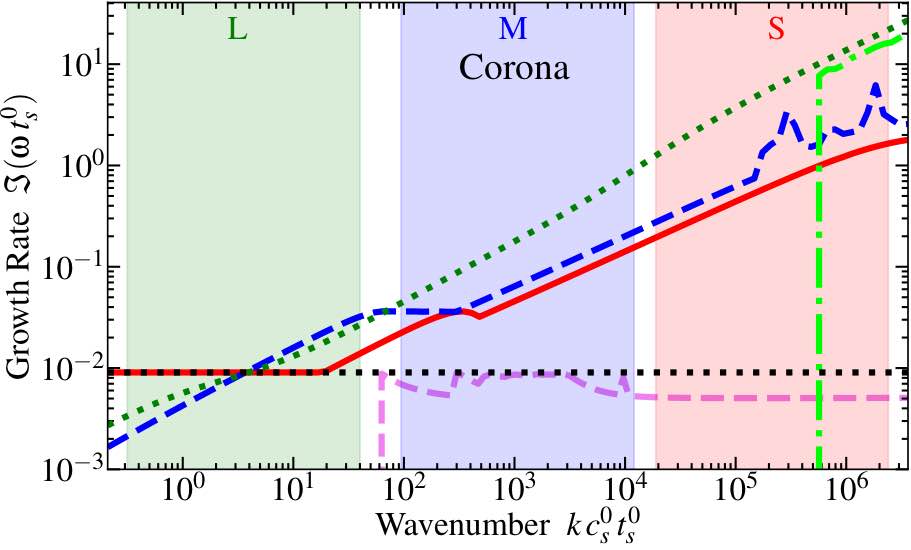}
\figthreeL{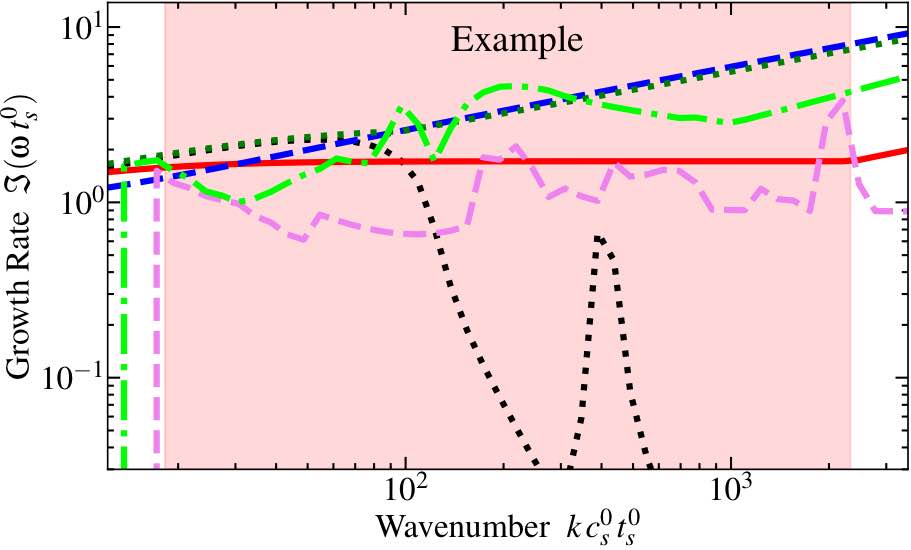}\\
\figthreeL{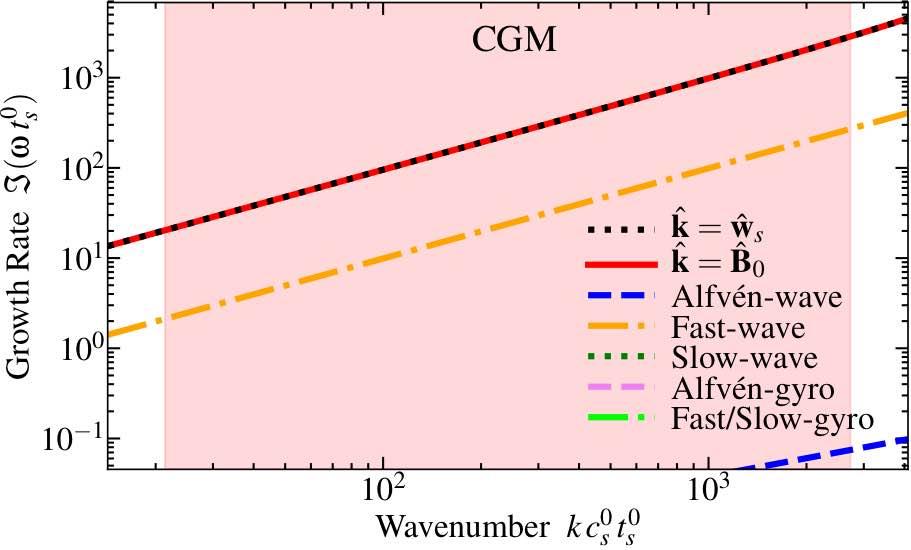}
\vspace{-0.2cm}
\caption{Maximal growth rates predicted by linear theory for different modes in the simulations. Each panel corresponds to one simulation set from \tref{table:sims} and \fref{fig:parameter.space} (fixed $\tau$, $\mu$, $\beta$, $|\initvallower{\hat{\B}}\cdot\hat{\acc}|$, $|\initvalupper{\driftvel}|/\initvalupper{\cs}$, $\gamma$, etc.), with shaded regions showing the range of wavelengths covered by each simulation box (S/M/L) with different ${\Lbox}/{\initvalupper{ \cs}\, \initvalupper{\ts }}$. We plot predicted linear growth rates versus wavenumber $k = |{\bf k}| = 2\pi/\lambda$. Different lines correspond to different mode directions $\hat{\bf k}$: parallel to the drift ($\hat{\bf k} = \driftvelhat$), parallel to the magnetic field ($\hat{\bf k} = \hat{\B}$), and directions which satisfy the conditions for different RDIs (the \Alf-wave, slow and fast magnetosonic-wave RDIs, and the \Alf-gyro and slow/fast-gyro RDIs). For the RDIs, we plot the maximum growth rate marginalized over mode angles ($\hat{\bf k}$) which satisfies the resonant condition at each $k$. Note the fast-wave RDI can only exist if $|\initvalupper{\driftvel}|/\initvalupper{\cs} >1$, and the gyro RDIs only exist above some $k$ and $\tau$ (slow and fast-gyro RDIs have degenerate solutions so are plotted as one line)$^{\ref{foot:gyromodes}}$ -- see \paperone\ for details. Generically, {\em all} wavelengths are unstable, and smaller-scale modes grow faster, although {\em which} modes dominate is scale-and-parameter dependent.\label{fig:rates}}
\epanelfigL
%
%

Almost all astrophysical, planetary, and atmospheric fluids are laden with grains of dust, which play a central role in many astrophysical processes including in planet and star formation; in the attenuation and extinction of observed light; in cool-star, brown dwarf, and planetary evolution; in atmospheric dynamics; in astro-chemistry;  in feedback and outflow-launching from star-forming regions, cool stars, and active galactic nuclei (AGN); and in inter-stellar cooling and heating \citep[see][for reviews]{draine:2003.dust.review,dorschner:dust.mineralogy.review,apai:dust.review}. The dynamical interactions between dust and gas therefore are of fundamental importance in astrophysics.

Recently, \citet{squire.hopkins:RDI} showed that dust-gas mixtures are unstable to a broad class of instabilities, which they referred to as  ``Resonant Drag Instabilities'' (RDIs). The \citet{squire.hopkins:RDI} instabilities manifest whenever a fluid, gas, or plasma system contains dust streaming with non-zero drift velocity $\driftvel\equiv \dustvel - \gasvel$ relative to the gas (where $\dustvel$ and $\gasvel$ are the dust and gas velocities, respectively). Although a broad range of wavelengths are unstable, the resonances which produce the most rapidly-growing instabilities occur when the natural frequency of a linear gas mode (e.g.\ a sound wave, MHD wave, or epicyclic oscillation) matches the natural frequency of a dust mode (e.g.\ advection, with frequency $\driftvel \cdot {\bf k}$, or gyro oscillations).  Every such pair of modes produces an instability, with a unique growth rate, resonance, and linear mode structure. Since dust-gas drift may be caused by many external forces, such as radiative absorption or scattering by dust or gas, gravity in quasi-hydrostatic systems, centrifugal or coriolis forces in rotating systems, or large-scale hydrodynamic or pressure forces, these instabilities will develop in a range of astrophysical environments.

In a series of papers, \citet{hopkins:2017.acoustic.RDI}, \citet{squire:rdi.ppd}, and \citet{hopkins:2018.mhd.rdi} analytically explored various examples of these instabilities in some astrophysical systems. \citet{hopkins:2018.mhd.rdi} (hereafter \paperone) focused on the case of instabilities involving charged dust in magnetized gas, relevant in the warm interstellar medium (ISM), circum and inter-galactic medium (CGM/IGM), HII regions, supernovae (SNe) ejecta and remnants, the Solar and stellar coronae, cool-star winds, AGN outflows and obscuring torii, and giant molecular clouds (GMCs).  They showed that a variety of instabilities appear with different properties and growth rate scalings, even in the case of a homogeneous gas obeying ideal MHD (a good approximation in most of these regimes), with a single group of grains interacting via drag and Lorentz forces. However, their analysis was restricted to analytic, linear perturbation theory. \citet{moseley:2018.acoustic.rdi.sims} presented simulations of the un-magnetized and un-charged instabilities in the non-linear regime. \citet{seligman:2018.mhd.rdi.sims} presented a case study of one example in the magnetized regime, and found that the introduction of a magnetic field produced novel dust behaviors and outcomes in both the linear and non-linear regimes of the instability. That first study necessarily neglected much of the large parameter space. 

In this paper, we present a large survey of $\sim 40$ simulations\footnote{Animations and additional visualizations of the simulations here are available at \movieurl} that explore the non-linear regime of these instabilities in a representative range of the astrophysically relevant parameter space for charged dust in magnetized gas. These idealized experiments inform our understanding of the  mechanisms responsible for the growth and saturation of the instabilities, the non-linear structure of the dust and gas, and the potential theoretical and observational ramifications. They are complex because the instabilities depend on six dimensionless parameters, and as shown in \paperone, at any given wavenumber ${\bf k}$, the linear dispersion relation typically features $\sim 3-7$ unstable modes (each of which has growth rates that depend on the mode angle $\hat{\bf k}$, at a given $|{\bf k}|$). This inherent complexity further underscores the necessity for numerical simulations that explore different non-linear regimes. We show that a diverse variety of behaviors  arise, depending both on the physical parameters of the system and the spatial scales studied, all of which may have important astrophysical consequences. 

This paper is organized as follows. \sref{sec:methods} presents our methodology, and \sref{sec:params} discusses the parameter space surveyed (see also \tref{table:sims} and \fref{fig:parameter.space}). \sref{sec:results} presents several results from the simulations (e.g.\ morphologies, saturated fluctuation amplitudes and PDFs). \sref{sec:discussion} discusses these results in more detail and compares them to theoretical expectations, attempting to identify classes of saturation mechanisms. We summarize and conclude in \sref{sec:conclusions}.

\section{Methods \&\ Simulation Setup}
\label{sec:methods}

%
%
\bpanelfigL
\figthreeL{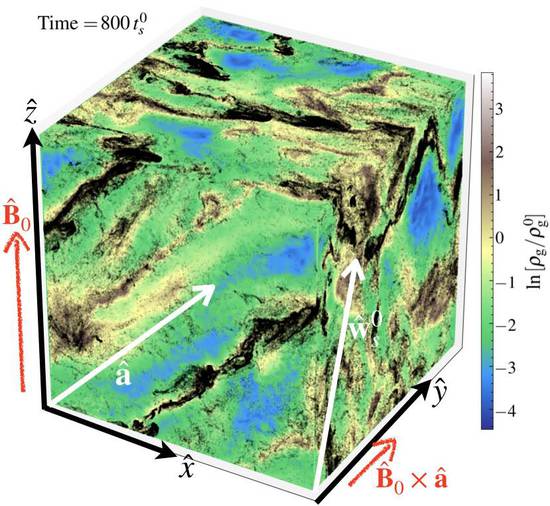}
\figthreeL{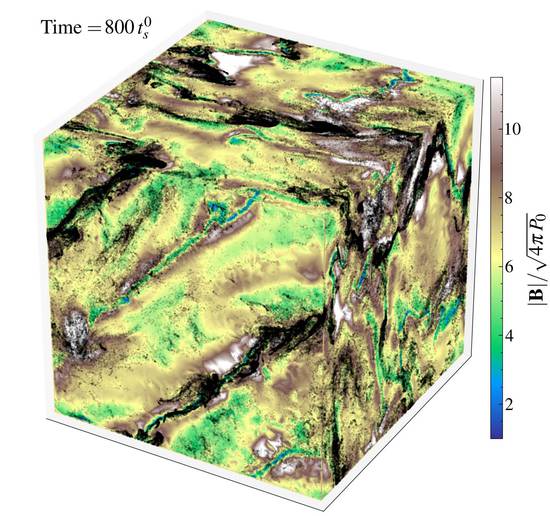} 
\figthreeL{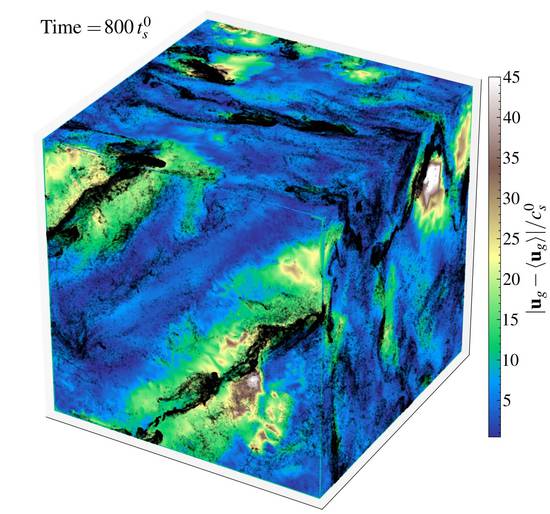} \\
\includegraphics[width=0.31\textwidth]{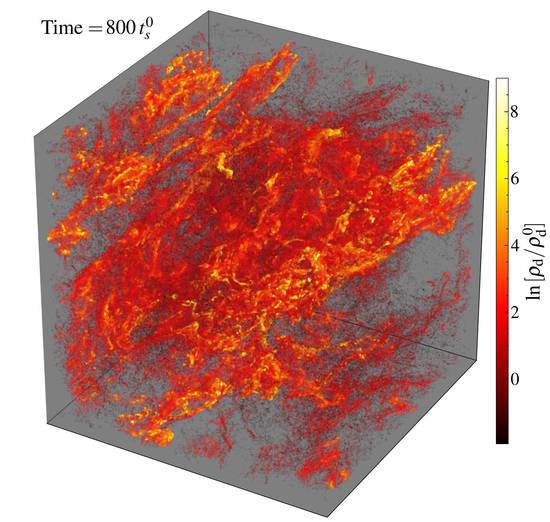}
\includegraphics[width=0.34\textwidth]{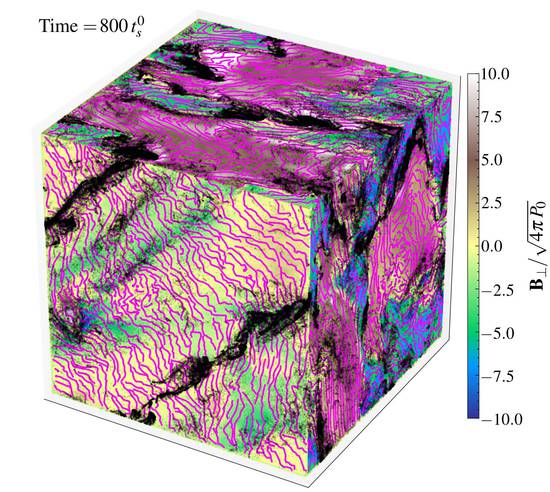}
\includegraphics[width=0.34\textwidth]{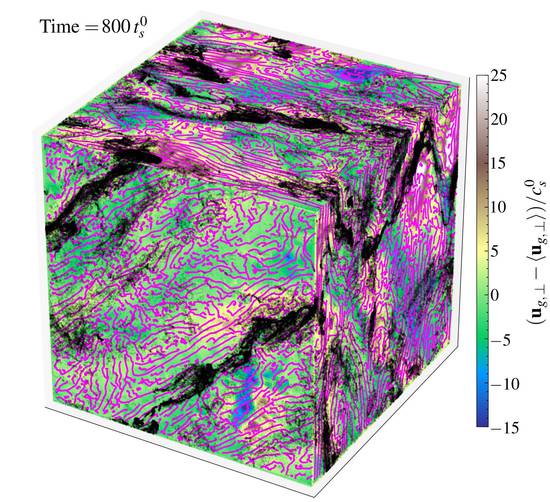}
\vspace{-0.25cm}
\caption{Images of our largest {\bf HII-near} box (box L), from \tref{table:sims} \&\ \fref{fig:rates}, after the  instabilities reach saturation. {\em Top:} Surface projection of slices through each axis of the box, at the time labeled (time in units of the initial grain stopping time $\initvalupper{\ts}$). Colors show gas quantities at the surface: $\gasden$, $\B$, $\gasvel$ (see colorbar). Black points show dust grains on the surface. Surfaces are oriented so $\initvallower{\hat{\B}}$ is the vertical axis ($\hat{z}$), $\hat{\acc}$ is oriented in the $\hat{x}$-$\hat{z}$ direction (front or width-height plane), and the mutually-perpendicular direction ($\hat{y} \propto \initvallower{\hat{\B}} \times \hat{\acc}$) is depth. {\em Bottom Left:} 3D isometric plot of all grain particles (colored by the local dust density), at the same time. {\em Bottom Middle and Right:} Surface plot of the vector $\B$ and $\gasvel$: lines show the field or flow lines of the projected component on the surface (i.e.\ $\B_{x}$, $\B_{y}$ in the $xy$ plane) while colors show the magnitude of the normal component ($\B_{z}$ in the $xy$ plane). 
The instabilities drive dramatic clumping/density structure in the dust and strong saturated \Alf{ic} turbulence in the gas. This occurs quickly relative to other timescales: the time plotted is just $\approx 0.9\,\Lbox/v_{f,\,0}$ (where $v_{f,\,0}^{2} \equiv (\initvalupper{\cs})^{2} + (\initvalupper{\vA})^{2}$ is the fastest gas wavespeed), i.e.\ less than one sound or \Alf\ box-crossing time.\label{fig:multiprop.demo}}
\epanelfigL
%
%

\subsection{Numerical Methods \&\ Equations Solved}
\label{sec:equations}

The numerical methods adopted here have been described in detail in \citet{moseley:2018.acoustic.rdi.sims,seligman:2018.mhd.rdi.sims}, and we briefly summarize them here. Our simulations were run with the code {\small GIZMO} \citep{hopkins:gizmo},\footnote{A public version of the code, including all methods used in this paper, is available at \href{http://www.tapir.caltech.edu/~phopkins/Site/GIZMO.html}{\url{http://www.tapir.caltech.edu/~phopkins/Site/GIZMO.html}}} using the Lagrangian ``meshless finite volume'' (MFV) method for MHD, which has been extensively tested on problems involving multi-fluid MHD instabilities, MRI, shock-capturing, and more \citep{hopkins:mhd.gizmo,hopkins:cg.mhd.gizmo,hopkins:gizmo.diffusion,su:2016.weak.mhd.cond.visc.turbdiff.fx}. Grains are integrated using the ``super-particle'' method \citep[see, e.g.][]{carballido:2008.grain.streaming.instab.sims,johansen:2009.particle.clumping.metallicity.dependence,bai:2010.grain.streaming.vs.diskparams,pan:2011.grain.clustering.midstokes.sims}, whereby the motion of each dust ``particle'' in the simulation follows Eq.~\eqref{eq:eom} below, but each represents an ensemble of dust grains with similar size, mass and charge (denoted $\grainsize$, $\grainmass$, $\graincharge$, respectively). The numerical methods used for the integration are described and tested in \citet{hopkins.2016:dust.gas.molecular.cloud.dynamics.sims,lee:dynamics.charged.dust.gmcs}. The back-reaction is accounted for as in \citet{moseley:2018.acoustic.rdi.sims} (see App.~B) and the the Lorentz forces is evolved using a Boris integrator.

Each individual grain (dust super-particle) in the code obeys
\begin{align}\label{eq:eom}
\Dt{\dustvel} &=  -\frac{\driftvel}{\ts} - \frac{\driftvel\times\Bhat}{\tL} + \acc_{\rm ext,\, dust}\, ,
\end{align}
where $\acc_{\rm ext,\, dust}$ is a constant external acceleration, $\ts$ is the drag coefficient or ``stopping time'' and $\tL$ the gyro or Larmor time. The gas obeys the ideal MHD equations -- the standard advection equation, $\partial \gasden/\partial t = -\nabla \cdot( \gasden\,\gasvel)$, for the gas density $\gasden$, and the standard induction equation, $\partial \B/\partial t = \nabla \times (\gasvel \times \B)$, for the magnetic field $\B$ -- with the addition of a back-reaction force from the grains in the momentum equation. In particular, whenever drag or Lorentz forces exert a force $\grainmass\,\mathrm{d}\dustvel/\mathrm{dt}$ on a grain within a given gas cell, an equal-but-opposite force is applied to the gas. This is treated as a usual momentum flux within {\small GIZMO}, which numerically guarantees {exact} force balance and total momentum conservation. The gas momentum equation reads
\begin{align}\label{eq:gas.mhd}
\nonumber \gasden \left( \frac{\partial}{\partial t} + \gasvel\cdot\nabla \right)  \gasvel =&\, -{\nabla P} - \frac{\B \times (\nabla \times \B) }{4\pi} + \gasden\,\acc_{\rm ext,\,gas} \\
 &- \int \mathrm{d}^{3}\dustvel\,f_{d}({\bf x},\,\dustvel)\,\acc_{\rm gas \mhyphen dust}(\dustvel,\,...),
\end{align}
where $f_{d}({\bf x},\,\dustvel)$ is the phase-space density distribution of dust (i.e.\ differential mass of grains per element $\mathrm{d}^{3}{\bf x}\,\mathrm{d}^{3}\dustvel$) and $\acc_{\rm ext,\, gas}$ is an external gas acceleration that we set  to zero in the simulations here.
The gas obeys an exactly polytropic equation of state with thermal pressure $P$ and sound speed $\cs$: 
\begin{align}
\gaspressure &= \initvallower{\gaspressure}\,\left( \frac{\gasden}{\initvalupper{\gasden}} \right)^{\gamma} \ , \\
\cs^{2} &\equiv \frac{\partial P}{\partial \gasden} \ .
\end{align} 

In our default simulations we assume Epstein drag, which can be approximated to very high accuracy with the expression (valid for both sub and super-sonic drift)
\begin{align} \label{eq:ts}
\ts &\equiv \sqrt[]{\frac{\pi \gamma}{8}}\frac{\internaldensity \,\grainsize}{\gasden\,\cs}\, \bigg( 1+\frac{9\pi\gamma}{128} \frac{|\driftvel|^{2}}{\cs^{2}} \bigg)^{-1/2},
\end{align}
where $\internaldensity$ and $\grainsize$ are the \textit{internal} grain density and radius, respectively. The Larmor time is:
\begin{align} \label{eq:tL} 
\tL &\equiv \frac{\grainmass\,c}{|\graincharge\,\B |} = \frac{4\pi\,\internaldensity\,\grainsize^{3}\,c}{3\,e\,|\grainchargeZ\,\B |}
\end{align}
where $\grainmass$ and $\graincharge = \grainchargeZ\,e$ are the grain mass and charge.

\subsection{Initial Conditions}
\label{sec:ics}

We initialize a periodic, cubic box of side-length $\Lbox$ with uniform gas density $ \initvalupper{\gasden} \equiv M_{\rm gas,\,box}/\Lbox^{3}$ and dust density $ \initvalupper{\dustden} \equiv \mu\, \initvalupper{\gasden}$ (dust-to-gas ratio $\mu$), gas velocity $ \initvalupper{\gasvel} = 0$, and dust drift $ \initvalupper{\driftvel} = |\acc|\, \initvalupper{\ts}\,(1+\mu)^{-1}\,(1+\tau^{2})^{-1}\,[\hat{\acc} - \tau\,(\hat{\acc} \times  \initvallower{\Bhat} ) + \tau^{2}\,(\hat{\acc}\cdot\initvallower{\Bhat})\,\initvallower{\Bhat}]$ (see \paperone, section 3.1). Here 
\begin{align} 
\initvallower{ X } \equiv \langle X(\initvalupper{ \gasden },\,\initvalupper{ \driftvel },\,...,\,t=0) \rangle
\end{align}
 is the initial homogeneous value of some variable $X$, $\acc \equiv \acc_{\rm ext,\,dust}-\acc_{\rm ext,\,gas}$ is the difference between dust and gas accelerations, and $\tau \equiv \initvalupper{ \ts } / \initvalupper{ \tL }$ parameterizes the grains' magnetization. The homogenous, steady-state equilibrium solution preserves this quasi-equilibrium while the whole box uniformly accelerates\footnote{As shown explicitly in \citet{hopkins:2017.acoustic.RDI} (see \S~2.1 and Appendix~B therein) and \paperone, the dynamics of this local (un-stratified) problem are manifestly identical in the stationary and free-falling or uniformly-accelerating frames moving with the homogeneous solution, or (by extension) if we add an equal-and-opposite {\em mean} acceleration on the gas such that the mean acceleration of the entire system vanishes. The problem is trivially invariant to any uniform velocity boost. We therefore will perform all our analysis in the co-moving (free-falling) frame. We have also verified (for numerical testing) that our results are identical up to machine error (as they should be given our Lagrangian code) if we instead add an explicit uniform acceleration and/or boost to gas and dust to ensure the homogenous $\gasvel(t) = \bf{0}$ in the lab frame.} with $\gasvel({\bf x},\,t) = \langle \gasvel(t)  \rangle = \acc_{\rm ext,\,gas}\,t + \acc\,\mu\,t / (1 + \mu)$.

We can make the coupled dust-gas equations  dimensionless by working in units of the equilibrium sound speed $\initvalupper{ \cs }$, gas density $\initvalupper{ \gasden }$, and ``weighted grain size'' $\internaldensity\,\grainsize/\initvalupper{\gasden}$. Then, for a given equation-of-state, the dynamics of the problem (at infinite numerical resolution) are entirely determined by six dimensionless parameters: (1) the acceleration $\tilde{a} \equiv |\acc|\,\internaldensity\,\grainsize /((\initvalupper{\cs})^{2}\,\initvalupper{ \gasden })$, (2) the box size or grain ``size parameter'' $\tilde{\alpha} \equiv \internaldensity\,\grainsize / \initvalupper{ \gasden }\, \Lbox$, (3) the grain ``charge parameter'' $\tilde{\phi} \equiv 3\,\initvalupper{ \grainchargeZ } \,e / (4\pi\,c\,\grainsize^{2}\,(\initvalupper{ \gasden })^{1/2})$, (4) the dust-to-gas ratio $\mu \equiv \initvalupper{\dustden}/\initvalupper{\gasden}$, (5) the plasma $\beta \equiv \initvallower{P} / (|\initvallower{\B}|^{2}/8\pi)$, and (6) the angle $|\cos{\theta_{\B\acc}}| \equiv |\initvallower{\Bhat}\cdot\hat{\acc}|$ between the initial field direction $\initvallower{\Bhat}$ and $\hat{\acc}$. Note that in our linear theory perturbation analysis we chose to work with a different, but mathematically equivalent, set of dimensionless variables: (1) $|\initvalupper{ \driftvel } |/ \initvalupper{ \cs }$, (2) $\Lbox/ \initvalupper{ \cs} \, \initvalupper{\ts }$, (3) $\tau$, (4) $\mu$, (5) $\beta$, (6) $\theta_{\B\acc}$.

Our default simulations adopt $N_{\rm gas}=128^{3}$ gas resolution elements and an equal number of dust elements, $\mu=0.01$ (the ISM mean), and an isothermal ($\gamma=1$) equation of state (appropriate for most ISM/CGM/HII region conditions of interest). But we vary all of this below. For simplicity, we assume throughout that grains are all of the same size and charge, and that  the grain charge $\graincharge$  is fixed during the simulation (as appropriate for large grains in isothermal gas) with grain  Larmor time $\tL \equiv \grainmass\,c/|\graincharge\,\B|$. The ``non-default'' simulations discussed in \sref{sec:results:eos} relax some of these restrictions, exploring different equations of states, dust-to-gas ratios, resolutions, and allowing  $\graincharge$ to vary with local gas parameters.

For convenience, throughout we adopt the Cartesian ($xyz$) axis  convention with $\hat{\bf z} \propto \initvallower{\B}$, $\hat{\bf x} \propto \acc_{\bot}$ (i.e.\ the $x-z$ plane is defined to contain $\acc$, so $\acc=\acc_{\bot}+\acc_{\|}={\rm a}_{\bot}\hat{\bf x}+{\rm a}_{\|}\hat{\bf z}$), and $\hat{\bf y} \propto \initvallower{\B} \times \acc$ (the mutually-perpendicular direction). In our 3D visualizations, the width/depth/height dimensions correspond to $x/y/z$. 

%
%
\bpanelfigL
\figthreeXLlab{t0/im_3d2dProj_v_high_terrain_0025}{S: Early NL}{c0/im_3d2dProj_b_high_terrain_0030}{M: Early NL}{k0/im_3d2dProj_rho_high_terrain_0020}{L: Early NL}
\figthreeXL{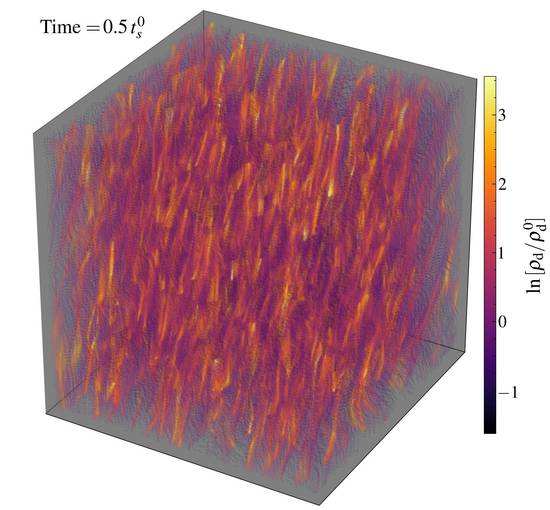}{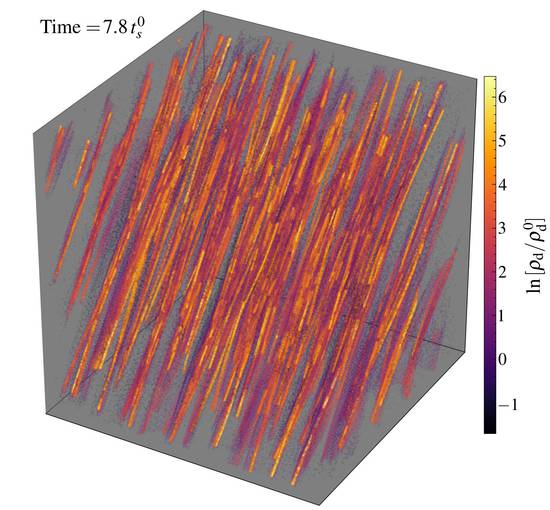}{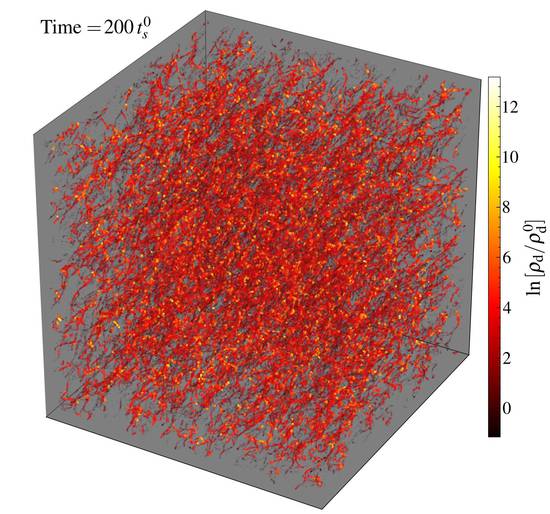}
\figthreeXLlab{t0/im_3d2dProj_v_high_terrain_0067}{S: Saturated}{c0/im_3d2dProj_b_high_terrain_0330}{M: Saturated}{k0/im_3d2dProj_rho_high_terrain_0080}{L: Saturated}
\figthreeXL{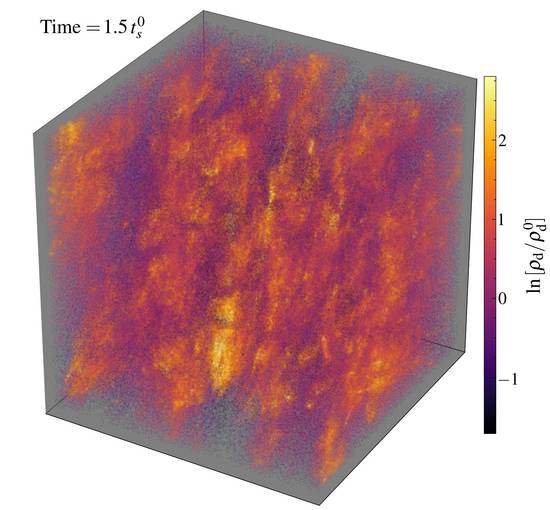}{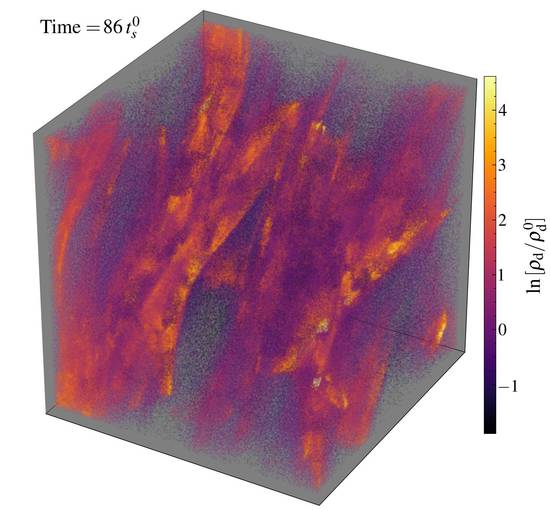}{k0/im_3dDust_hot_0080}
\caption{Images of {\bf HII-near}, one of the default simulation ``sets'' from \tref{table:sims} \&\ \fref{fig:rates}. Columns show the three different box sizes S/M/L (increasing left-to-right). 
{\em Top \&\ Second Row:} Gas+dust surface maps ({\em top}), and 3D dust maps ({\em second}), as \fref{fig:multiprop.demo}, at time labeled (during early non-linear evolution). 
{\em Third \&\ Bottom Row:} Same, but at a later time (in saturated state). 
Each box (moving left-to-right) is $\sim 300\times$ larger than the previous (so an entire box at the left is approximately $\sim 1/2$ of a pixel/element in the box to its immediate right). Dust and turbulence clearly exhibit structure on all scales. Smaller-scale modes drive weaker \&\ less-compressible gas turbulence, as predicted  (\sref{sec:discussion}). Slightly different physical parameters (compare \fref{fig:HIIfar}), or different scales with the same parameters (boxes here) can produce wildly different morphologies (and resonant angles), even with different dimensionality of dust structures (e.g.\ point-like clumps, 1D filaments, 2D sheets). This owes to complicated wavelength-dependence of the dominant modes (\fref{fig:rates}). The morphologies present in the L simulation (right-hand panels) provide an example of the ``clumped'' saturation mode (see \sref{sec:sub.clumped}).\label{fig:HIInear}}
\epanelfigL
%
%

\begin{table*}
\begin{center}
 \begin{tabular}{| l c c c c c c c c c c c |} 
 \hline
Name & $\frac{\delta {\bf u}_{g,\,x}}{\initvalupper{\cs}}$ &  $\frac{\delta {\bf u}_{g,\,y}}{\initvalupper{\cs}}$ &  $\frac{\delta {\bf u}_{g,\,z}}{\initvalupper{\cs}}$ & 
$\frac{\delta {\bf v}_{d,\,x}}{\initvalupper{\cs}}$ &  $\frac{\delta {\bf v}_{d,\,y}}{\initvalupper{\cs}}$ &  $\frac{\delta {\bf v}_{d,\,z}}{\initvalupper{\cs}}$ & 
$\frac{\delta \B_{x}}{(4\pi\initvallower{P})^{1/2}}$ & $\frac{\delta \B_{y}}{(4\pi\initvallower{P})^{1/2}}$ & $\frac{\delta \B_{z}}{(4\pi\initvallower{P})^{1/2}}$ & 
$\delta \ln{\left[ \frac{\gasden}{\initvalupper{\gasden}} \right]}$ & 
$\delta \ln{\left[ \frac{\dustden}{\initvalupper{\dustden}} \right]}$ \\ 
\hline\hline
{\bf Example:} & & & & & & & & & & & \\
Default & 0.075 & 0.025 & 0.017 & 0.086 & 0.063 & 0.32 & 0.075 & 0.025 & 6.4e-3 & 0.013 & 1.2 \\
$\beta$=1 & 0.067 & 0.015 & 6.6e-3 & 0.095 & 0.086 & 0.20 & 0.067 & 0.013 & 3.1e-3 & 3.7e-3 & 1.7 (1.4) \\
$\gamma$=5/3 & 0.088 & 0.039 & 0.020 & 0.12 & 0.11 & 0.33 & 0.089 & 0.037 & 0.013 & 8.2e-3 & 1.2 (1.3) \\
$\mu$=1e-3 & 4.3e-3 & 1.8e-3 & 6.7e-3 & 2.6e-3 & 3.2e-3 & 0.035 & 5.1e-3 & 1.8e-3 & 4.1e-3 & 4.9e-3 & 2.7 \\
$\mu$=0.1 & 0.18 & 0.15 & 0.11 & 0.32 & 0.30 & 0.41 & 0.17 & 0.12 & 0.056 & 0.055 & 0.89 \\
\hline
{\bf AGB:} & & & & & & & & & & & \\
S & 0.052 & 0.050 & 0.014 & 6.9e-4 & 9.4e-4 & 7.6e-4 & 8.6e-3 & 9.4e-3 & 0.049 & 0.052 & 0.75 \\
M & 0.031 & 0.010 & 0.038 & 0.10 & 6.8e-3 & 0.11 & 0.066 & 4.2e-3 & 0.071 & 0.074 & 0.90 \\
L & 1.0 & 0.50 & 1.2 & 1.7 & 0.50 & 1.9 & 1.2 & 0.51 & 1.2 & 0.85 & 0.88 \\
XL & 19 & 4.3 & 18 & 20 & 3.2 & 20 & 11 & 2.6 & 12 & 1.1 & 1.1 \\
\hline 
{\bf HII-near:} & & & & & & & & & & & \\
S & 0.060 & 0.057 & 0.013 & 3.5e-3 & 3.6e-3 & 3.7e-3 & 5.1e-3 & 4.7e-3 & 0.023 & 0.080 & 1.2 \\
M & 0.15 & 0.13 & 0.20 & 0.76 & 0.80 & 0.65 & 0.15 & 0.11 & 0.14 & 0.058 & 1.4 \\
L & 3.3 & 1.8 & 3.6 & 13 & 15 (31) & 12 & 3.7 & 1.9 & 3.5 & 1.2 & 1.5 (2) \\
L:$\tau$=10 & 4.7 (5.5) & 1.6 (2.1) & 4.8 (5.7) & 9.9 (8.0) & 3.0 (3.3) & 10 (8.6) & 4.1 & 1.9 & 4.0 & 1.3 (1.2) & 2.0 (2.4) \\
L:CC & 2.9 (13) & 2.6 (12) & 2.7 (12) & 29 (200) & 30 (200) & 26 (170) & 2.5 & 1.6 & 2.6 & 1.6 (0.9) & 1.8 (2.5) \\
L:PE & 6.9 (60) & 7.0 (61) & 6.3 (51) & 80 (1200) & 81 (1200) & 63 (940) & 4.3 & 4.0 & 5.0 & 2.0 (1.1) & 2.2 (2.2) \\
L:70$^{\circ}$ & 7.2 (8.5) & 2.1 (2.4) & 4.7 (5.7) & 15 (11) & 3.4 (3.1) & 8.5 (7.4) & 4.8 & 2.2 & 3.0 & 1.2 & 2.2 \\
L:20$^{\circ}$ & 2.8 (3.1) & 1.5 (1.9) & 7.1 (7.7) & 4.4 (4.2) & 2.7 (2.9) & 10.7 (9.5) & 3.0 & 1.7 & 6.8 & 1.3 & 1.9 (2.5) \\
L:$\mu$=1e-3 & 1.1 & 0.36 & 1.2 & 3.0 (2.3) & 1.2 & 3.2 (2.5) & 0.88 & 0.28 & 0.95 & 0.56 & 2.4 (3.2) \\
L:$\mu$=0.1 & 14 (17) & 3.4 (5.6) & 14 (17) & 15 (24) & 7.4 (24) & 15 (22) & 12 & 3.5 & 12 & 1.7 & 1.7 \\
\hline
{\bf HII-far:} & & & & & & & & & & & \\
S & 1.8e-3 & 1.8e-3 & 2.0e-3 & 4.7e-5 & 1.4e-5 & 2.1e-5 & 1.4e-3 & 1.4e-3 & 1.4e-3 & 4.3e-4 & 0.63 \\
M & 2.9e-3 & 2.7e-3 & 4.5e-3 & 1.5e-3 & 1.6e-3 & 3.5e-3 & 2.4e-3 & 1.9e-3 & 2.3e-3 & 7.4e-4 & 1.3 (2.5) \\
L & 0.55 & 0.20 & 0.58 & 1.1 & 0.90 & 1.0 & 0.43 & 0.19 & 0.44 & 0.28 & 1.5 (2.2) \\
\hline
{\bf WIM:} & & & & & & & & & & & \\
S & 9.8e-4 & 9.8e-4 & 1.1e-3 & 6.4e-4 & 6.4e-4 & 9.4e-6 & 9.7e-4 & 9.7e-4 & 1.1e-3 & 7.2e-4 & 0.49 \\
S:LoV & 9.8e-4 & 9.8e-4 & 1.1e-3 & 7.0e-4 & 7.0e-4 & 4.3e-6 & 9.7e-4 & 9.7e-4 & 8.6e-4 & 7.2e-4 & 0.42 \\
M &  9.9e-4 & 9.8e-4 & 1.1e-3 & 1.4e-3 & 1.4e-3 & 1.2e-4 & 9.7e-4 & 9.7e-4 & 9.4e-4 & 7.2e-4 & 0.36 \\
M:LoV & 1.0e-3 & 1.0e-3 & 1.1e-3 & 6.0e-4 & 6.0e-4 & 2.0e-4 & 9.7e-4 & 9.7e-4 & 8.5e-4 & 7.3e-4 & 0.25 \\
L & 0.032 & 0.021 & 0.088 & 5.3 & 5.3 & 0.085 & 3.6e-3 & 3.6e-3 & 3.9e-3 & 3.5e-3 & 0.37 \\
L:LoV & 1.5e-3 & 1.3e-3 & 2.3e-3 & 1.1e-3 & 7.3e-3 & 1.2e-3 & 1.1e-3 & 1.1e-3 & 8.0e-4 & 8.0e-4 & 0.25 \\
\hline
{\bf Corona:} & & & & & & & & & & & \\
S & 0.011 & 0.011 & 0.013 & 0.012 & 0.012 & 3.5e-5 & 0.011 & 0.011 & 0.013 & 0.060 & 0.63 \\
M & 0.012 & 0.012 & 0.014 & 0.62 & 0.63 & 0.010 & 0.011 & 0.011 & 0.022 & 0.060 & 0.86 \\
L & 0.27 & 0.24 & 0.086 & 24 & 24 & 3.1 & 0.076 & 0.069 & 0.15 & 0.066 & 0.57 \\
L:$\tau$=100 & 0.14 & 0.05 & 0.21 & 0.17 & 0.13 & 0.21 & 0.023 & 0.023 & 0.013 & 0.085 & 0.6 (0.8) \\
\hline
{\bf CGM:} & & & & & & & & & & & \\
Default & 0.83 & 0.82 & 0.63 & 23 & 23 & 22 & 0.040 & 0.040 & 0.042 & 0.34 & 0.28 \\
$\gamma$=5/3+PE & 2.5 & 2.5 & 2.0 & 63 & 63 & 63 & 0.14 & 0.14 & 0.13 & 0.56 & 0.25 \\
$\tau_{\rm low}$ & 0.031 & 0.031 & 0.036 & 2.0 & 2.0 & 1.9 & 0.024 & 0.021 & 0.019 & 0.021 & 0.27 \\
$\tau_{\rm low}$+PE & 0.047 & 0.046 & 0.057 & 2.9 & 2.9 & 2.9 & 0.035 & 0.032 & 0.023 & 0.025 & 0.27\\
\hline
\end{tabular}
\end{center}\vspace{-0.35cm}
\caption{Dispersion in various gas and dust quantities in the simulations, during the saturated state. We measure the rms dispersion $\delta X$ in quantity $X$, averaged over the last several snapshots in time for each run. We show gas velocity $\delta\gasvel^{x,\,y,\,z}$ in each direction (see \fref{fig:multiprop.demo} for axis convention), dust velocity $\delta\dustvel$, magnetic field $\delta\B$, gas density $\delta\ln{\gasden}$ and dust density $\delta\ln{\dustden}$. The dispersions shown are {\em mass-weighted} for $\gasvel$ and $\dustvel$ (so that the kinetic energy of gas is just $(1/2)\,M_{\rm gas}\,|\delta\gasvel|^{2}$, and likewise for dust), while dispersions for $\B$ and $\gasden$, $\dustden$ are volume weighted. Usually the mass and volume weights give similar values; where they differ substantially, the value in parenthesis gives the other (for more details, see Figs.~\ref{fig:pdfs.example}-\ref{fig:pdfs.HIInear.dustgas}). For each parameter set, we list the ``default'' boxes S/M/L/XL as defined in \tref{table:sims}, as well as the variants discussed in \ref{sec:results:eos}. For {\bf Example}, we show variations in $\beta$, $\gamma$, and $\mu$ from Figs.~\ref{fig:example.physics} and \ref{fig:example.dustgas}. For {\bf HII-near} L, we compare variations in the dust charge (``L:$\tau$=10''; $q_{\rm grain}$ increased by a factor $\sim4$), as well as different charging models, including un-saturated collisional charging (``L:CC'', $q_{\rm grain} \propto T$ with $\gamma=5/3$) and photo-electric charging (``L:PE''; $q_{\rm grain}\propto T/\rho^{1/2}$, $\gamma=1$), from \fref{fig:HIInear.physics.variations}. We also compare variations in the angle, $\initvallower{\B}\cdot\acc =  \cos\theta_{\B\acc}$ with $\theta_{\B\acc} = (70^{\circ},\,20^{\circ})$ (``L:70$^{\circ}$'', ``L:20$^{\circ}$'', respectively) from \fref{fig:HIInear.angle.variations}, and different dust-to-gas ratios $\mu$ from \fref{fig:HIInear.dustgas}. For {\bf WIM}, we compare variations of S/M/L with 5$\times$ lower $\acc$ and $\driftvel$ (``LoV'') from \fref{fig:WIM.lowVdust}. For {\bf CGM}, we show runs with lower $\tau$ ($\tau\approx 4500$ implying $\graincharge$ is $\sim 30\times$ lower; labeled ``$\tau_{\rm low}$''), and runs with un-saturated photo-electric (PE) charging and $\gamma=5/3$  (for both the default and low $\tau$ cases), from \fref{fig:CGM}.}\vspace{-0.65cm}
\label{table:saturation}
\end{table*}


%
%
\bpanelfigL
\figthreeXLlab{l0/im_3d2dProj_b_high_terrain_0026}{S: Early NL}{s0/im_3d2dProj_b_high_terrain_0035}{M: Early NL}{e0/im_3d2dProj_b_high_terrain_0160}{L: Early NL}
\figthreeXL{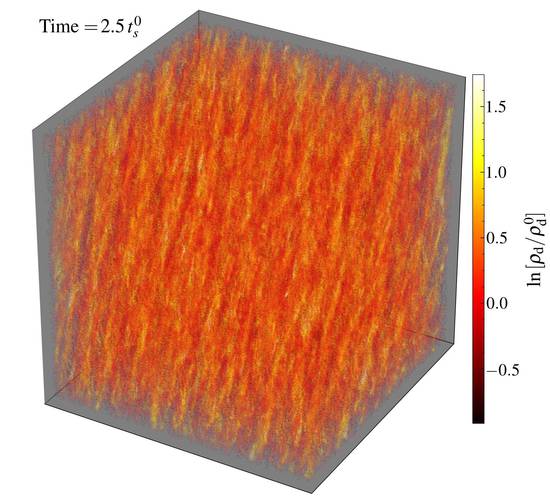}{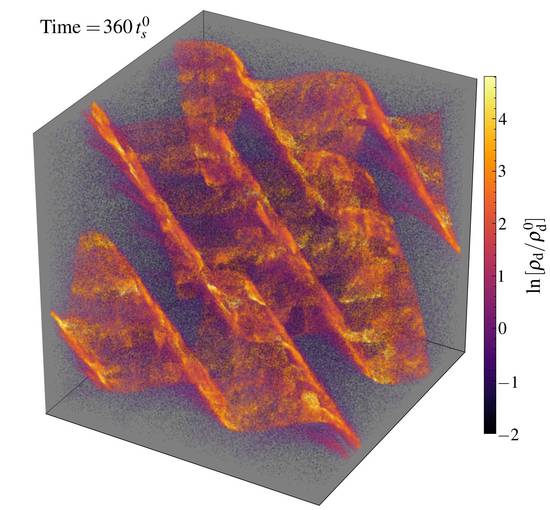}{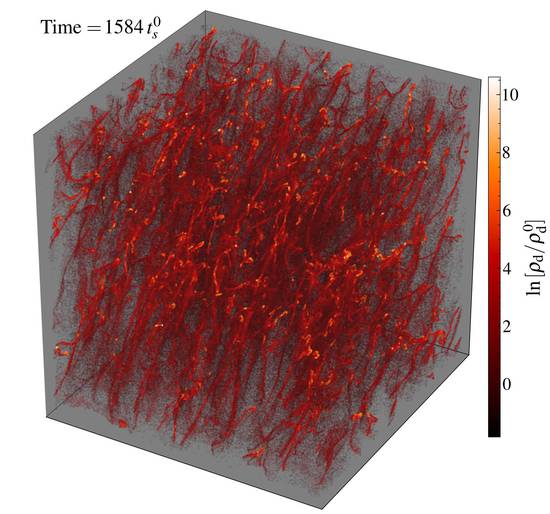} 
\figthreeXLlab{l0/im_3d2dProj_b_high_terrain_0066}{S: Saturated}{s0/im_3d2dProj_b_high_terrain_0060}{M: Saturated}{e0/im_3d2dProj_b_high_terrain_0260}{L: Saturated}
\figthreeXL{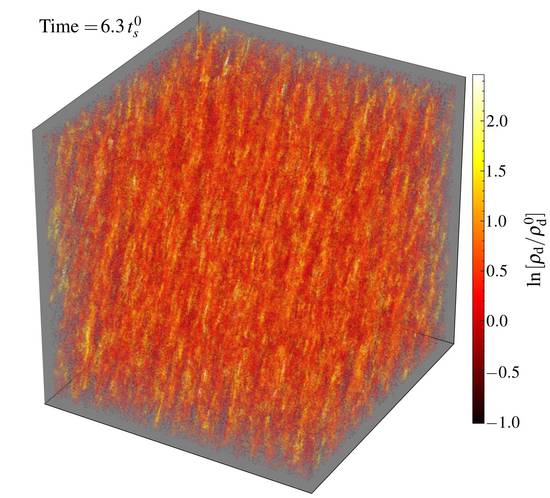}{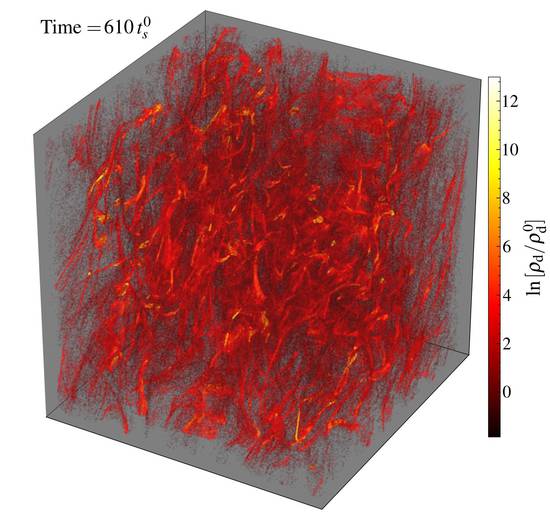}{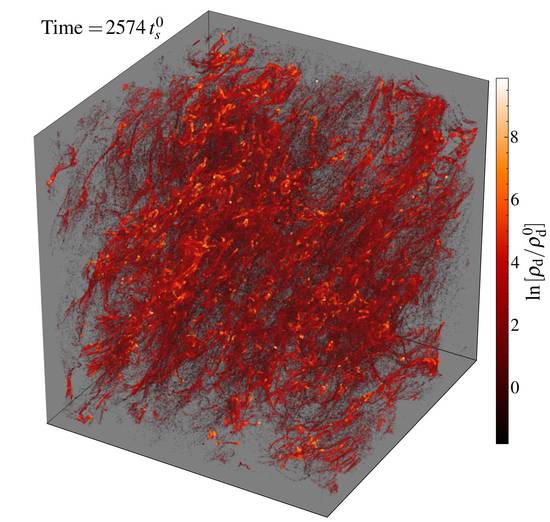} 
\caption{As \fref{fig:HIInear}, for {\bf HII-far}. Despite the drift being sub-sonic (no ``fast mode'' or acoustic resonances are possible), dramatic fluctuations still evolve on broadly similar timescales. The dust exhibits strong clumping even in cases (S/M) where the gas is nearly incompressible. Different scales again exhibit distinct structure: L is dominated by the drift-aligned quasi-sound and pressure free modes (direction $\driftvelhat - (\driftvelhat\cdot\hat{\B})\hat{\B}$); M by the slow and \Alf\ MHD-wave modes (direction $\hat{\B} - (\driftvelhat\cdot\hat{\B})\driftvelhat$), S by the slow-gyro modes (see Fig.~\ref{fig:rates}).\label{fig:HIIfar}}
\epanelfigL
%
%

%
%
\bpanelfigL
\figthreeXLlab{v0/im_3d2dProj_b_high_terrain_0052}{M: Early NL}{u0/im_3d2dProj_b_high_terrain_0050}{L: Early NL}{x0/im_3d2dProj_rho_high_terrain_0030}{XL: Early NL}
\figthreeXL{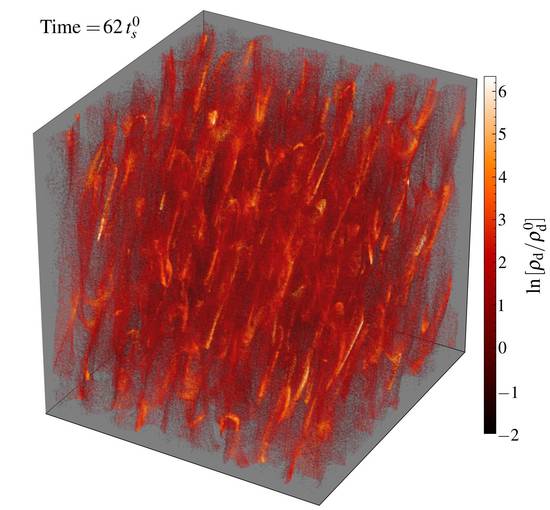}{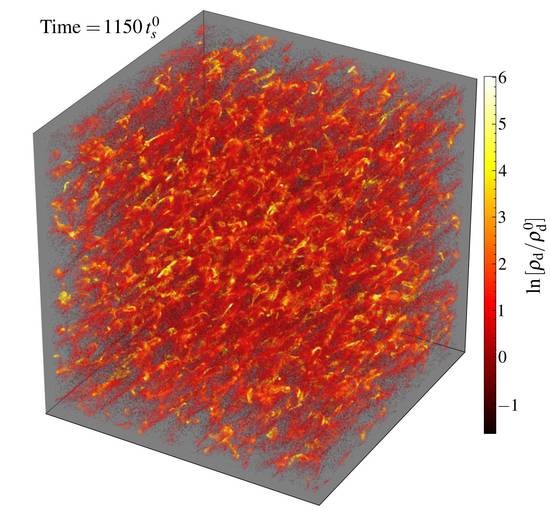}{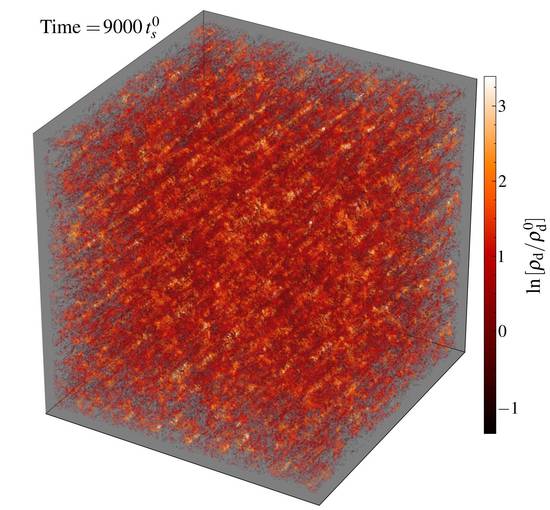}
\figthreeXLlab{v0/im_3d2dProj_b_high_terrain_0089}{M: Saturated}{u0/im_3d2dProj_b_high_terrain_0115}{L: Saturated}{x0/im_3d2dProj_rho_high_terrain_0100}{XL: Saturated}
\figthreeXL{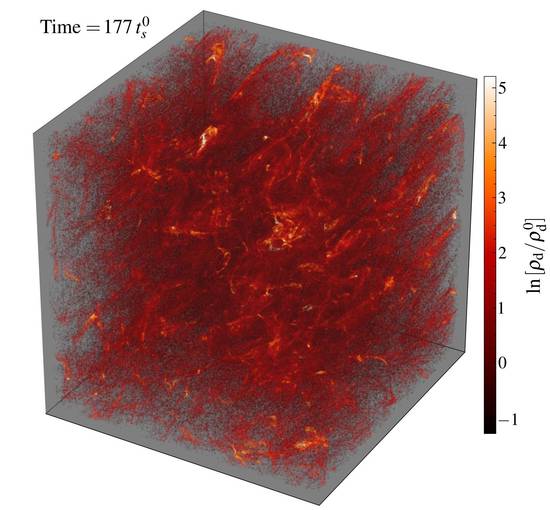}{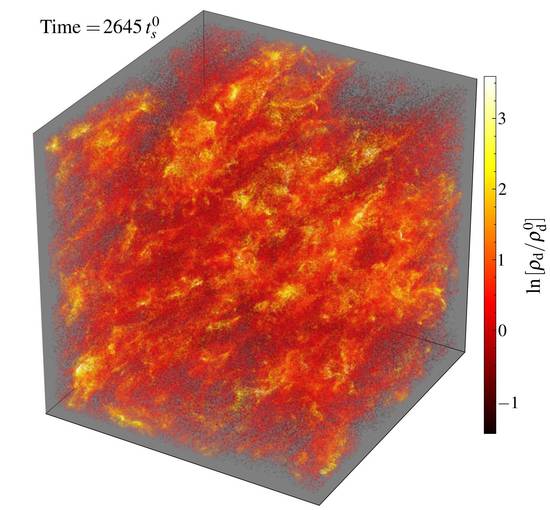}{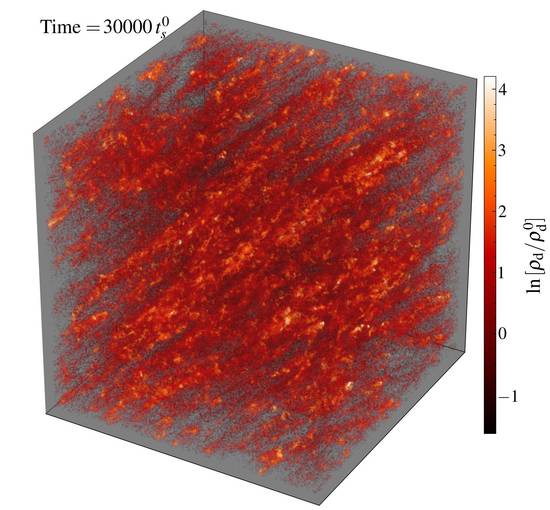} 
\caption{As \fref{fig:HIInear}, for {\bf AGB} (M/L/XL shown; S is morphologically similar to {\bf HII-near} S). Because this initial condition features weak Lorentz forces on dust ($\tau \ll 1$), the mode structure is simpler (dominated by the fast-MHD resonance at all $k$) and does not vary as dramatically with scale. Note here and in all similar figures, the colorbar for dust density only shows a fraction of the full dynamic range (empty regions reach lower densities, and small patches/clusters reach higher densities).\label{fig:AGB}}
\epanelfigL
%
%

%
%
\bpanelfigL
\figthreeXLlab{g0_5x/im_3d2dProj_b_high_terrain_0020}{S: Early NL}{q0_5x/im_3d2dProj_vz_high_terrain_0040}{M: Early NL} {f0_5x/im_3d2dProj_v_high_terrain_0150}{L: Early NL}
\figthreeXL{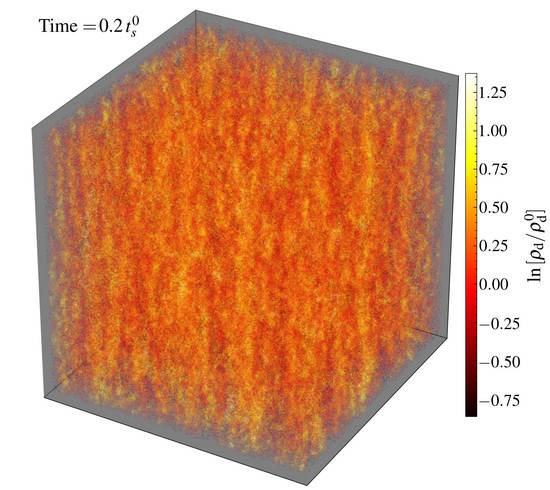} {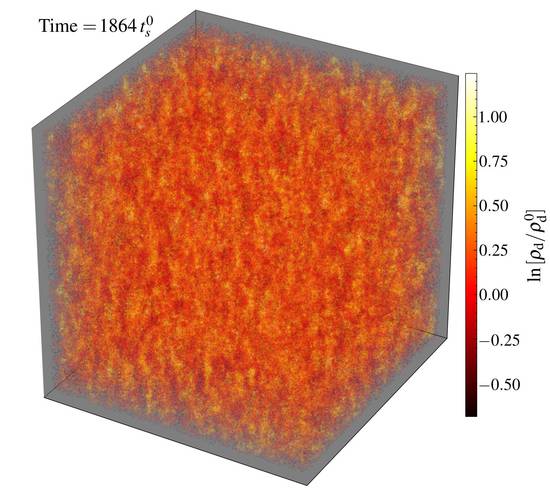} {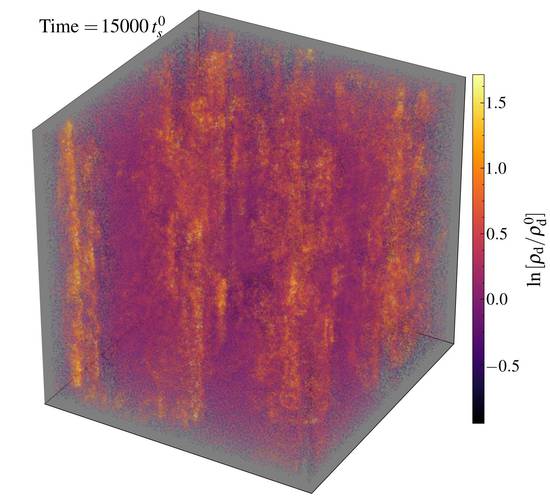} 
\figthreeXLlab{g0_5x/im_3d2dProj_b_high_terrain_0065}{S: Saturated}{q0_5x/im_3d2dProj_vz_high_terrain_0086}{M: Saturated}{f0_5x/im_3d2dProj_v_high_terrain_0193}{L: Saturated}
\figthreeXL{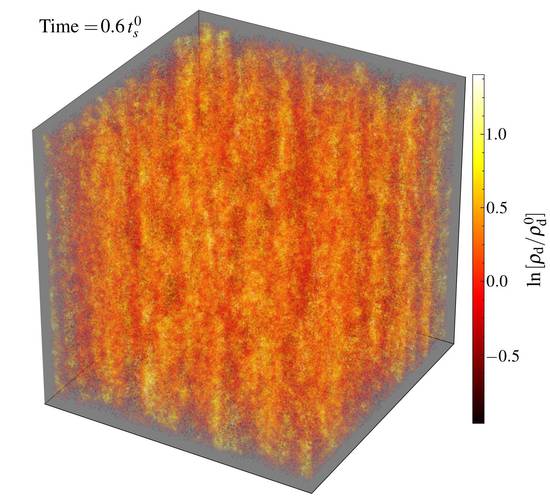} {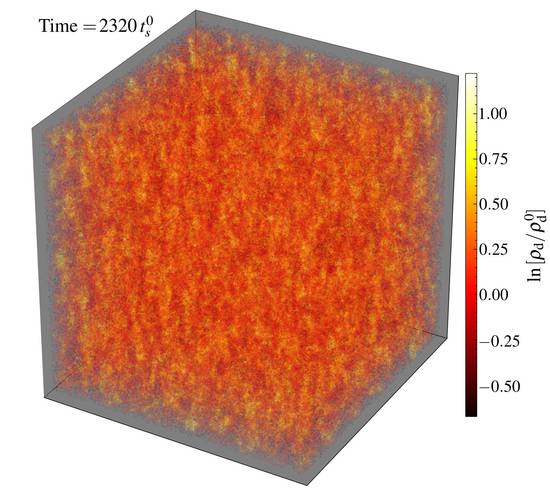} {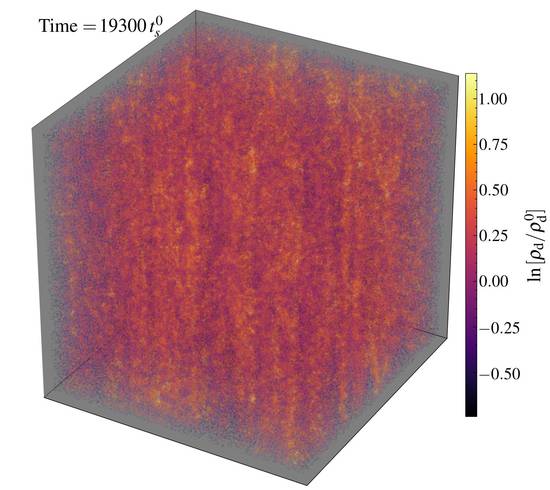} 
\caption{As \fref{fig:HIInear}, for {\bf WIM}. In box S the strong slow-gyro resonance produces rapid growth of dust ``columns'' parallel to the field, which collect in a ``granular'' nature in the direction perpendicular to $\B$. Box L features a slow-growing, almost-laminar mixing of large scale modes via the \Alf\ resonance, with the dust confined along $\B$ (moving the $\B$-fields nearly-incompressibly in the $xy$ plane). Box M features the weakest growth and saturated turbulence. \label{fig:WIM}}
\epanelfigL
%
%

%
%
\bpanelfigL
\figthreeXLlab{m0/im_3d2dProj_b_high_terrain_0070}{S: Early NL}{r0/im_3d2dProj_b_high_terrain_0139}{M: Early NL}{h0/im_3d2dProj_b_high_terrain_0010}{L: Early NL}
\figthreeXL{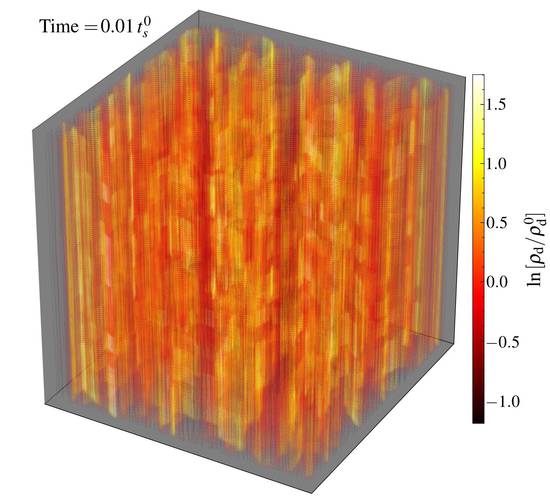}{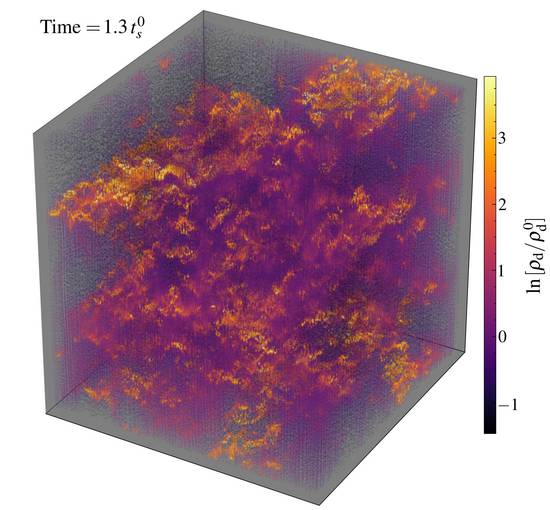}{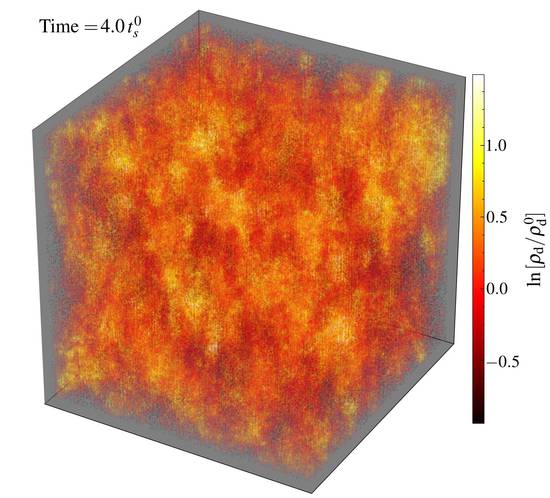} 
\figthreeXLlab{m0/im_3d2dProj_b_high_terrain_0177}{S: Saturated}{r0/im_3d2dProj_b_high_terrain_0175}{M: Saturated}{h0/im_3d2dProj_b_high_terrain_0089}{L: Saturated}
\figthreeXL{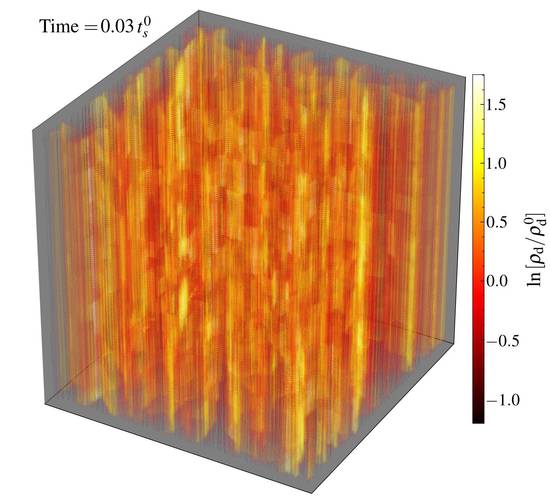}{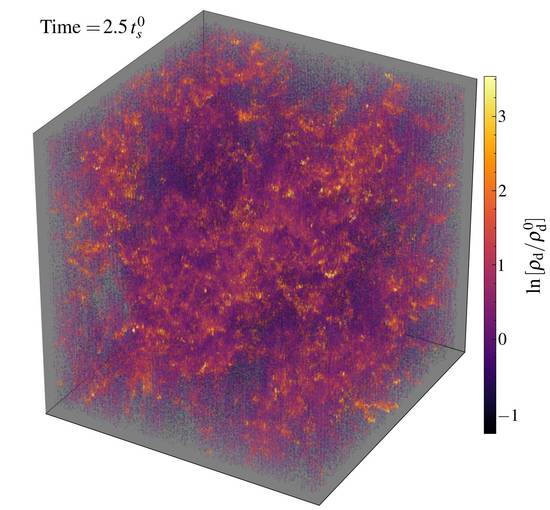}{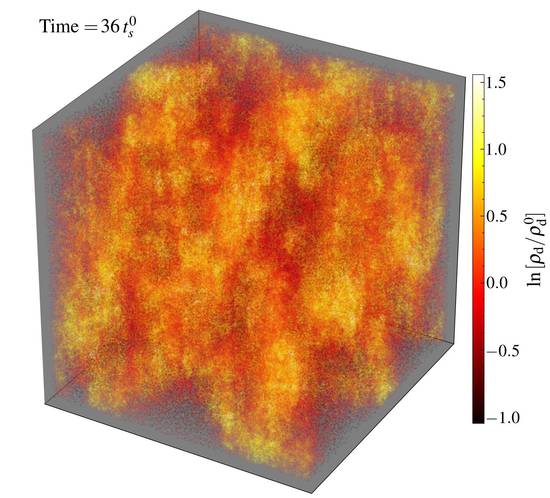} 
\caption{As \fref{fig:HIInear}, for {\bf Corona}. The strong magnetization means almost all grain motion is tightly bound to field lines. This produces the $\hat{\B}$-aligned ``granular'' structure in box S, and sheets of dust perpendicular to $\hat{\B}$ in M/L. Boxes M/L non-linearly transition to the ``disperse'' mode (see \ref{sec:sub.dispersed}), where the dust is driven to large isotropic velocity dispersions in the plane perpendicular to $\hat{\B}$ (representing large gyro orbits). \label{fig:Corona}}
\epanelfigL
%
%

\section{Parameter Space Explored}
\label{sec:params}

Because the possible parameter space is enormous (the six dimensions above, plus the choice of equation-of-state, drag law, and charge law), we do not attempt to survey it systematically, but instead choose several unique parameter combinations motivated by values one might expect in different astrophysical systems. For more information, see  \S~9 of  \paperone, which discusses  each of these physical regimes extensively (\fref{fig:parameter.space} is adapted from this). The baseline parameters for our ``default'' simulation set are given in \tref{table:sims} and illustrated in \fref{fig:parameter.space}. We give a brief description of each system in the following paragraphs.

The parameters {\bf HII-near} and {\bf HII-far} (from Fig.~7 in \paperone) correspond to plausible parameters in a massive HII region at two different radii from the star(s). Specifically, {\bf HII-near} corresponds to parameters expected for HII regions with $\sim 0.1\,\mu{\rm m}$ grains at a distance $r\sim0.1\,$pc from an OV-star or group or stars with luminosity $\sim 10^{6}\,L_{\sun}$ (providing the radiation pressure on the grains), local gas density $\sim 10^{4}\,{\rm cm^{-3}}$, temperature $\sim 10^{4}\,$K, plasma $\beta\sim10$, $\ts$ calculated including both Epstein and Coulomb drag terms, and the grain charge calculated including photo-electric and collisional charging in that radiative environment (and accounting for saturation of the grain charge). {\bf HII-far} takes the same system, and re-calculates all properties assuming a distance $r\sim1\,$pc from the star (assuming the gas density falls $\propto r^{-2}$ and $\beta$ is constant). The important difference, for our purposes, is that the equilibrium drift velocity $\driftvel$ is super-sonic in {\bf HII-near} (where the radiation field is stronger), and sub-sonic in {\bf HII-far}. 

Likewise {\bf AGB} is chosen to represent grains near the base (at $r \sim 100\,R_{\odot}$) of a dust-driven wind from a cool giant star, with a steady-state wind mass-loss rate $\dot{M}_{w} \sim 10^{-6}\,M_{\odot}\,{\rm yr^{-1}}$, wind velocity $v_{w} \sim 10\,{\rm km\,s^{-1}}$, temperature $T\sim 2000\,$K, $\beta\sim 1$, stellar luminosity $\sim 10^{3}\,L_{\odot}$ providing the grain acceleration, and $\sim 0.1\,\mu\,{\rm m}$ grains with a similar calculation of the charge and drag parameters. The distinguishing feature of this case is that the high gas densities ($\sim 10^{12}\,{\rm cm^{-3}}$) mean drag (collisional) grain coupling strongly dominates over Lorentz forces, so $\tau \ll 1$.\footnote{\label{foot:gyromodes} As shown in \paperone, at low $\tau \ll 1$, the gyro RDIs are formally present, but are generally stabilized, except at either very large $k$ or very specific angles where they become degenerate with the MHD-wave RDIs. For this reason we do not plot them in Fig.~\ref{fig:parameter.space} for the {\bf AGB} case.}

{\bf WIM} represents a random patch of the diffuse warm interstellar medium, with $\beta\sim1$, $T\sim 10^{4}\,$K, gas density $\sim 1\,{\rm cm^{-3}}$. We assume that the radiation energy density accelerating grains is comparable to the thermal energy density. Here $\tau \gg 1$ because the gas is diffuse, and the drift is highly sub-sonic.

{\bf Corona} has parameters similar to those expected near the base of the solar Corona: $r\sim R_{\odot}$, $L\sim L_{\odot}$ (with both gravity and radiation contributing to the drift), $n \sim 10^{8}\,{\rm cm^{-3}}$, $\beta\sim 0.001$, $T\sim 10^{6}\,$K, for $\sim 0.1\,\mu{\rm m}$ grains. Here $\beta \ll 1$ makes this regime distinct; this also means the drift is super-sonic but sub-\Alf{ic}, and $\tau \gg 1$. 

{\bf CGM} represents parameters that could be present in the circum-galactic medium at $r \sim 100\,$kpc from a bright quasar with $L\sim 10^{13}\,L_{\odot}$, and $n\sim 10^{-2}\,{\rm cm^{-3}}$, $T\sim 10^{6}\,$K, $\beta \sim 1000$. The low density means $\tau \sim 10^{3}-10^{7}$ is large, while the high luminosity provides a super-\Alf{ic} equilibrium drift, producing a distinct mode structure. 

{\bf Example} is not chosen to match a particular system, but is the case examined in \citet{seligman:2018.mhd.rdi.sims}, including a resolution study. It lies between {\bf HII-near} and {\bf WIM}, so we include it here for comparison.

For each of these parameter sets, \fref{fig:rates} shows the linear growth rates of the instabilities as a function of wavelength. For a given $k=|{\bf k}|$, these are calculated by choosing a mode angle $\hat{\bf k}$, which is either parallel to the drift or magnetic field, or satisfies one of the various resonant conditions at which a natural dust oscillation frequency matches one of the gas (e.g.\  dust advection and \Alf\ waves). For each class of resonance, we plot the maximum growth rate for each $\hat{\bf k}$. More information, including how to calculate the resonant angles, is given in \paperone.

Although all scales are unstable, different modes have growth rates that depend on the scale differently. In the physical systems explored above, the dynamic range between the largest global scales (where our local box treatment would be inappropriate) and the smallest scales where the instability operates is so large that it is impossible to resolve in a single box. Therefore, we construct a series of boxes for most cases here,\footnote{We only consider a single box-size in {\bf Example}, as the results resemble {\bf HII-far}, and {\bf CGM}, as the mode structure does not change over a large range of wavelengths.} each of which resolves a different range of wavelengths. 

The parameters presented above are plausible, but will vary within and between different astrophysical regimes. Parameters that depend on highly uncertain grain chemistry or physical structure, such as $\tau$ (which depends on the grain charge), are particularly uncertain. Other quantities held fixed in our study -- e.g.\ the gas equation-of-state, or the dust-to-gas ratio -- might vary between regimes. For this reason we consider a number of variations in gas and dust physics compared to the ``default'' simulations. These are noted in  \tref{table:saturation} and discussed in detail in \sref{sec:results:eos}.

The simulations presented here are scale free, and can represent any system that has the same dimensionless parameters. They are not strictly tied to the one specific physical system described above, but rather provide a well-motivated starting point for our study. For example, as discussed in \paperone, at different stages in the evolution of a supernova remnant (SNR), the SNR will pass through extended periods with parameters broadly similar to the {\bf AGB}, {\bf Corona}, and {\bf HII-near} regimes above (although details will differ). Similarly, many regions of GMCs and the obscuring dusty torus around AGN will feature parameters resembling the {\bf AGB} case. Some further intuition can be gained from \fref{fig:parameter.space} (although note that there are other important parameters that cannot be easily shown on a two-dimensional plot)

We broadly classify the behavior of the saturated simulations into two regimes. The first is the  ``clumped'' regime, where the dust becomes strongly  concentrated; this occurs, for example, in {\bf HII-near} (\fref{fig:HIInear}), {\bf HII-far} (\fref{fig:HIIfar}), {\bf Example} (\fref{fig:example.dustgas}), and {\bf AGB} (\fref{fig:AGB}). The second is the ``disperse'' regime, where the dust is expelled from certain regions at high velocity but remains relatively homogeneous; this occurs in (some scales in) {\bf WIM} (\fref{fig:WIM}), {\bf CGM} (\fref{fig:CGM}) and {\bf Corona} (\fref{fig:Corona}).



%
%
\begin{figure*}
\begin{center}
\includegraphics[width=0.75\textwidth]{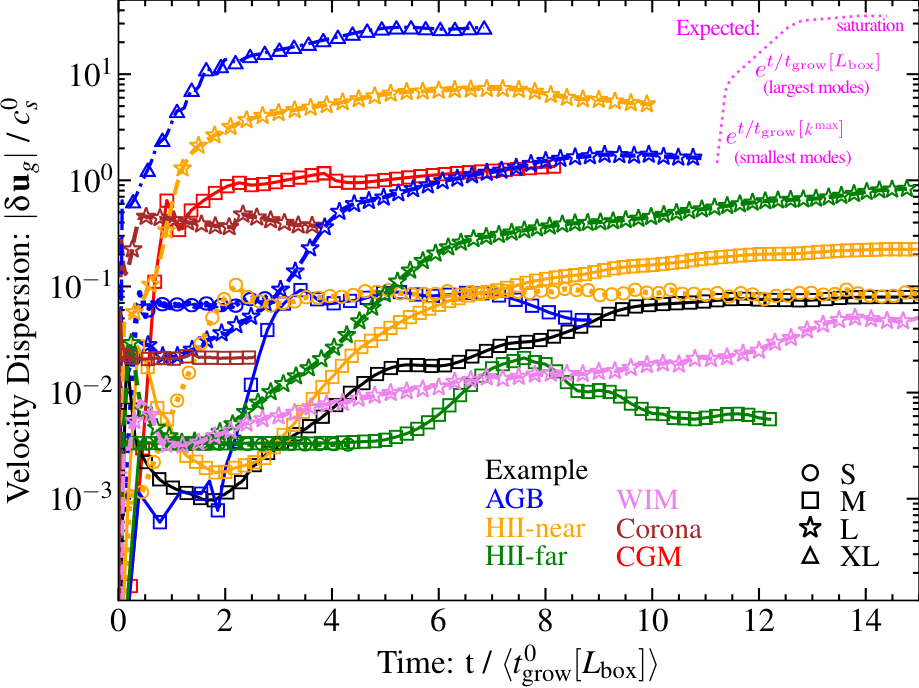}\vspace{-0.2cm}
\caption{Instability growth vs.\ time. 
We plot the (volume-weighted) gas velocity dispersion ($|\delta \gasvel |$, in units of sound speed $c_{s}^{0}$) as a function of time ($t$) for boxes in \tref{table:sims} (lines with points; as labeled). 
Times $t$ are normalized by the analytic linear growth time of the fastest-growing mode at the largest (box) scales: $\langle t_{\rm grow}^{0}[\Lbox] \rangle \equiv 1/{\rm MAX}(\omega[k=2\pi/\Lbox])$, calculated from the initial conditions at $|{\bf k}|=k^{\rm min}=2\pi/\Lbox$ (see \fref{fig:rates}). 
The smallest modes in the box ($\sim 100\times$ larger $k$) typically have $\sim10\times$ larger growth rates ($t_{\rm grow}[k^{\rm max}] \sim 0.1\,t_{\rm grow}[L_{\rm box}]$): these are generally the fastest-growing resolved modes and produce the initial rapid growth but saturate at smaller amplitude, so large-scale modes take over and grow at approximately the predicted rate (even though the box is often non-linear already), until global saturation.
Magenta dotted line illustrates the predicted qualitative behavior for this sequence. 
There are some exceptions, e.g.\ {\bf CGM} where small-scale modes grow so violently non-linear that there is no obviously ``large-scale'' mode domain. 
Nonetheless, noting  that $\langle t_{\rm grow}^{0}[\Lbox] \rangle$ varies by factors $\gtrsim 10^{6}$ in the examples plotted, and  that there is no unique growth rate for a given box (or even at a given $k$), we see  surprisingly robust agreement with the linear predictions for both initial ($\sim k^{\rm max}$-dominated) and late-time ($\sim k^{\rm min}$-dominated) growth rates.
\label{fig:growth}}
\end{center}
\end{figure*}
%
%

%
%
\begin{figure*}
\begin{center}
\includegraphics[width=\textwidth]{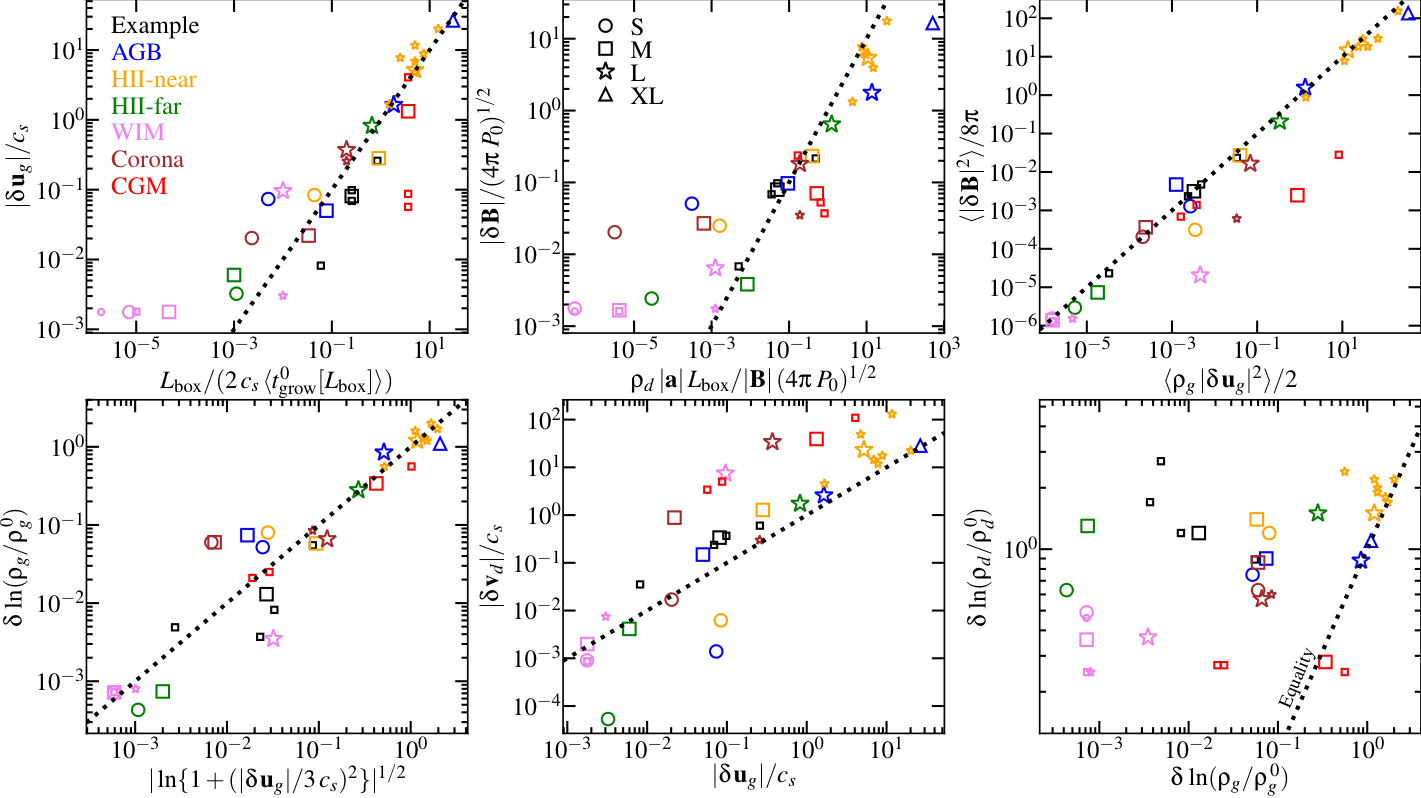}
\vspace{-0.35cm}
\caption{Correlations between the saturated fluctuations in gas and dust quantities. For each box (points as \fref{fig:growth}), we plot the fluctuations from \tref{table:saturation}, and compare simple analytic scalings (dotted lines). Small points denote the physics/parameter variations within each box from \tref{table:saturation}.
{\em Top Left:} Gas velocity dispersions vs.\ those predicted from quasi-linear theory (equating eddy turnover and linear growth times at the box scale). We neglect anisotropy and just take $|\delta \gasvel |$; here $\omega(\Lbox)=1/t_{\rm grow}^{0}[\Lbox]$ as defined in \fref{fig:growth}. While imperfect (e.g.\ this systematically under-predicts the turbulence in the ``S'' boxes) the scaling is order-of-magnitude plausible, with a couple notable outliers ({\bf CGM} $\tau_{\rm low}$ and {\bf WIM} S/M), all of which are in the ``disperse'' mode described in \S~\ref{sec:discussion}.
{\em Top Middle:} Magnetic fluctuations vs.\ expectation if magnetic tension balances the force from gas on dust. this provides an order-of-magnitude plausible scaling but also with significant scatter and additional outliers. 
{\em Top Right:} Kinetic and magnetic ``energy'' of fluctuations (considering {\em only} the fluctuating terms). These agree to within a factor of $\sim 3$ over most of the dynamic range.
{\em Bottom Left:} Gas density vs.\ velocity fluctuations. A scaling analogous to isothermal pure gas turbulence, $\delta \ln{(\gasden/\initvalupper{\gasden})} = \sqrt{\ln{[1 + (b\,|\delta \gasvel|/\cs)^{2}]}}$ provides a reasonable fit, with the range $b\approx0.2-1$. 
{\em Bottom Center:} Dust vs.\ gas velocity fluctuations. There is a clear correlation, but with very large ($\sim\pm1$\,dex) scatter. Usually $\delta \dustvel > \delta \gasvel$, with the notable exceptions of {\bf HII-far/near} S and {\bf AGB} S. 
{\em Bottom Right:} Dust vs.\ gas density fluctuations. There is essentially no correlation, except to require $\delta \dustden > \delta \gasden$. We see $\delta \dustden$ can be large even when the gas is nearly incompressible ($\delta\gasden < 10^{-3}$). 
See text for more discussion of each case (\sref{sec:discussion}).
\label{fig:correlations}}
\end{center}
\end{figure*}
%
%

\section{Results}
\label{sec:results}

\subsection{Default Simulation Set}
\label{sec:results:default}

We evolve each box well into its non-linear growth phase. As a first example, \fref{fig:multiprop.demo} shows visualizations\footnote{The gas+dust visualization is constructed by interpolating the continuous gas properties onto the $x/y/z$ axis surfaces, and taking all dust ``super-particles'' that lie within a thin slice of width equal to roughly the median inter-particle separation and projecting them onto the surface. The dust-density visualization is constructed by first calculating the dust density in the vicinity of each dust particle element using a kernel density estimator as in \citet{moseley:2018.acoustic.rdi.sims}, then plotting each dust particle in the projected 3D space, color-coded by the density (with a constant transparency). The range on the color-bars is scaled to include some fraction (typically $>90\%$) of the plotted elements, in order to show contrast, but there are always some elements with higher/lower values.} of the dust and gas in the large {\bf HII-near} simulation box, at a time well into the  non-linear phase of the evolution. Large fluctuations and coherent structure are visible in all gas quantities and in the dust. 

\begin{figure}
\begin{center}
\includegraphics[width=0.49\columnwidth]{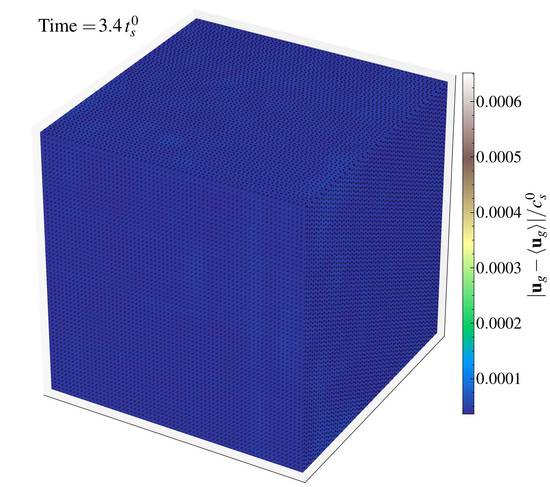} 
\includegraphics[width=0.49\columnwidth]{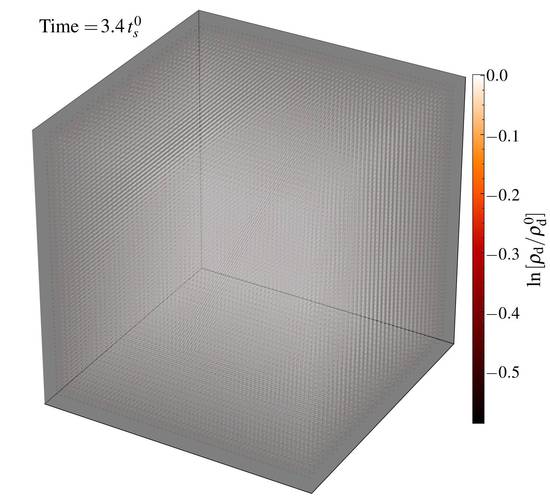} 
\caption{Visualizations for a variation of the {\bf Example} run,  removing all grain charge (Lorentz forces on grains). As shown in \citet{seligman:2018.mhd.rdi.sims}, the linear growth rate becomes multiple orders of magnitude smaller in this limit. We see here that at similar times to those when the default {\bf Example} run has gone strongly non-linear (\fref{fig:example.dustgas}), no structure has developed.
\label{fig:example.uncharged}}
\end{center}
\end{figure}

Figures \ref{fig:HIInear}, \ref{fig:HIIfar}, \ref{fig:AGB}, \ref{fig:WIM}, and \ref{fig:Corona} show the morphologies of gas and dust in the default {\bf HII-near}, {\bf HII-far}, {\bf AGB}, {\bf WIM}, and {\bf Corona} sets, respectively. We show gas and dust properties during two different simulation times, corresponding to the early non-linear and saturated phases of evolution. For each parameter set, we show the boxes of different sizes in parallel columns. Sequential simulations from left-to-right are  $\sim 300\times$ larger in size scale;  an entire box at the left is approximately $\sim 1/2$ of a pixel/element in the box to its immediate right. We only show the gas quantities that exhibit the most obvious morphological structure. \fref{fig:example.physics}, \ref{fig:CGM}, \ref{fig:HIInear.physics.variations}, and \ref{fig:example.uncharged} show the same for the {\bf Example}, {\bf CGM} and {\bf HII-near} sets. Parallel columns now compare compare physical variations (e.g.\ changing the equation of state of the gas, or dust charge scaling). These are discussed further in \sref{fig:example.physics}.

Figure \ref{fig:growth} show the magnitude of fluctuations in dust velocity versus time for each simulation. All dust fluctuations are plotted in units of the expected linear-theory growth timescales, which allows us to compare all of the simulations on the same time axis. Since there is no single growth time in a simulation (see \fref{fig:rates}),  we consider the maximum growth rate of modes at the box scale.\footnote{Specifically we find the minimum ``box scale growth time'' $t_{\rm grow}(\Lbox) \equiv 1/\Im{(\omega_{\rm max})}$ where $\omega_{\rm max}$ corresponds to the mode with $|{\bf k}| = 2\pi/\Lbox$ that has the largest positive value of $\Im(\omega)$ (marginalizing over direction $\hat{\bf k}$).} In most cases, modes near the resolution scale ($\Lbox/N$) grow $\sim 10\times$ faster  than those at the box scale,\footnote{This estimate is not always accurate and depends on details of the dispersion relation. E.g.\ in some cases, a different mode starts to dominate at some mid-range scale.} and so the initial rapid growth is dominated by these small-scale modes. However, these modes  generally saturate at a lower amplitude, so the later growth is then dominated by the box-scale modes, at rates in agreement with linear theory. This is surprising given the ambiguity of defining the total growth rate, and the fact that small-scale modes have already become non-linear. All of the boxes eventually reach saturation, with the fluctuations in all properties in a statistical steady-state. The cases in \fref{fig:growth} that continue off the plot have been evolved to longer times to ensure that they are approximately saturated.\footnote{We have verified that all simulation quantities within a given simulation saturate on a similar timescale, as shown in \fref{fig:growth}. Explicit demonstrations of this, as well as plots of the full PDFs of the salient quantities vs.\ time, for representative simulations, are shown in both \citet{moseley:2018.acoustic.rdi.sims} and \citet{seligman:2018.mhd.rdi.sims}.}

In  \tref{table:saturation}, we provide the magnitude of the time-averaged saturated fluctuations\footnote{We define the fluctuation $\delta X$ in quantity $X$ as the rms ($1\sigma$) deviation, i.e.\ $\delta X \equiv \langle (X - \langle X \rangle)^{2} \rangle^{1/2}$, where $\langle U \rangle \equiv ( \int U\,\varpi\,d^{3}{\bf x} ) / (\int \varpi\,d^{3}{\bf x})$ and $\varpi({\bf x})$ is a weight. For $\gasvel$, $\dustvel$, and $\B$, it is most physical to relate the fluctuations to the energy in each component. So, for $\gasvel$ we use the gas-mass-weighted average ($\varpi = \gasden$, such that the total kinetic energy of fluctuations is $(1/2)\,M_{\rm gas}^{\rm box}\,|\delta \gasvel|^{2}$), for $\dustvel$ we likewise use the dust-mass-weighted average ($\varpi = \dustden$, so dust kinetic energy is $(\mu/2)\,M_{\rm gas}^{\rm box}\,|\delta \dustvel|^{2}$), and for $\B$ we use the volume average ($\varpi=1$, so magnetic energy is $V_{\rm box}\,|\delta \B|^{2}/8\pi$). For $\gasden$, $\dustden$, and $\dustden/\gasden$ we quote the volume-averaged fluctuations ($\varpi=1$). The differences between mass and volume weighting are discussed further below. These are all defined in the center-of-mass (i.e.\ co-moving or free-accelerating) frame.} in each component of $\gasvel$, $\dustvel$, $\B$, $\gasden$, $\dustden$. Figure \ref{fig:correlations} plots these statistically averaged quantities against one another in various forms, and against some expectations from quasi-linear theory for the saturated state. This is discussed further in \sref{sec:discussion} below.

Figures \ref{fig:example.physics}, \ref{fig:CGM}, \ref{fig:WIM.lowVdust}, \ref{fig:HIInear.physics.variations}, \ref{fig:HIInear.angle.variations}, \ref{fig:example.uncharged}, \ref{fig:example.resolution}, consider further comparisons of different physics and parameter variations, as discussed below. 

Figures \ref{fig:pdfs.example}, \ref{fig:pdfs.agb}, \ref{fig:pdfs.HIInear}, \ref{fig:pdfs.HIIfar}, \ref{fig:pdfs.Corona},  \ref{fig:pdfs.WIM},  \ref{fig:pdfs.CGM} examine various statistics of each run in more detail, plotting the probability distribution functions (PDFs) of  fluctuations of different quantities  in the saturated state.\footnote{Given our Lagrangian numerical method, at our default simulation resolution, dust under-densities $\rho_{d}/\rho_{d}^{0} \ll 10^{-6}$ cannot be resolved, but these are much smaller than any plotted in Figures \ref{fig:pdfs.example}-\ref{fig:pdfs.CGM} and much smaller than the typical fluctuations in \tref{table:saturation}. There is no formal upper limit to the maximum resolved concentration, but the dust becomes increasingly over-resolved relative to the gas at very large $\rho_{d}/\rho_{g} \gg 10^{6}$ -- however this is precisely the regime where we expect other physics to dominate, discussed below.}

%
%
\bpanelfigL
\figthreeXLlab{jonodefault_b0_128/im_3d2dProj_v_high_terrain_0110}{Default: Early}{b0_N128_gamma53/im_3d2dProj_v_high_terrain_0110}{$\gamma=5/3$: Early} {b0_N128_beta1traditionalbeta/im_3d2dProj_v_high_terrain_0160}{$\beta=1$: Early}
\figthreeXL{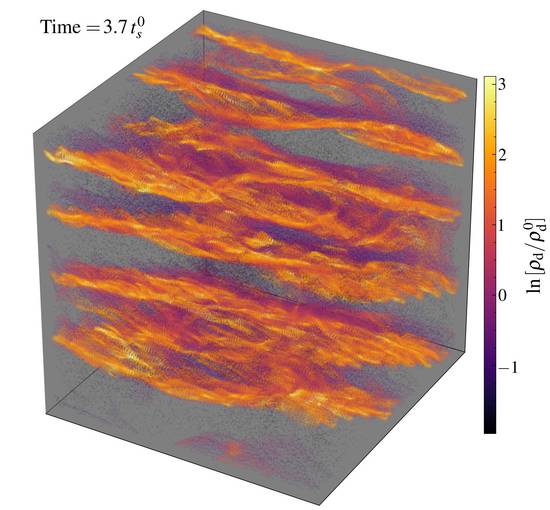} {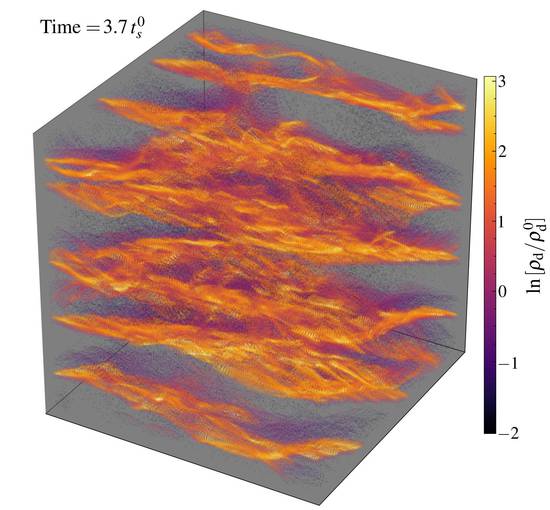} {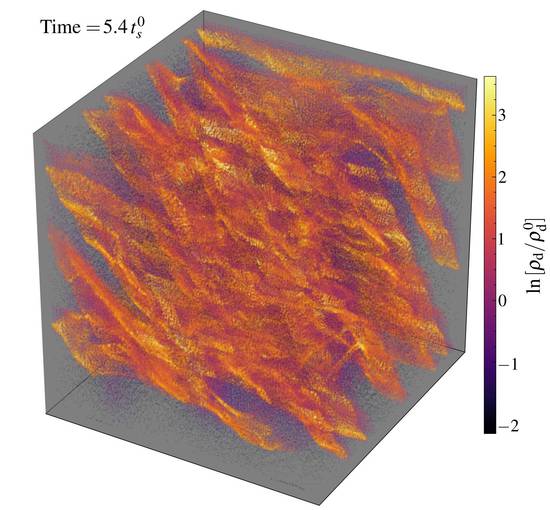} 
\figthreeXLlab{jonodefault_b0_128/im_3d2dProj_v_high_terrain_0300}{Default: Saturated}{b0_N128_gamma53/im_3d2dProj_v_high_terrain_0300}{$\gamma=5/3$: Saturated}{b0_N128_beta1traditionalbeta/im_3d2dProj_v_high_terrain_0360}{$\beta=1$: Saturated}
\figthreeXL{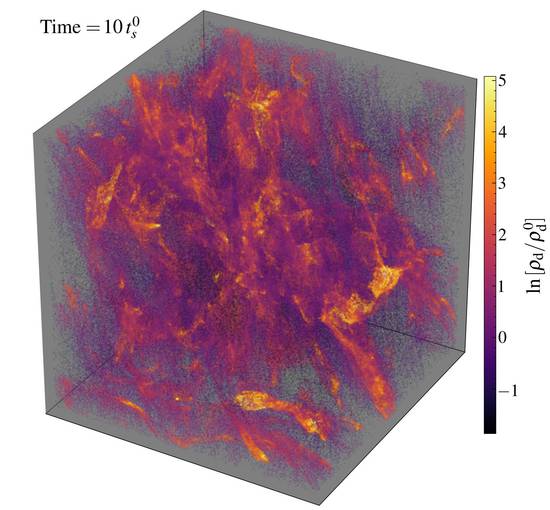} {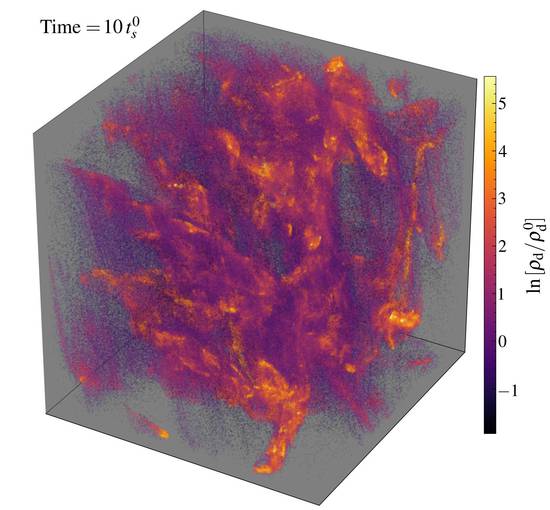} {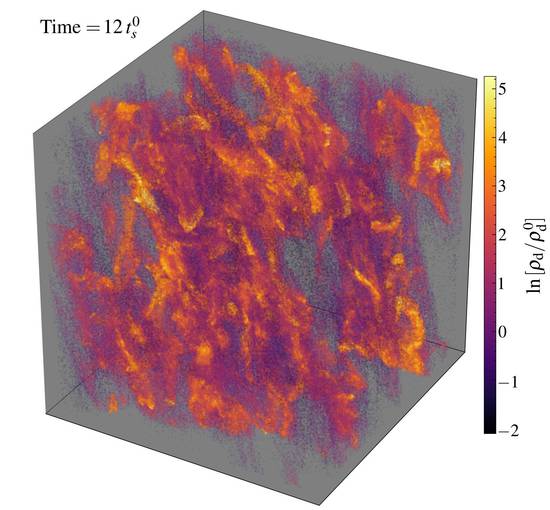} 
\caption{Dust and gas in our {\bf Example} runs, as \fref{fig:HIInear}, but comparing physics variations instead of box sizes. We compare the ``Default'' ({\em left}) run (with parameters in \tref{table:sims} and $\gamma=1$), to one with $\gamma=5/3$ ({\em center}), and one with $\beta = 1$ ({\em right}), all with otherwise equal parameters. Because the turbulence is only weakly-compressible in the default ($\gamma=1$) case, $\gamma=5/3$ has little effect. Shifting $\beta$ shifts the resonant mode angle, and gives slightly lower growth rates (hence the slightly later times plotted), but the behavior is qualitatively similar.\label{fig:example.physics}}
\epanelfigL
%
%

%
%
\bpanelfigL
\figthreeXLlab{k0_tauHi/im_3d2dProj_rho_high_terrain_0075}{Large $\graincharge$} {k0_pelecscalecharge/im_3d2dProj_rho_high_terrain_0033}{$\graincharge\propto T$} {k0_pelecscalecharge_2/im_3d2dProj_rho_high_terrain_0040} {$\graincharge\propto T^{1/2}/\rho$}
\figthreeXL{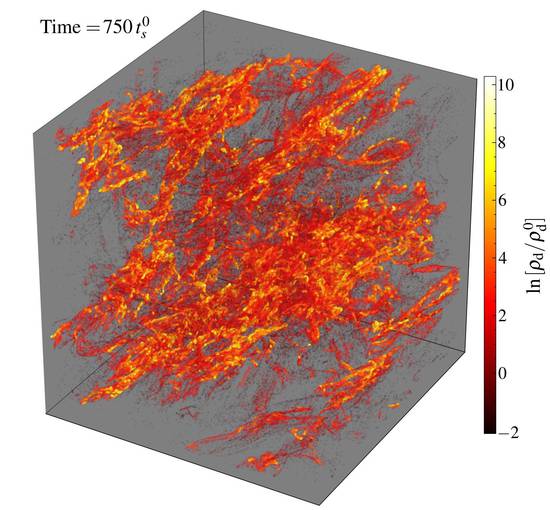} {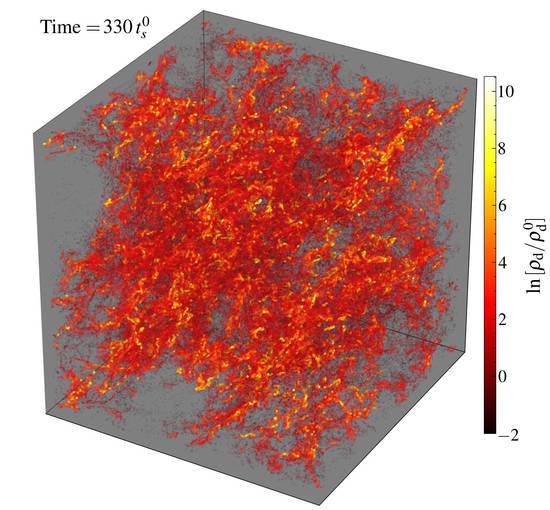} {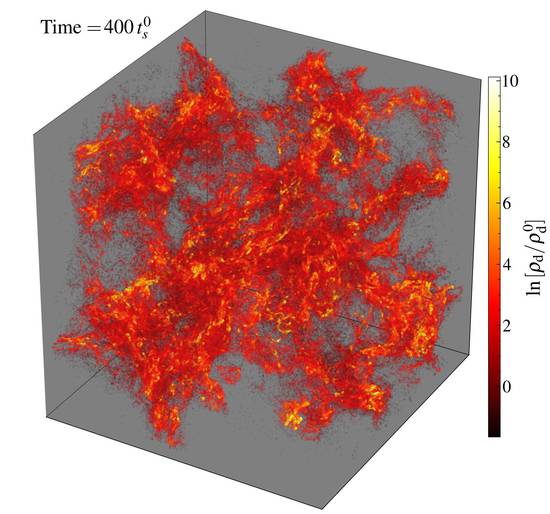} 
\caption{Comparing variations of the dust charge and charge law in the saturated turbulence of our {\bf HII-near} L box (see bottom panels of \fref{fig:HIInear} and \tref{table:saturation}). We compare (1) increasing the grain charge by an arbitrary factor $\sim4$ ($\gamma=1$, $\tau\approx 10$, constant/saturated charge; {\em left}); (2)  assuming adiabatic gas (so the temperature can change) and un-saturated collisional charging  ($\gamma=5/3$; $q_{\rm grain} \propto T$; {\em middle}); (3) isothermal gas assuming un-saturated photo-electric charging dominates ($q_{\rm grain} \propto T^{1/2}/\rho$; {\em right}). The linear and early non-linear stages are very similar. Increasing charge  leads to denser, more compact dust structures. In the late/saturated state, the run with the ``photo-electric'' charging law produces more coherent dust structures with a distinct topology, and gas more obviously ``dragged along'' to higher densities and velocities where the dust is overdense. All these  runs produce stronger dust clumping in the saturated state, and develop fully non-linear structure more rapidly, than the default {\bf HII-near} L simulation.\label{fig:HIInear.physics.variations}}
\epanelfigL
\bpanelfigL
\figthreeXLlab{j0_tauHi/im_3d2dProj_rho_high_terrain_0005}{Default}{j0_tauHi_gamma53_pelecscalecharge/im_3d2dProj_rho_high_terrain_0005}{Photo-electric}{j0_boris/im_3d2dProj_rho_high_terrain_0012}{Low $\graincharge$}
\figthreeXL{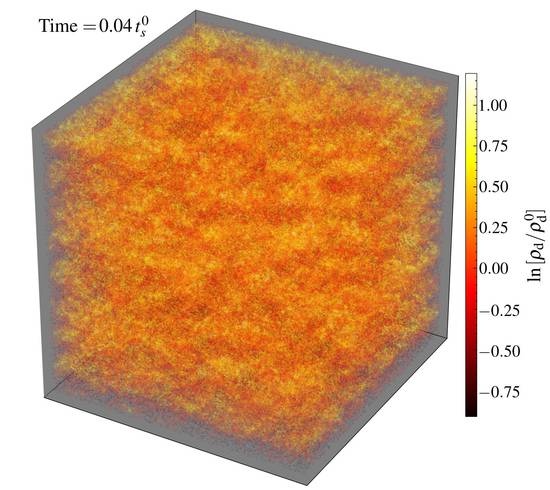}{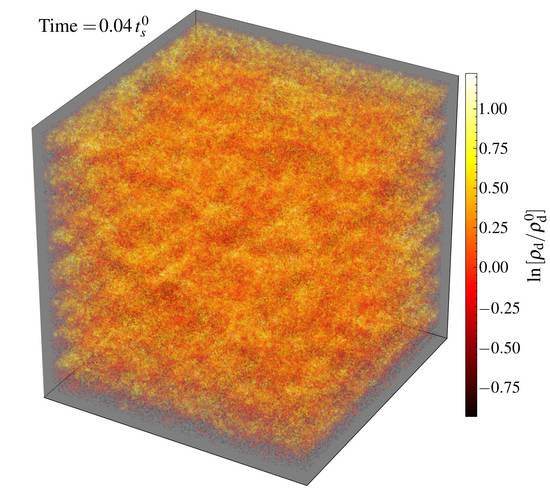}{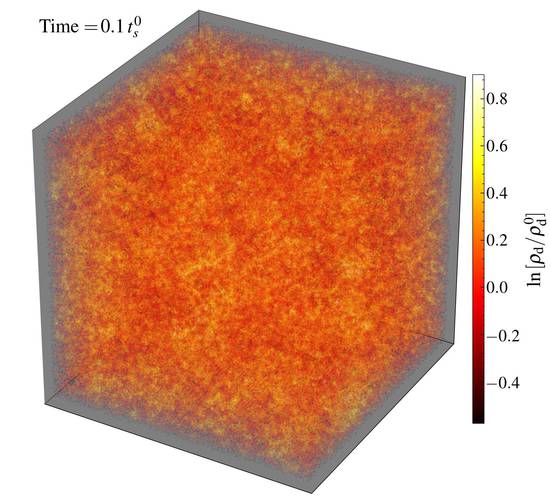}
\caption{Comparison of  {\bf CGM} runs with different physics choices in the early non-linear state (as in top panels of Figs.~\ref{fig:HIIfar}--\ref{fig:Corona}). Left panels show  our ``default'' case ($\gamma=1$, constant/saturated charge); middle panels show a simulation with an adiabatic EOS and the charge following the un-saturated, photo-electric-charging-dominated expectation ($\gamma=5/3$, $q_{\rm grain} \propto T^{1/2}/\gasden$); and right panels show a case with lower grain charge ($q_{\rm grain}$ and $\tau$ $30\times$ lower). We have also run the low-$\tau$ case with the photo-electric charge scaling, but its visual morphology is very similar (not shown). The qualitative behavior in each case is similar, though the ``sheets'' arising initially from the parallel mode are more evident in the low-$\tau$ cases since the dust (with weaker Lorentz forces) is accelerated less-strongly to large $\delta\dustvel$. In each case the dust saturates in the ``disperse'' mode with nearly-isotropic $\delta\dustvel$ (and $\delta \gasvel$), and $|\delta\dustvel| \sim |\initvalupper{\driftvel} |$.  \label{fig:CGM}}
\epanelfigL
%
%

%
%
\bpanelfigL
\figthreeXLlab{g0/im_3d2dProj_b_high_terrain_0077}{{\bf WIM}-S:LoV} {f0/im_3d2dProj_v_high_terrain_0257}{{\bf WIM}-L:LoV} {h0_tauLo/im_3d2dProj_vz_high_terrain_0142}{{\bf Corona}:$\tau=100$}
\figthreeXL{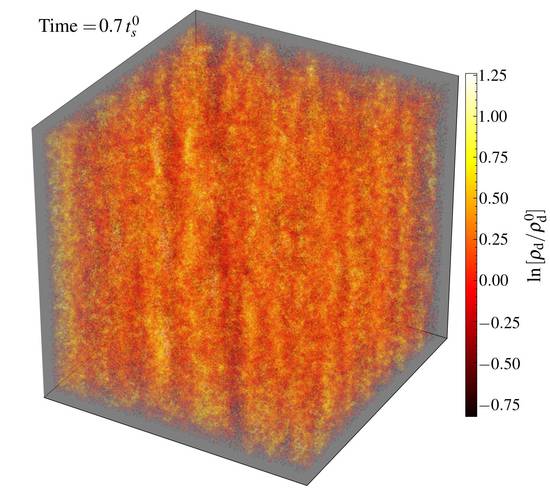} {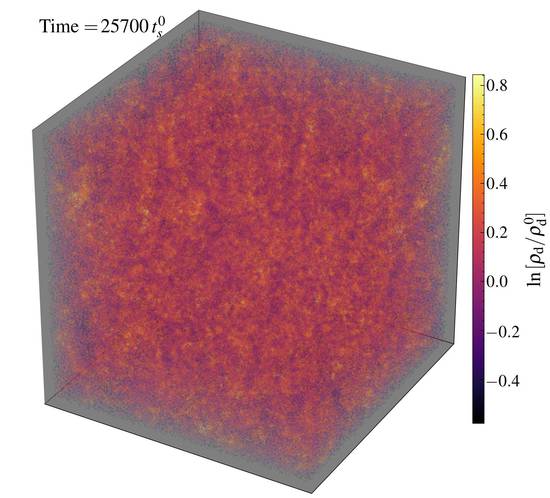} {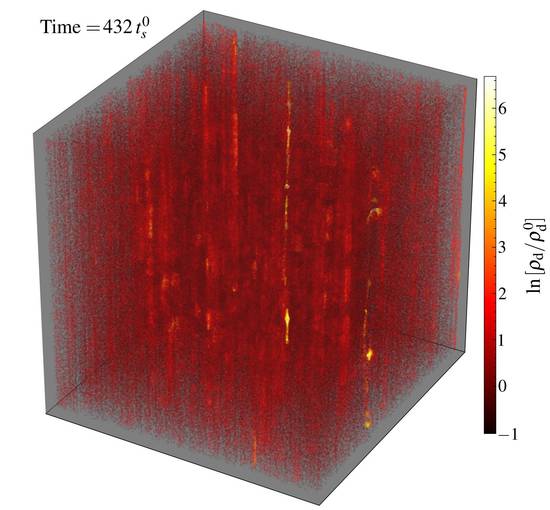} 
\caption{{\em Left \&\ Middle:} As in the bottom panels of \fref{fig:WIM}, showing  the  {\bf WIM}-S:LoV ({\em left}) and {\bf WIM}-L:LoV ({\em middle}) variations at the {\bf WIM} parameters in the saturated state. These variations have lower accelerations $\tilde{a}$ by a factor of $5$ and, correspondingly, lower drift velocity, $|\initvalupper{\driftvel}|/\initvalupper{\cs}=0.01$. The same qualitative features are evident. The linear growth rate and granular structure in box S are nearly identical in both cases.  Box L exhibits similar features, with large-scale, almost laminar mixing modes, but their saturated amplitude in the gas is weaker by a factor $\sim 25$ at the same time. Note that box M has $\sim 5\times$ lower growth rate and very weak turbulence, so is not shown. {\em Right:} As in \fref{fig:Corona}, showing box {\bf Corona}-L:$\tau$=100, which is the same as {\bf Corona}-L but with $\sim 30\times$ lower $q_{\rm grain}$ and $\tau$. The $\tau$ and  box size are similar to box {\bf WIM}-L, and although the drift is super-sonic it is significantly sub-\Alf{ic}. This causes the emergent morphology to qualitatively resemble {\bf WIM}-L more than {\bf Corona}-L. However note the much denser ``nodes'' of dust that appear. 
\label{fig:WIM.lowVdust}} 
\epanelfigL
%
%

%
%
\bpanelfigL
\figthreeXLlab{k0_angle1/im_3d2dProj_rho_high_terrain_0080}{$\theta_{\B\acc}=70^{\circ}$} {k0/im_3d2dProj_rho_high_terrain_0080}{$\theta_{\B\acc}=45^{\circ}$} {k0_angle2/im_3d2dProj_rho_high_terrain_0100} {$\theta_{\B\acc}=20^{\circ}$} 
\figthreeXL{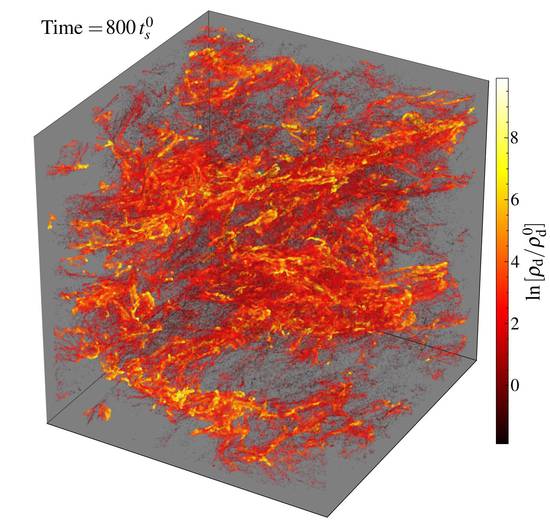} {k0/im_3dDust_hot_0080} {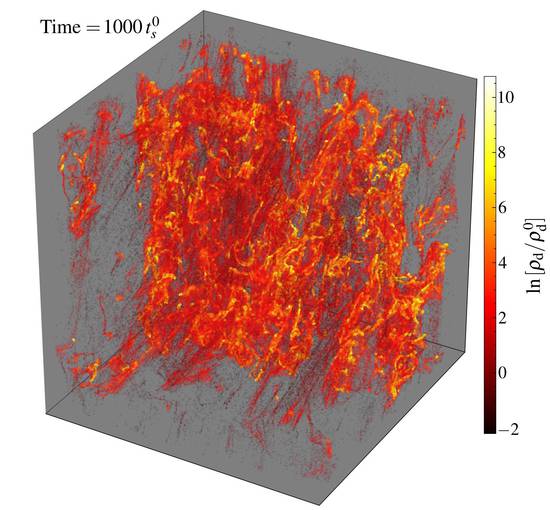} 
\caption{As in bottom panels of \fref{fig:HIInear},  comparing variations of the acceleration direction in the {\bf HII-near} L box in the saturated state. We take our default run (shown in the {\em middle} panels) and, keeping all other parameters fixed, vary the direction of the external net acceleration between dust and gas ($\acc$) so $\theta_{\B\acc}=(70^{\circ} , 45^{\circ} , 20^{\circ} )$ at ({\em left}, {\em middle}, {\em right}). The case with $\acc$ closer to parallel $\initvallower{\B}$ ({\em right}) produces slightly lower initial growth rates, so grows more slowly, even though the equilibrium drift velocity $|\driftvel |/\cs$ is slightly larger. However, it saturates with slightly stronger fluctuations. The case with $\acc$ closer to perpendicular $\B$ grows slightly faster (despite smaller $|\driftvel |/\cs$). The qualitative structure is nonetheless broadly similar in each case. \label{fig:HIInear.angle.variations}}
\epanelfigL
%
%

%
%
\bpanelfigL
\figthreeXLlab{b0_mu/im_3d2dProj_v_high_terrain_0029_alt}{Early: $\mu=0.001$}{jonodefault_b0_128/im_3d2dProj_v_high_terrain_0110}{Early: $\mu=0.01$}{b0_mu2/im_3d2dProj_v_high_terrain_0018}{Early: $\mu=0.1$}
\figthreeXL{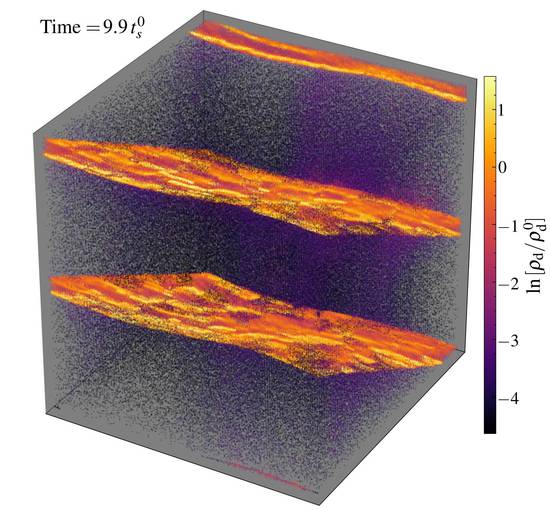}{jonodefault_b0_128/im_3dDust_inferno_0110}{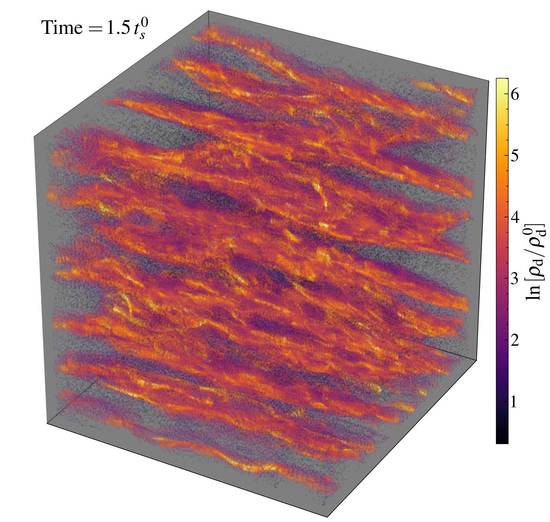}
\figthreeXLlab{b0_mu/im_3d2dProj_v_high_terrain_0169_alt}{Sat: $\mu=0.001$}{jonodefault_b0_128/im_3d2dProj_v_high_terrain_0300}{Sat: $\mu=0.01$}{b0_mu2/im_3d2dProj_v_high_terrain_0050}{Sat: $\mu=0.1$}
\figthreeXL{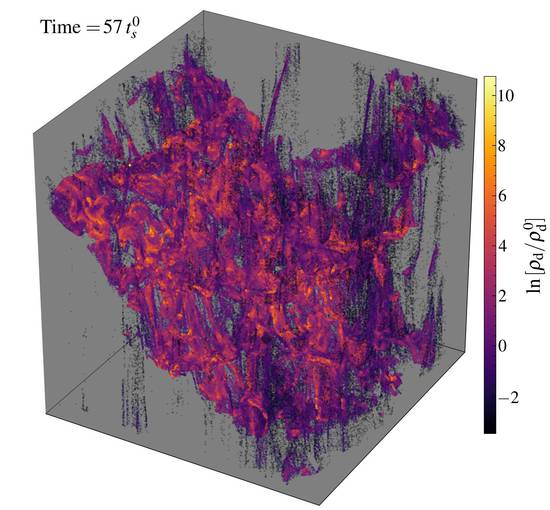}{jonodefault_b0_128/im_3dDust_inferno_0300}{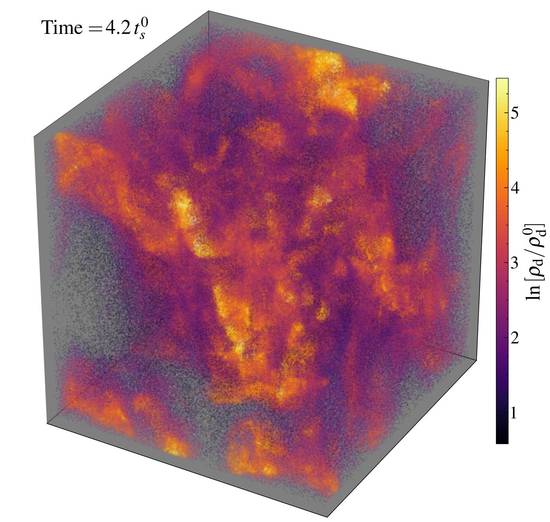}
\caption{Like \fref{fig:example.physics}, comparing our {\bf Example} run with three different dust-to-gas ratios $\mu=0.001$ ({\em left}), $0.01$ (our default; {\em middle}), $0.1$ ({\em right}). The broad form of the instabilities is similar in each case, and the growth timescale scales as $\sim \mu^{1/2}$, as expected from linear theory. At early times the wavenumber (number of ``sheets'') corresponding to the most visually obvious mode is lower for lower-$\mu$: this corresponds to the fastest-growing wavelength where the ``aligned'' mode dominates in Fig.~\ref{fig:rates.example.dustgas}, so again is predicted by linear theory. The saturated gas turbulence is stronger with increasing $\mu$, as  expected because the forcing from dust is stronger). However the low-$\mu$ case exhibits extremely small, dense sheets and filaments that do not turn into less-dense or ``fuzzy'' patches at late times, as occurs in the high-$\mu$ case. In fact most, in each run, of the dust mass resides at approximately the same dust-to-gas ratio (i.e.\ the typical ``overdensity'' is systematically larger at lower-$\mu$), and the highest dust-to-gas ratios appear in the lowest-$\mu$ box.\label{fig:example.dustgas}}
\epanelfigL
%
%

%
%
\bpanelfigL
\includegraphics[width=0.33\textwidth]{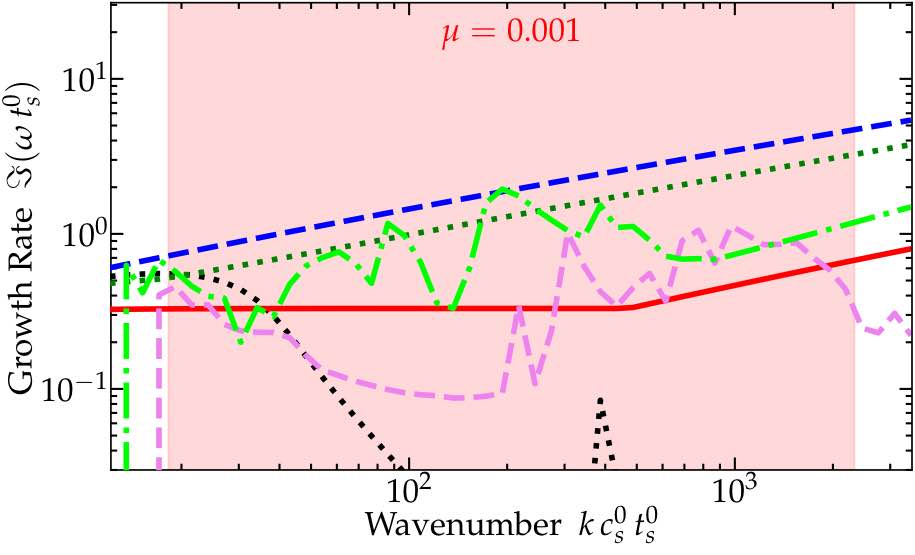}\includegraphics[width=0.33\textwidth]{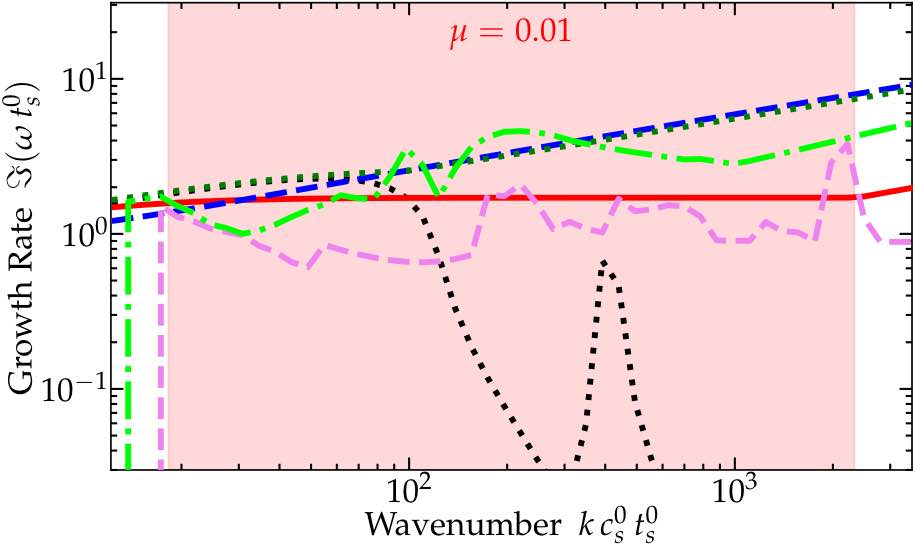}\includegraphics[width=0.33\textwidth]{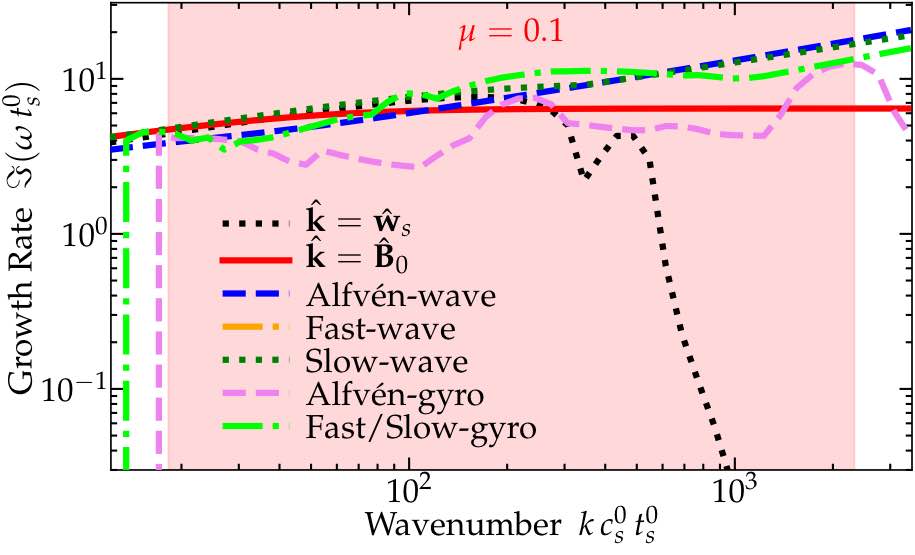}\\\vspace{-0.25cm}
\caption{Linear growth rates, as in \fref{fig:rates}, for the variants of {\bf Example} with varied dust-to-gas ratio $\mu$ in \fref{fig:example.dustgas} (as labeled). The growth rate normalization scales approximately as $\sim \mu^{1/2}$, explaining the relative timescales in \fref{fig:example.dustgas}. The wavenumber of the ``sheets'' that dominate the early non-linear phase   corresponds to the fastest-growing (highest) wavenumbers $k$ where the aligned modes (with $\hat{\bf k}=\hat{\B}$ or $\hat{\bf k}=\driftvelhat$; red and dotted black lines) have growth rates comparable to or greater than the other modes. The MHD-wave (\Alf\ and slow) RDIs generate the ``corrugation'' of the sheets -- the oscillations with $\hat{\bf k}$ nearly perpendicular to the ``aligned'' modes of the sheets. 
\label{fig:rates.example.dustgas}}
\epanelfigL
%
%

%
%
\bpanelfigL
\figthreeXLlab{k0_mu/im_3d2dProj_rho_high_terrain_0023}{$\mu=0.001$}{k0/im_3d2dProj_rho_high_terrain_0080}{$\mu=0.01$}{k0_mu2/im_3d2dProj_rho_high_terrain_0100}{$\mu=0.1$}
\figthreeXL{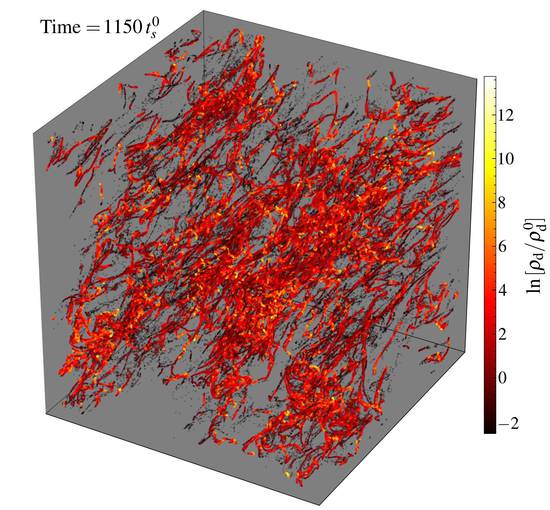}{k0/im_3dDust_hot_0080}{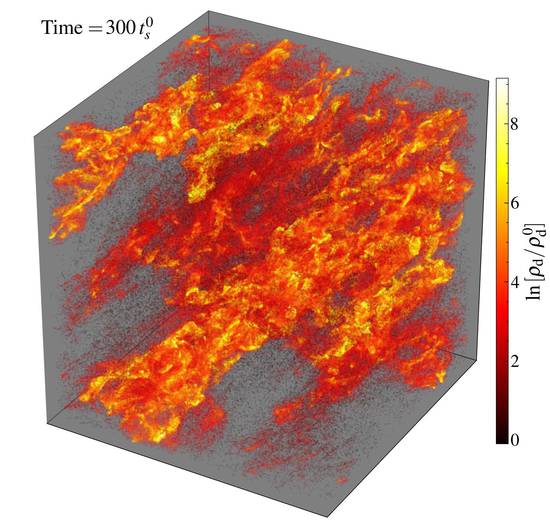} 
\caption{Like \fref{fig:example.dustgas} (bottom panels), comparing variants of the {\bf HII-near} L run with three different dust-to-gas ratios $\mu=0.001$ ({\em left}), $0.01$ (default; {\em middle}), $0.1$ ({\em right}) in the saturated state. The conclusions are similar to \fref{fig:example.dustgas}: the growth times and wavenumbers scale with predictions of linear theory as $\mu^{1/3\rightarrow1/2}$ and the gas turbulence is stronger at high-$\mu$, but the low-$\mu$ case retains  extremely dense filaments or clumps that do not get ``broken up'' or ``spread out'' at later times. The dust reaches extremely high over-densities in the low-$\mu$ case.
\label{fig:HIInear.dustgas}}
\epanelfigL
%
%

%
%
%
%


\subsection{Different Equations-of-State, Drag Laws, and Grain Charge Scalings}
\label{sec:results:eos}

In this subsection, we discuss the effects of additional physical variations in the simulations. We discuss the physical applicability of different drag laws in different regimes (expanding the discussion from \paperone), and explore different choices. We outline three idealized, physically motivated regimes of the grain charge scaling, and compare the effects of each of these in the simulations. We justify our usage of an isothermal equation of state for the majority of our simulations, and discuss astrophysical situations where different equations of state are appropriate. We also  investigate how changes to the magnetic field, drift geometries, and additional grain parameters alters the non-linear evolution of the instabilities. 

\subsubsection{Drag law}\label{sec:subsub:drag.law}
The scalings and physical applicability of different drag laws are discussed in detail in \paperone. There it is shown that for any regime where the MHD RDIs are important, the drag is dominated by either Epstein or Coulomb drag (e.g.\ Stokes drag, important only when the grain is larger than the gas mean free path and Reynolds numbers are less than unity, is never relevant for the physical grain sizes in magnetized media considered here). Our default simulations adopt the Epstein drag scalings, which, for fixed grain properties, give 
\begin{align}
\ts^{-1} &\propto \gasden\,\cs\,(1+a_{E}\,\driftvelmag^{2}/\cs^{2})^{1/2} &\hfill ({\rm Epstein})
\end{align} 
($a_{E}=9\pi\,\gamma/128$). Likewise, Coulomb drag gives 
\begin{align}
\ts^{-1} &\propto \gasden\,\cs\,(1+a_{C}\,\driftvelmag^{3}/\cs^{3})\,f(T,\,U) &\hfill ({\rm Coulomb}) \ ,
\end{align}
 where $a_{C}=(2\gamma^{3}/9\pi)^{3}$ and $f(T,U) = f_{\rm ion}\,(z_{i}\,U / T)^{2} \ln{\Lambda}$ is a function of the temperature $T$, ionization fraction $f_{\rm ion}$, grain electrostatic potential $U$, gas ion charge $z_{i}$, and a Coulomb logarithm $\ln{\Lambda}$. 

It is difficult to evaluate Coulomb drag without specifying the physical system (to calculate $f_{\rm ion}$, $T$, $\Lambda$, etc.), whereas Epstein drag is determined by scale-free hydrodynamic quantities. Epstein drag dominates at high  $\driftvelmag/\cs$, scaling as $\sim\driftvelmag/\cs$, while Coulomb drag  becomes weaker, scaling as $(\driftvelmag/\cs)^{-3}$. Therefore, Epstein drag always dominates in super-sonic cases (where $\driftvelmag/\cs \gtrsim 1$ or $|\delta \dustvel|/\cs \gtrsim 1$). Moreover, Epstein drag often dominates in sub-sonic cases if the ionization fraction is small or if the grains reach charge saturation (see \paperone). We therefore expect Epstein drag to be a good approximation in many cases. Even in sub-sonic regimes where Coulomb drag may dominate, if the temperature and grain potential $U$ are fixed (as they are in our default runs), then Epstein and Coulomb drag scale identically (as $\ts^{-1} \propto \rho$). Because the normalization of $\ts$ is arbitrary (it simply goes into the dimensionless parameter $\cs\,\ts/\Lbox$),  our predictions therefore apply equally well to either drag law in this sub-sonic regime, provided one re-scales the simulations dimensionless  parameters  appropriately.

\subsubsection{Grain charge}

Scalings of grain charge remain uncertain and depend on a variety of environmental and local factors (see \paperone\ and, e.g.\ \citealp{draine:1987.grain.charging,weingartner:2001.pah.model,weingartner:2001.grain.charging.photoelectric,tielens:2005.book}). However, if we assume that external environmental properties are fixed (e.g.\ a UV radiation field) and grain material properties are fixed (e.g.\ compositions, sizes), then three regimes emerge, which we define as:
\begin{align}
Z_{\rm grain} &\propto 
\begin{cases}
{\rm constant} & \hfill ({\rm a}) \\ 
T & \hfill ({\rm b}) \\ 
T^{1/2}\,\gasden^{-1} & \hfill ({\rm c}) \\ 
\end{cases}
\end{align}
(a) ``Fixed Charge'': if the grain charge is saturated (maximal/minimal), or the charge-aggregation/equilibration times are long compared to the timescales we evolve, then the charge is approximately fixed. (b) ``Collisional Charging'': if the charging is dominated by collisional processes, is sufficiently fast, and is un-saturated, it depends solely on the temperature as $\grainchargeZ \propto T$. (c) ``Photo-Electric Charging'': if the charging is dominated by photo-electric processes (and again is also un-saturated and sufficiently rapid), then $\grainchargeZ \propto T^{1/2}\,\gasden^{-1}$. Our default simulations assume regime (a), i.e.\ fixed charge, which is common for large grains in well-ionized environments.\footnote{We assume throughout that we have large grains with $|\langle \grainchargeZ \rangle | \gg 1$, so charge can be treated as a continuous, smooth quantity.} In an isothermal gas, our default case, regime (b), collisional charging, is identical (because $\grainchargeZ$ depends only on $T$ for specified grain properties). Regime (c), photo-electric charging, is only different in the isothermal case if the gas is highly-compressible so that fluctuations in $\rho$ are large. 

In order to explore the non-linear effects of Regime (c), we  re-run {\bf HII-near} L, but adopt the photo-electric scaling $\tL^{-1} \propto \grainchargeZ \propto T^{1/2}\,\gasden^{-1}$ with the same initial homogeneous value of $\grainchargeZ$ such that the constants in \tref{table:sims} are identical (i.e.\ only the scaling is changed). We also compare a run with $\tL^{-1} \propto T$ corresponding to collisional charging, adopting $\gamma=5/3$ so that $T$ will actually vary. These {\bf HII-near} L comparisons are shown in \fref{fig:HIInear.physics.variations}. It is evident that the photo-electric charging law produces dust structures with distinct topologies. The dust is more coherent, and the gas is dragged along into regions with higher dust densities and higher velocities.

Given the (significant) physical uncertainties in predicting grain charge, we also explore the effects of re-scaling the assumed charge from our default assumptions. The results of this is shown in the left column of \fref{fig:HIInear.physics.variations} for {\bf HII-near}, the right column of \fref{fig:CGM} for {\bf CGM}, and the right column of \fref{fig:WIM.lowVdust} for {\bf Corona}. We find that increasing the charge (holding all else equal) produces faster-growing, more violent instabilities.

\subsubsection{Gas equation of state}

Our default simulations adopt an isothermal equation of state ($\gamma=1$). This is a good approximation to many astrophysical cases of interest (e.g.\ the WIM, HII regions, the Solar corona), where cooling is efficient. Moreover, the majority of our runs are only weakly-compressible, so changing $\gamma$ (with all other parameters fixed) only has a weak effect. In Figure \ref{fig:example.physics}, we present a comparison of {\bf Example} with $\gamma=5/3$ (with the initial pressure and all other parameters identical to the ``default'' run). It is apparent that changing the equation of state in the weakly compressible limit has very little effect on the non-linear evolution of the instability. This near-independence of the instabilities from changes to the equation of state also suggests that different charge scalings or drag scalings (i.e.\ a different dependence on $T$, $\rho$, etc., discussed above) will produce similar results to our default assumption (constant charge + Epstein drag), because gas properties do not vary dramatically regardless of $\gamma$. On smaller scales than examined here, different equation of states are more appropriate because the growth time of the instabilities may become short compared to the cooling time.

 The few of our runs that are more strongly compressible in the gas all have $|\delta \dustvel|/\cs \gg 1$ (as they must, since compressibility requires $|\delta \gasvel | /\cs \gg 1$ and typically $|\delta \gasvel | \lesssim |\delta \dustvel |$). This implies that Epstein drag should better approximate dust drag forces (see \sref{sec:subsub:drag.law}). It is difficult to identify any parameter regime where Coulomb drag both dominates over Epstein drag, and has the potential to significantly alter our findings (due  to its temperature   or charge dependence in a  non-isothermal EOS). Of our compressible simulations, we note that most represent parameters motivated by the very regions we expect to have $\gamma \approx1$ on the scales of interest. The only case where one might expect $\gamma \ne 1$ among this set is run {\bf CGM}, representative of the CGM around a bright source (where cooling is inefficient so $\gamma\approx5/3$ is more appropriate). We therefore re-run  this simulation with $\gamma=5/3$ and Epstein drag (see \fref{fig:CGM}). In this case the physically realistic charge scaling is not obvious, although it is probably saturated (see \paperone). For the sake of exploring different regimes we  assume un-saturated photo-electric charging, so $\tL^{-1} \propto Z \propto T^{1/2}/\rho \propto \rho^{(\gamma-3)/2}$. The resulting differences with the default parameters are quite small (cf.\ left and middle panels in \fref{fig:CGM})

\subsubsection{Other parameter variations: Magnetic field orientation and strength, Drift geometry and speed, and Dust-to-gas ratios}

Aside from these variations in drag-law, gas properties, and grain charge, we consider  variations of other parameters compared to the default simulations in  Figs.~\ref{fig:example.physics}-\ref{fig:example.uncharged}.

We examine how changes to the magnetic and drift geometries and magnitudes effect the evolution. We decrease $\beta$ by a factor of $2$ (from $\beta=2$ to $\beta=1$) in {\bf Example}, keeping all other parameters fixed. This has only minor effects on the qualitative saturation behavior, but it does shift the resonant angles, effectively rotating the box (see the right column of  \fref{fig:example.physics}). We compare {\bf WIM} S/M/L with very low drift velocity $|\driftvel|/\cs\sim 0.01$ ($5\times$ lower than our default case). This makes the growth rates of some modes prohibitively long, requiring too long an evolution to capture; nonetheless, we see many similar qualitative features emerge as in the default case, but with weaker saturated amplitudes (see \fref{fig:WIM.lowVdust}). In \fref{fig:HIInear.angle.variations}, we re-run {\bf HII-near} L varying the angle between $\B$ and $\acc$. This does not qualitatively change the behavior, but does shift the angle of the streams of dust in the saturated state (which tend to align with $\acc$), and produces slightly faster growth rates as $\acc$ becomes more orthogonal to $\B$ (this is predicted by linear theory, but is not obvious, as $\driftvel$ actually becomes smaller). 

We also investigate how changes to the dust properties effect the evolution. As shown in \citet{seligman:2018.mhd.rdi.sims}, removing dust charge in our {\bf Example} box gives a linear growth rate approximately 2-3 orders of magnitude lower. We show the result of simulating this at similar times to our charged cases in \fref{fig:example.uncharged}, to show that no instability has grown. This also makes it clear that our numerical scheme does not introduce any  artificial dust clumping. Finally, we simulate {\bf Example} and {\bf HII-near} with lower and higher dust-to-gas ratios ($\mu=0.001$ and $\mu=0.1$), which are shown in Figs.~\ref{fig:example.dustgas} and \ref{fig:HIInear.dustgas}). The lower-$\mu$ cases are the most interesting: as expected, the growth rate decreases by a factor $\sim 3$ (see \fref{fig:rates.example.dustgas}) and the gas density and turbulent velocity fluctuations decrease since the forcing from the dust on the gas is weaker. Surprisingly however, lowering $\mu$ appears to enhance the strength of the saturated dust density fluctuations. This may occur because the weaker gas turbulence is less efficient at disrupting dense dust structures.

%
%
\bpanelfig
\figthreeXSlab{b0_N64/im_3d2dProj_v_high_terrain_0110}{ $2\times 64^{3}$} {jonodefault_b0_128/im_3d2dProj_v_high_terrain_0110} { $2\times 128^{3}$}{b0_N256/im_3d2dProj_v_high_terrain_0011} { $2\times 256^{3}$}
\figthreeX{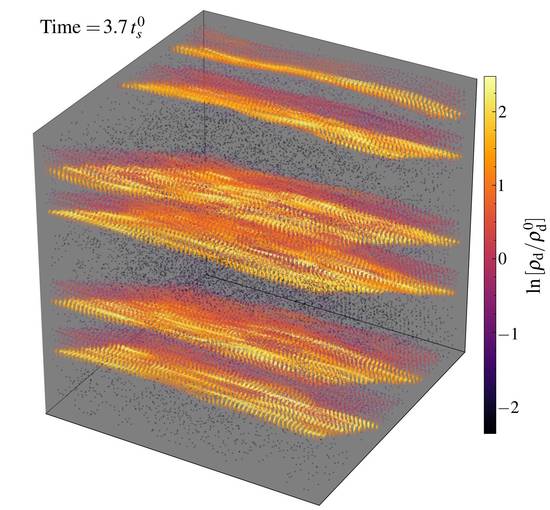} {jonodefault_b0_128/im_3dDust_inferno_0110} {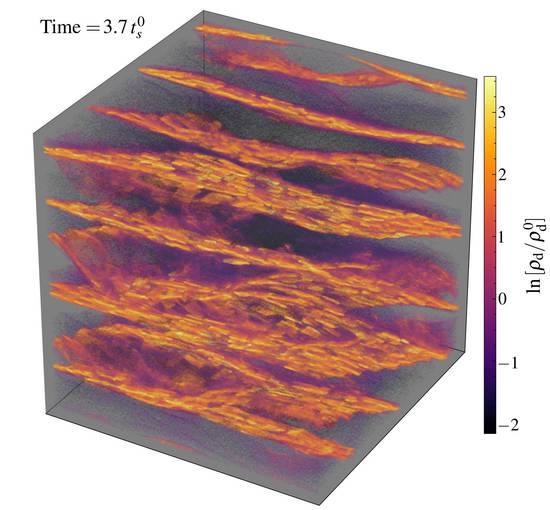} 
\figthreeX{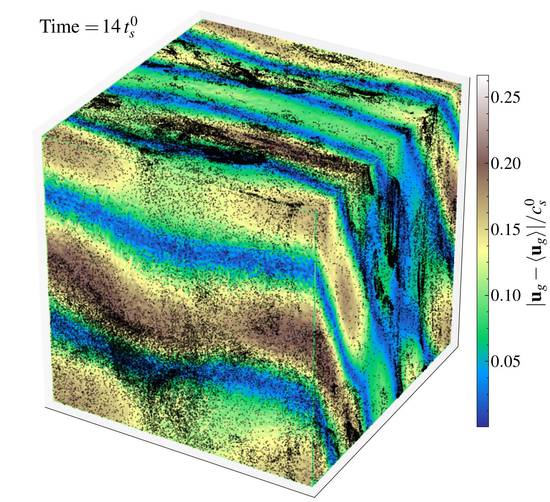} {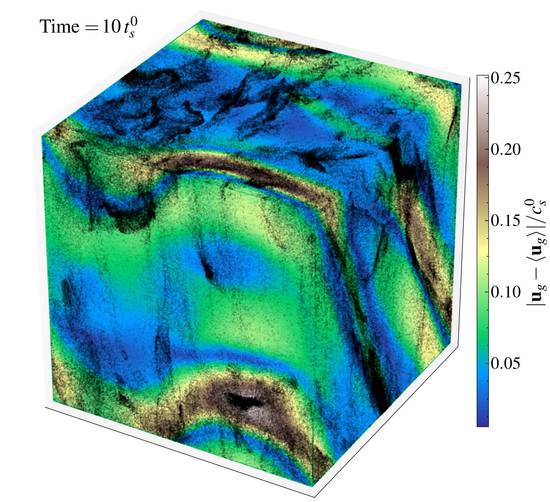} {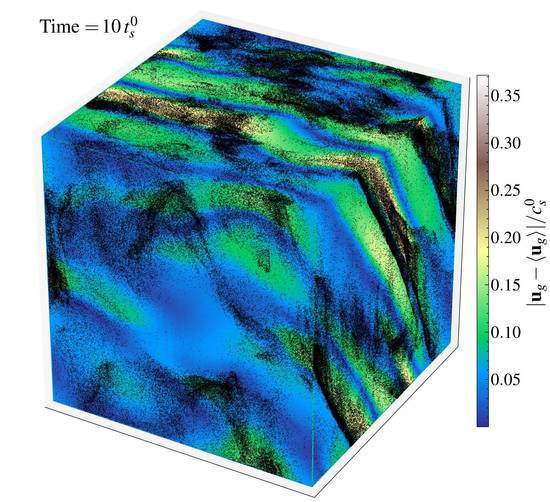} 
\figthreeX{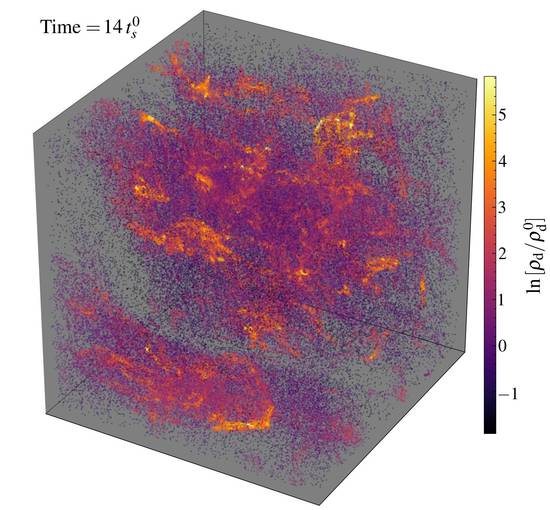} {jonodefault_b0_128/im_3dDust_inferno_0300} {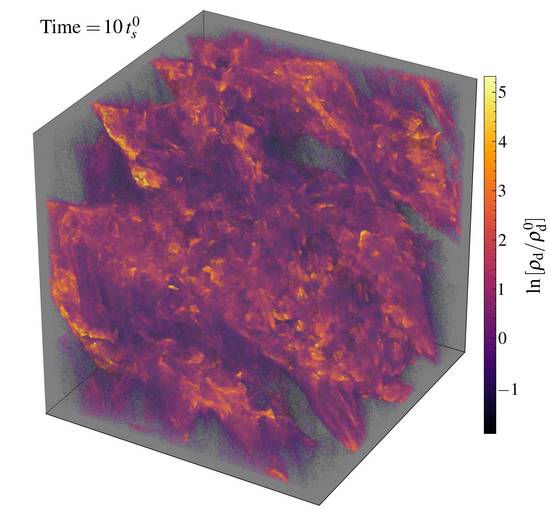} 
\caption{Comparison of the {\bf Example} run at different resolution levels: $2\times 64^{3}$ ({\em left}), $2\times128^{3}$ (our ``default'' case; {\em center}), and $2\times 256^{3}$ ({\em right}). The instabilities develop slightly more quickly at increasing resolution, owing to the fact that the growth rates increase almost monotonically with $k$ (\fref{fig:rates}). As expected, we can also see more detailed, finer, and sometimes denser structure in the dust at higher resolution (the gas remains relatively smooth). However the qualitative form of the instabilities and the saturated amplitude of the gas turbulence remains similar.\label{fig:example.resolution}}
\epanelfig
%
%

\subsection{Additional Numerical Tests}
\label{sec:results:numerical}

In addition to the generic code validation tests described in \sref{sec:methods}, previous papers have tested the numerical methods here \citep{carballido:2008.grain.streaming.instab.sims,johansen:2009.particle.clumping.metallicity.dependence,bai:2010.grain.streaming.vs.diskparams, pan:2011.grain.clustering.midstokes.sims,hopkins:mhd.gizmo,hopkins:cg.mhd.gizmo,hopkins:gizmo.diffusion,su:2016.weak.mhd.cond.visc.turbdiff.fx,hopkins.2016:dust.gas.molecular.cloud.dynamics.sims,lee:dynamics.charged.dust.gmcs,moseley:2018.acoustic.rdi.sims,seligman:2018.mhd.rdi.sims}. For example \citet{hopkins.2016:dust.gas.molecular.cloud.dynamics.sims} show that the ``finite-sampling'' effects in super-particle methods, which introduce some shot noise in the particle densities and divergence  between particle trajectories and gas (at the integration error level) in the perfectly-coupled limit \citep[see][]{price:2010.grid.sph.compare.turbulence,2012MNRAS.423.1450A,genel:tracer.particle.method,2017MNRAS.471L..52T}, introduce $\sim 0.01-0.05\,$dex scatter in the dust density in a super-sonically turbulent medium ($\delta\gasvel /\cs \gg1$). In both \citet{hopkins.2016:dust.gas.molecular.cloud.dynamics.sims} and \citet{moseley:2018.acoustic.rdi.sims} we show in a variety of tests that even the ``worst-case'' scatter of this nature is completely negligible for our predictions,\footnote{We note that \citet{2017MNRAS.471L..52T} contains an error in their comparison of ``one-fluid'' dust treatments (which cannot represent multi-valued dust velocity distribution functions, essential for the present study; \citealt{2019A&A...621A..96C,2019A&A...626A..96L}) to the ``tracer particle'' ($\mu=0$) method used in \citet{hopkins.2016:dust.gas.molecular.cloud.dynamics.sims}. The discussion in  \citet{2017MNRAS.471L..52T} incorrectly compares to simulations in \citet{hopkins.2016:dust.gas.molecular.cloud.dynamics.sims} which adopted different dust parameters ($\tilde{\alpha}$), a different dust drag law, and did not use the same weighting for the PDFs/dispersions measured. If one correctly compares the simulations from \citet{hopkins.2016:dust.gas.molecular.cloud.dynamics.sims} which have similar $\tilde{\alpha} =0.001-0.01$ to the cases studied in \citet{2017MNRAS.471L..52T}, and weights the PDFs identically, then the agreement is reasonably good, and in fact the simulations in \citet{hopkins.2016:dust.gas.molecular.cloud.dynamics.sims} predict slightly {\em smaller} fluctuations in the dust-to-gas ratio $\rho_{d}/\rho_{g}$ than those in \citet{2017MNRAS.471L..52T}, the opposite of their conclusion.} and that in sub-sonically turbulent (or laminar) media this drops to $\ll 0.01\,$dex scatter. In either case this is much smaller than the saturated dispersions in $\dustden$ here.

As in \citet{moseley:2018.acoustic.rdi.sims} and \citet{seligman:2018.mhd.rdi.sims}, we have re-run variants of several of our standard simulations with varied numerical choices, including: (1) a different hydrodynamic solver (for {\bf Example}, {\bf WIM} L, {\bf HII-far} S/M/L), we use the ``meshless finite mass'' (MFM) method from \citet{hopkins:gizmo}; (2) a different scheme for calculating gradients and reconstruction of the magnetic field at cell faces (for {\bf Example}, {\bf Corona} S), specifically the ``constrained gradient'' MHD scheme in \citet{hopkins:cg.mhd.gizmo}; (3) using a naive (non-manifestly energy-conserving) explicit leapfrog integrator instead of the Boris integrator for the Lorentz forces on grains (in {\bf Example}, {\bf HII-near} S/M/L {\bf Corona} S/M/L, {\bf WIM} M/L, {\bf CGM}); and (4) using different initial conditions, namely glass-like instead of lattice initial particle configurations (for {\bf Example}). Choices (2), and (4) have no significant effect on any results we explore. We find that choice  (1), the MFM method, moderately suppresses the initial maximum growth rates of the modes in the box, owing to its slightly less-accurate integration of highly-subsonic flows near the grid scale (effectively, the fastest-growing resolved modes in the simulation correspond to those at $\sim 4-6$ times the grid scale, instead of $\sim 2-3$ times the grid scale). Choice (3) has negligible effects on boxes {\bf Example} and {\bf HII-near} S/M/L, where the dust is strongly clumped and $\tau$ is not extremely large. However for boxes {\bf Corona}, {\bf WIM}, and {\bf CGM}, where $\tau \ge 100$ is larger and the systems often end up in the ``disperse'' mode, we find that it becomes important to use an integrator (like our default Boris scheme) that manifestly conserves Lorentz orbits. Otherwise, the dust tends to drift and spiral outwards and gain energy artificially, an effect that is well-known in standard particle-in-cell codes. 

We have also run a number of resolution tests. Our default runs here use $2\times128^{3}$ elements ($128^{3}$ each in gas and dust). All of our runs here have also been run at lower resolution ($2\times32^{3}$ and $2\times64^{3}$), and one at higher ($2\times256^{3}$). Figure~\ref{fig:example.resolution} compares the saturated state in this resolution series. The qualitative behavior is similar across  this range of resolutions. However, without some sort of physical isotropic viscosity, or non-fluid effects (e.g.\ the ion gyro radii), these instabilities are  linearly unstable at all wavelengths, with  growth rates that increase with $k$. Thus, increasing resolution and keeping all else fixed will always resolve new, faster-growing modes at smaller scales, so it is not possible to undertake a formal convergence analysis. A more formal resolution study, as well as a number of additional numerical validation tests and more extensive discussion, are presented in \citet{moseley:2018.acoustic.rdi.sims}.\footnote{Note that in \citet{moseley:2018.acoustic.rdi.sims}, where we studied the pure acoustic RDI, we noted that the high-$k$ modes (with $|\driftvel| \gg \cs$) were difficult to resolve in some cases because the growth rates become sharply-peaked around a narrow resonant angle of width $\delta \theta \propto \mu^{2/3}\,k^{-1/3}\,|\driftvel|^{-1}$. As shown in \paperone, the MHD modes have a much more complex but also broader resonant structure, especially at the lower $\driftvel$ and higher $\tau$ values studied here. This contributes to making our results significantly less sensitive to resolution.} For example, Fig.~B3 in \citet{moseley:2018.acoustic.rdi.sims} compares the PDFs of the same properties studied here (e.g.\ dust-to-gas ratio) as a function of the relative number of dust and gas elements (varying $N_{\rm dust}/N_{\rm gas}$ by a factor of $64$) -- as expected, the resulting PDFs are essentially identical up e.g.\ Poisson sampling in the tails of the distribution. We have verified this with our runs here as well. As shown in \cite{hopkins.2016:dust.gas.molecular.cloud.dynamics.sims,lee:dynamics.charged.dust.gmcs,moseley:2018.acoustic.rdi.sims}, these PDFs are also robust to the details of kernel density estimators for dust or gas.

The saturated amplitudes of the velocity and density fluctuations are relatively insensitive to resolution in run {\bf Example}. This suggests that most of the power is dominated by the well-resolved largest modes in the box, rather than the fastest-growing but smallest-scale modes. This is consistent with our power-spectrum analysis in \citet{seligman:2018.mhd.rdi.sims}, where we showed explicitly in run {\bf Example} that most of the power in the saturated state was in modes a factor $\sim 2$ around the box scale. This may not be true in some other runs, e.g.\ {\bf HII-far} L, where the dust  clumps on small scales.

%
%
\begin{figure*}
\begin{center}
\includegraphics[width=\textwidth]{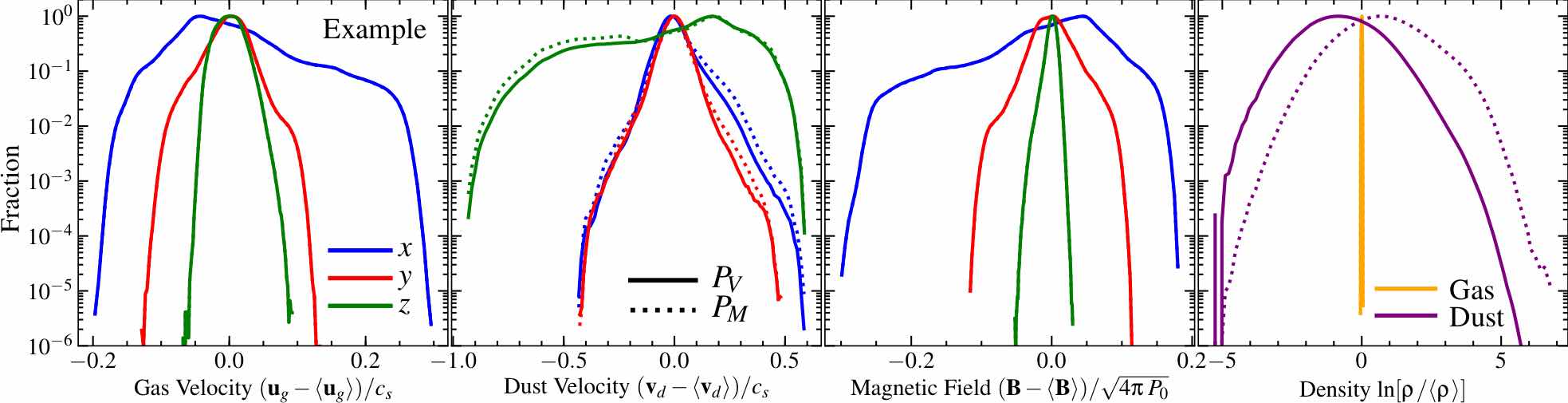} 
\vspace{-0.5cm}
\caption{Probability distribution functions (PDFs) of gas and dust velocities, and magnetic field fluctuations, in each direction $xyz$ (labeled), as well as the dust and gas densities. This is for the {\bf Example} run, averaging over the last ten snapshots (in saturation). For each PDF, solid lines show the volume-weighted PDF ($P_{V}$), while dotted lines show the mass-weighted PDF ($P_{M}$). PDFs are normalized to the same peak amplitude for convenience. Substantial structure is evident, related to coherent large-scale morphological structures in \fref{fig:example.physics}. The PDF tails often exhibit highly non-Gaussian statistics.}
\label{fig:pdfs.example}
\end{center}
\end{figure*}
%
%

%
%
\begin{figure}
\begin{center}
\includegraphics[width=\columnwidth]{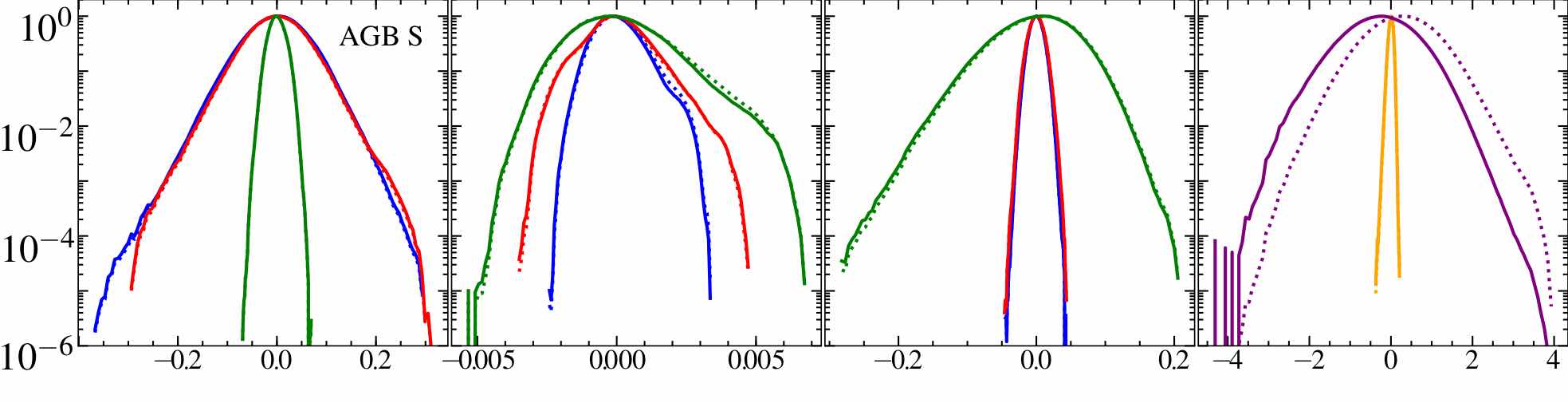}\vspace{-0.15cm} \\
\includegraphics[width=\columnwidth]{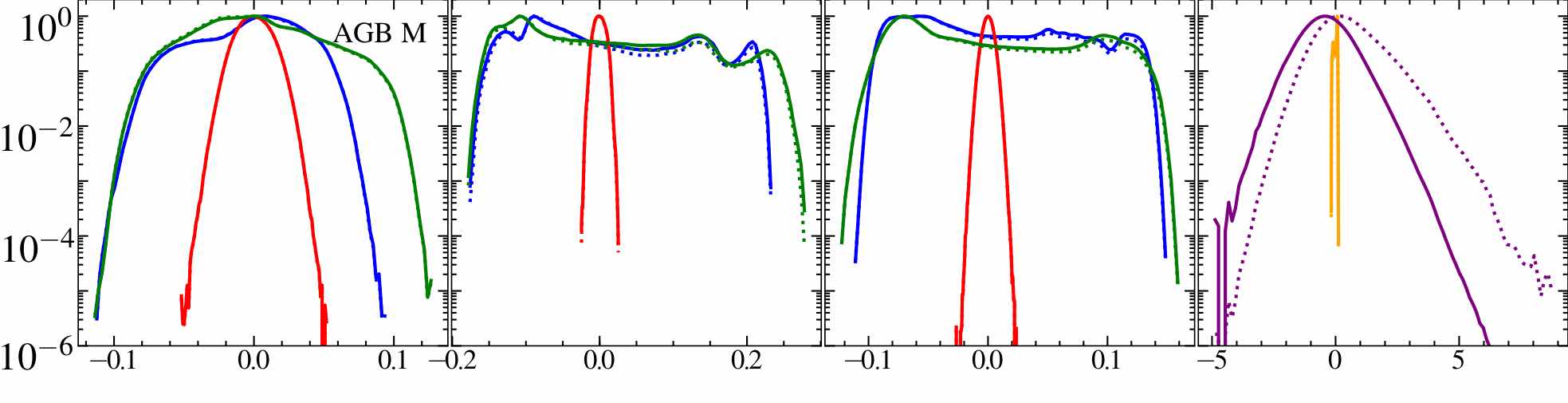}\vspace{-0.15cm} \\
\includegraphics[width=\columnwidth]{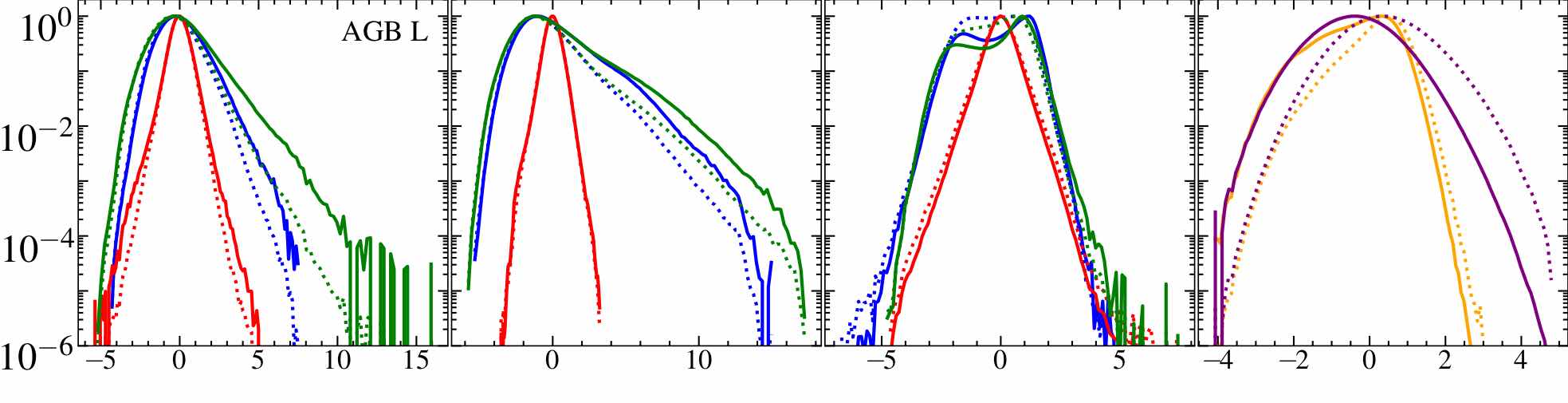}\vspace{-0.15cm} \\
\includegraphics[width=\columnwidth]{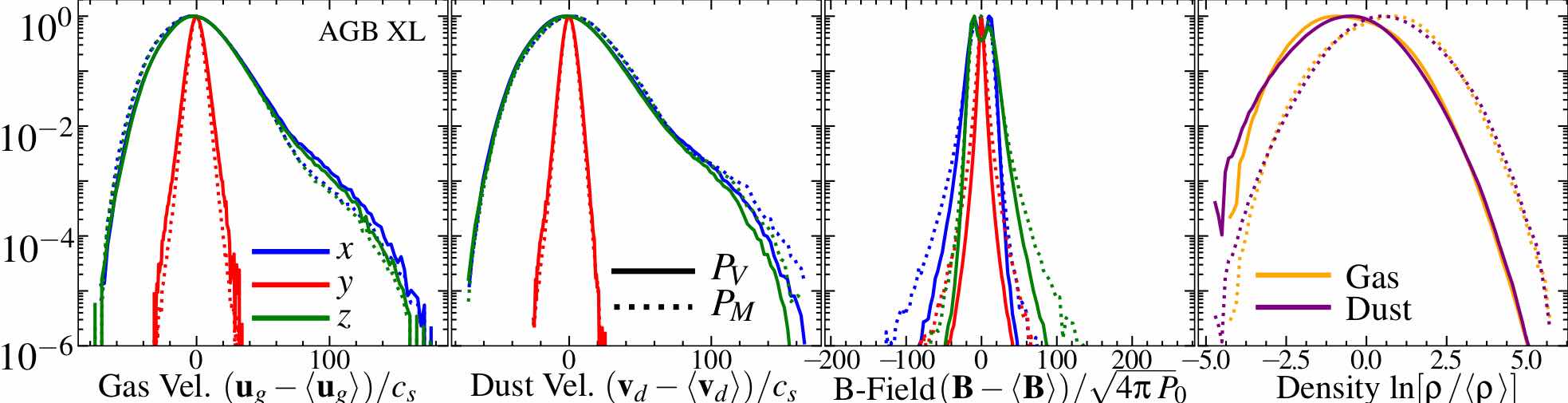} 
\vspace{-0.5cm}
\caption{PDFs for {\bf AGB} S/M/L/XL from top-to-bottom, as \fref{fig:pdfs.example}. 
All cases exhibit complicated structure in $\gasvel$, $\dustvel$, $\B$. Anisotropy is obvious: statistics here are similar in the direction of acceleration ($\hat{x}-\hat{z}$) and fluctuations are weaker in the perpendicular ($\hat{y}$) direction, except for the smallest box, which reverses this. The velocity and density PDFs typically exhibit exponential tails ($P \propto \exp{(-|v/v_{\ast}|)}$), discussed in the text. Dust density fluctuations are much larger than gas, except on the largest scales. Most of the dust {\em mass} lies in regions with above-average dust density, while most of the {\em volume} has lower-than-average dust density.
\label{fig:pdfs.agb}}
\end{center}
\end{figure}
%
%

%
%
\begin{figure}
\begin{center}
\includegraphics[width=\columnwidth]{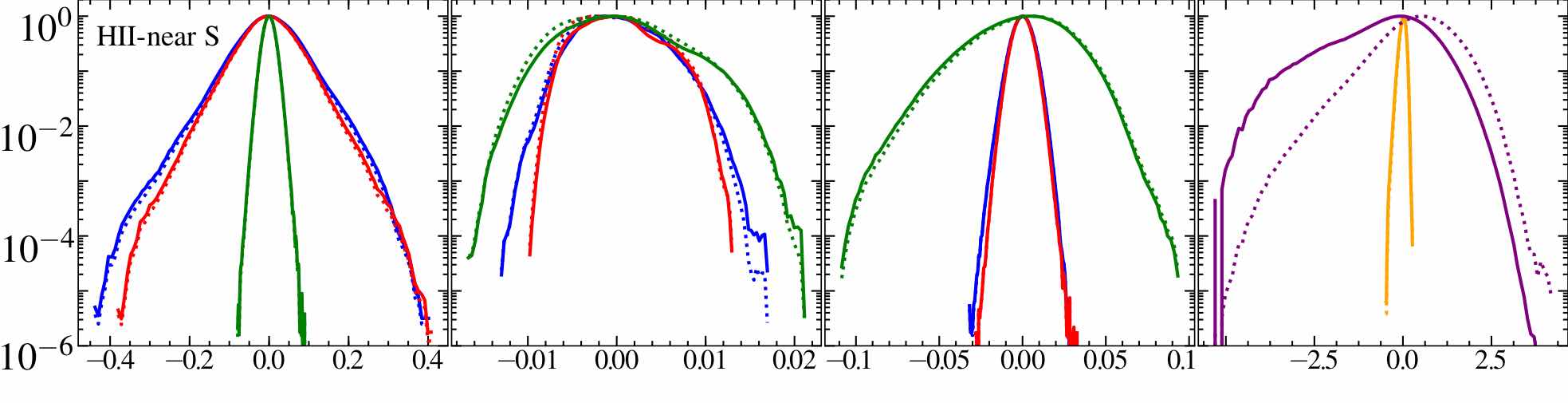}\vspace{-0.15cm} \\
\includegraphics[width=\columnwidth]{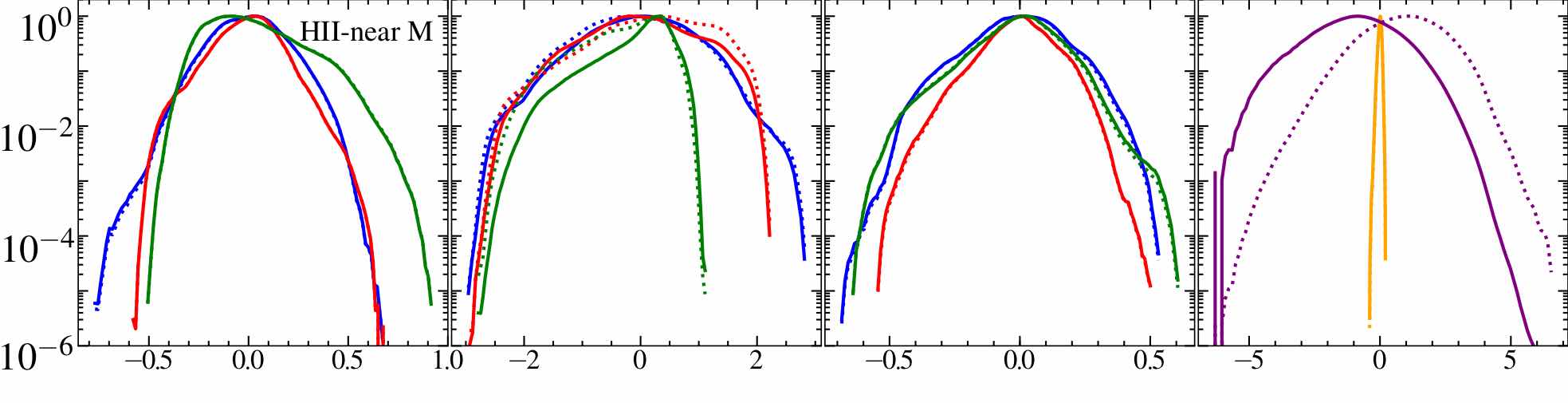}\vspace{-0.15cm} \\
\includegraphics[width=\columnwidth]{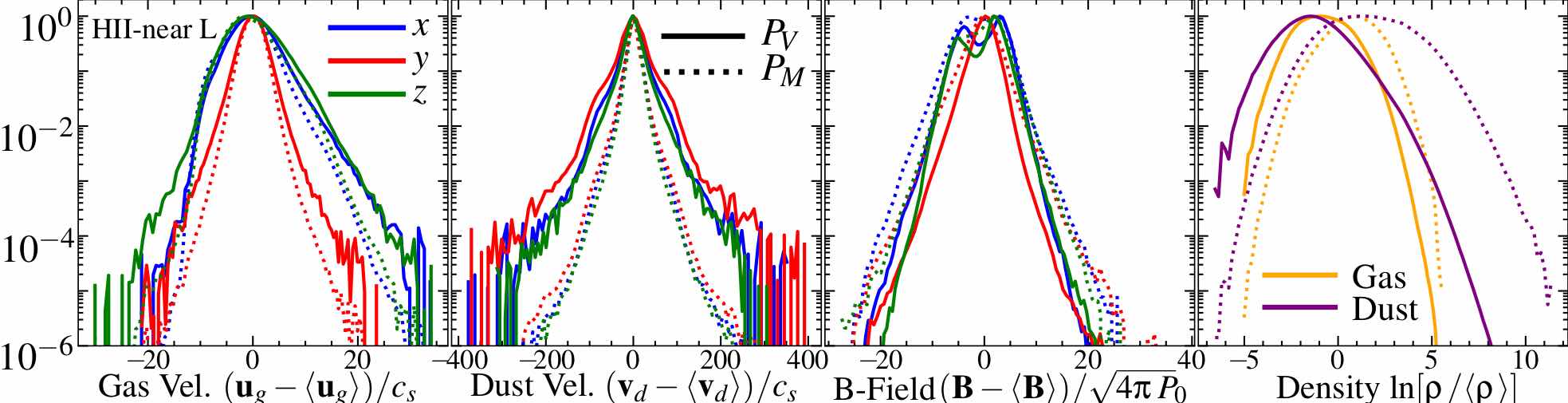} 
\vspace{-0.5cm}
\caption{PDFs for {\bf HII-near} S/M/L from top-to-bottom, as \fref{fig:pdfs.agb}. Sub-structure and anisotropy are common again. Some PDFs are close to Gaussian. Others (e.g.\ $\B$ in box L) are almost pure exponential. The $\dustden$ PDFs are highly skew, with large fluctuations so mass vs.\ volume weighting makes a large difference. The ``clumping factor'' $C = \langle \dustden^{2} \rangle/\langle \dustden\rangle^{2} \sim (3,\,6,\,100)$ in boxes (S,\,M,\,L).  In box L, $\dustvel$ exhibits a narrow ``core'' with a broad or ``wide'' stretched-exponential distribution super-posed (the PDF has ``fat tails''). These are all suggestive of strong intermittency (see \sref{sec:results:PDFs}).
\label{fig:pdfs.HIInear}}
\end{center}
\end{figure}
%
%

%
%
\begin{figure}
\begin{center}
\includegraphics[width=\columnwidth]{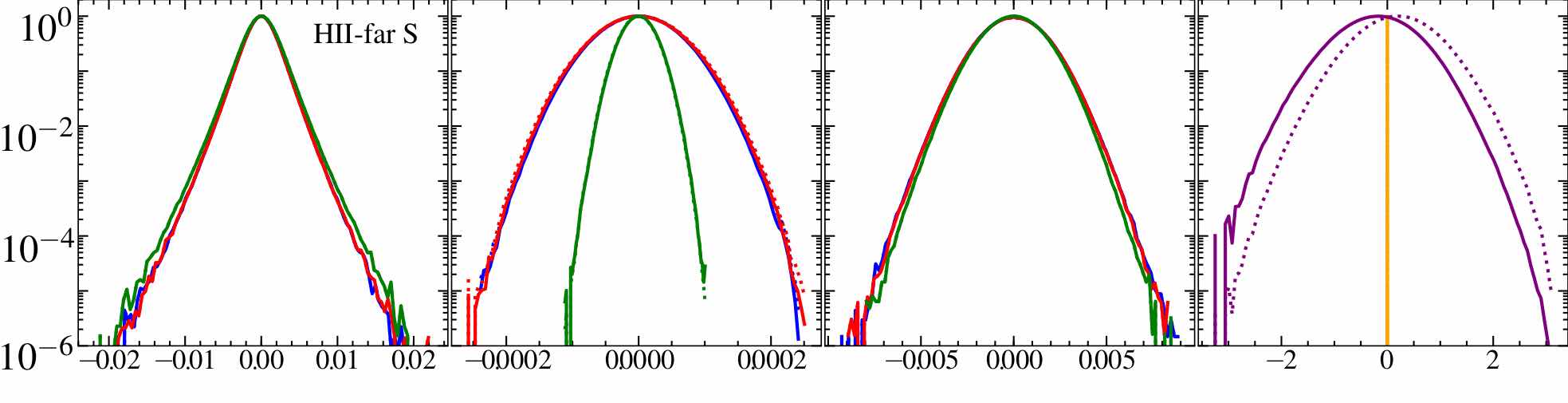}\vspace{-0.15cm} \\
\includegraphics[width=\columnwidth]{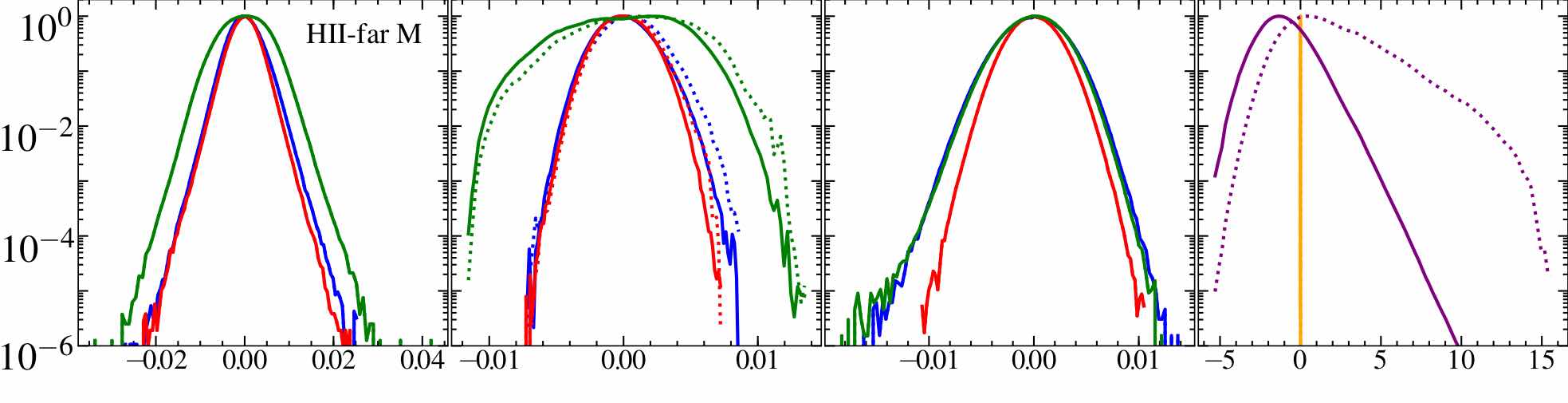}\vspace{-0.15cm} \\
\includegraphics[width=\columnwidth]{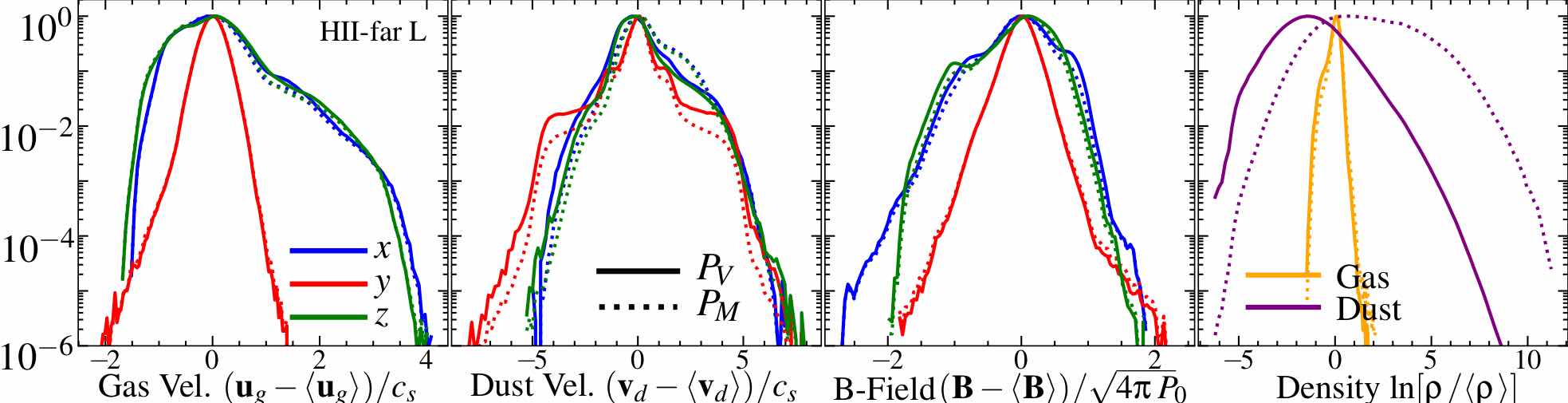} 
\vspace{-0.5cm}
\caption{PDFs for {\bf HII-far} S/M/L from top-to-bottom, as \fref{fig:pdfs.agb}. Again note sub-structure and ``fat tails'' in the PDFs (especially in box L, related to the large-scale coherent filaments in \fref{fig:HIIfar}); nearly-exponential PDFs for $\gasvel$ in boxes S, M; and large skew tails to high $\dustden$. While the gas is very weakly compressible here, the dust exhibits enormous and non-Gaussian density fluctuations: consider box M, where $|\delta \gasvel |/\cs \sim |\delta \dustvel |/\cs \sim |\delta \B|/\sqrt{4\pi\,\initvallower{\gaspressure}} \sim |\delta \ln{\gasden}| \sim 0.01$, but the dust $\delta \ln{\dustden} \sim 1.3$ (2.5) weighted by volume (mass). Moreover since the dust PDF is highly non-Gaussian, $\sim 0.1\%$ of the dust mass lies at densities $\dustden \gtrsim 10^{7}\,\langle \dustden \rangle$ ($>13\sigma$), and the clumping factor $C\sim 1500$, in box M. 
\label{fig:pdfs.HIIfar}}
\end{center}
\end{figure}
%
%

%
%
\begin{figure}
\begin{center}
\includegraphics[width=\columnwidth]{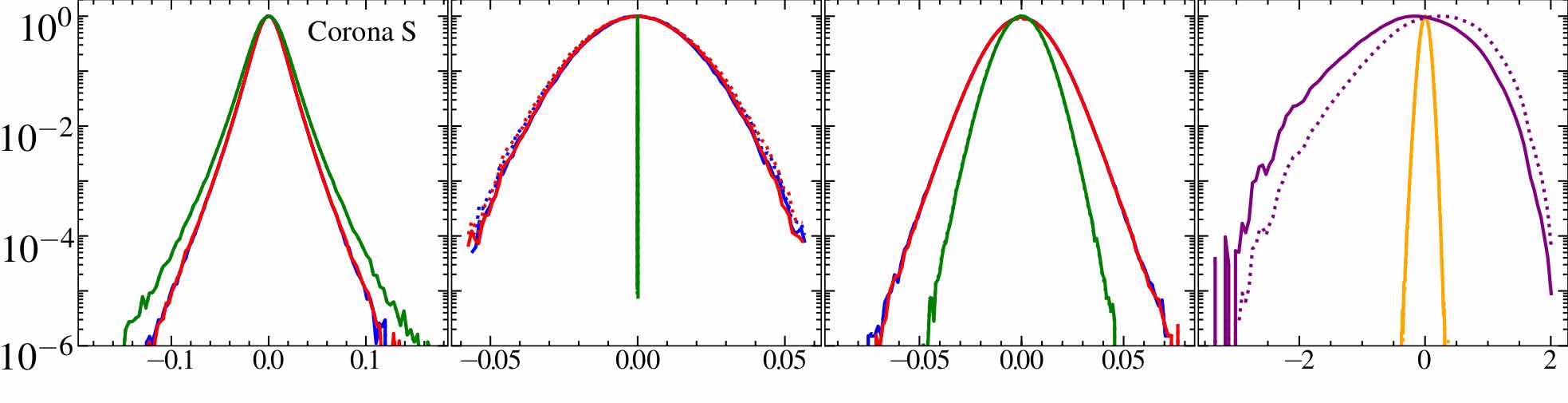}\vspace{-0.15cm} \\
\includegraphics[width=\columnwidth]{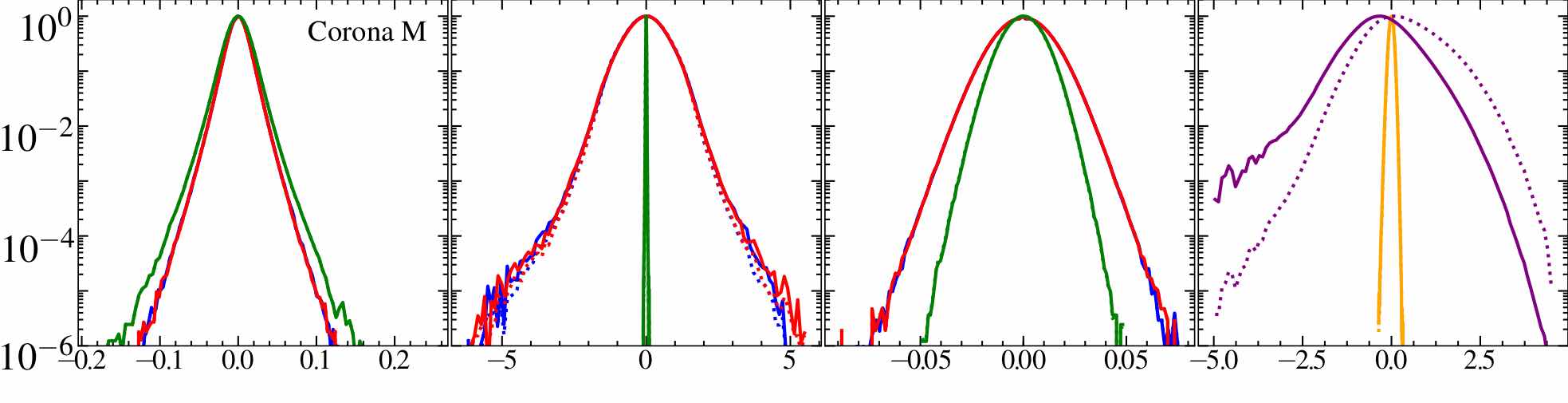}\vspace{-0.15cm} \\
\includegraphics[width=\columnwidth]{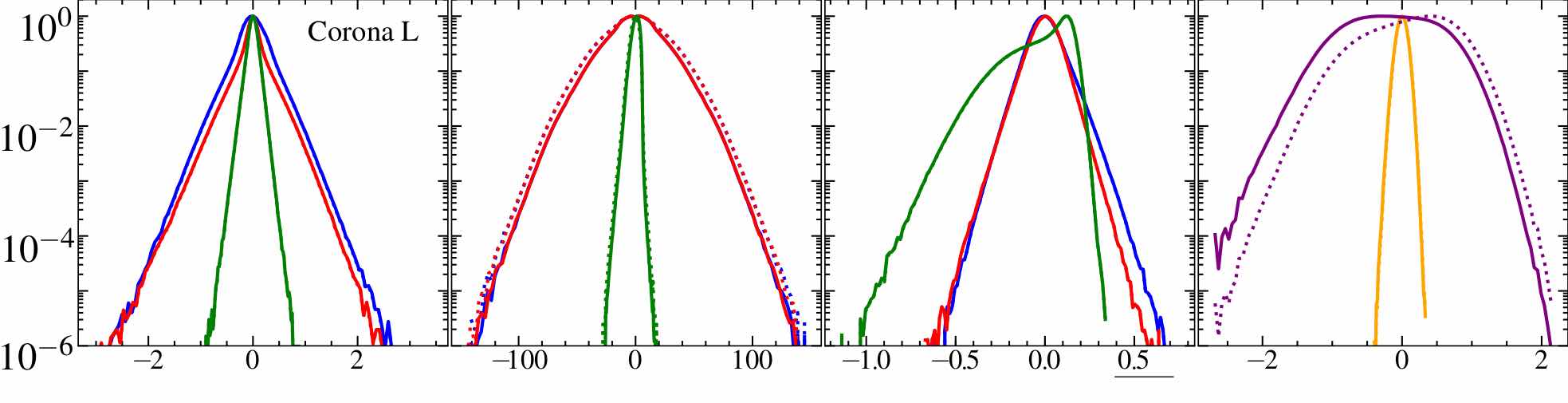}\vspace{-0.15cm} \\
\includegraphics[width=\columnwidth]{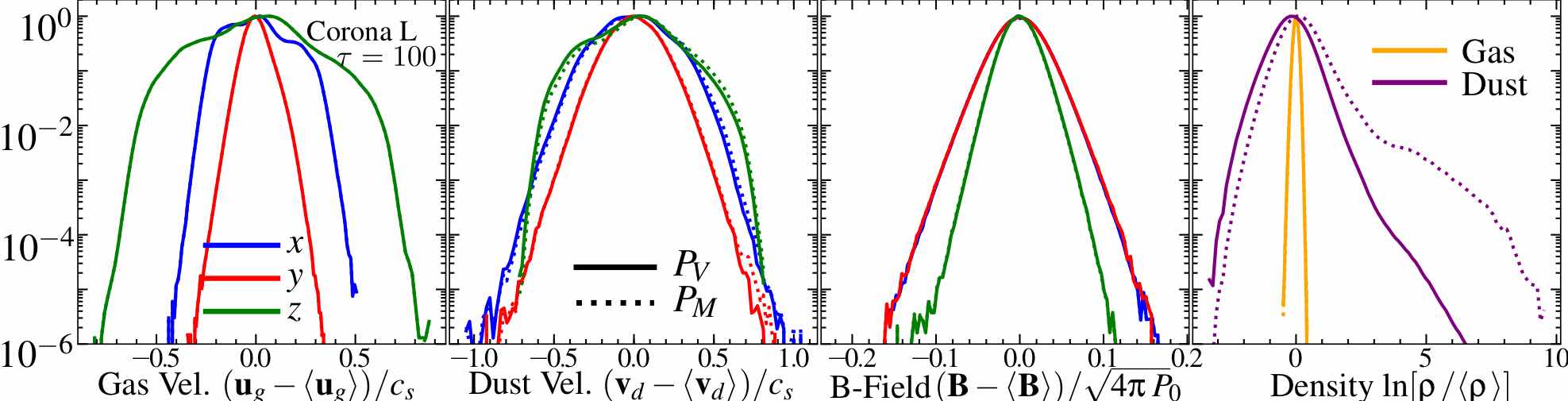} 
\vspace{-0.5cm}
\caption{PDFs for {\bf Corona} S/M/L (top/second/third), as \fref{fig:pdfs.agb} ({\em bottom} shows run L:$\tau$=100). The ``default'' runs all exhibit weak $\gasden$, and nearly-pure exponential $\gasvel$ fluctuations, with large $\dustvel$ fluctuations in the plane perpendicular to $\hat{\B}$ but small along $\hat{\B}$. The asymmetry in $\delta \B_{z}$ in box L stems from the large energetic cost of such fluctuations at low-$\beta$. The low-$\tau$=100 run exhibits qualitatively different behavior: the fluctuations in $\dustvel$ are $\sim 100\times$ smaller (and $\delta\gasvel$, $\delta\B$ are $\sim 10\times$ smaller), however while the ``core'' of the $\dustden$ PDF has similar width it exhibits a ``tail'' (dominated by the caustics in the folded dust ``sheets'' in \fref{fig:WIM.lowVdust}) to orders-of-magnitude larger $\dustden$.
\label{fig:pdfs.Corona}}
\end{center}
\end{figure}
%
%

%
%
\begin{figure}
\begin{center}
\includegraphics[width=\columnwidth]{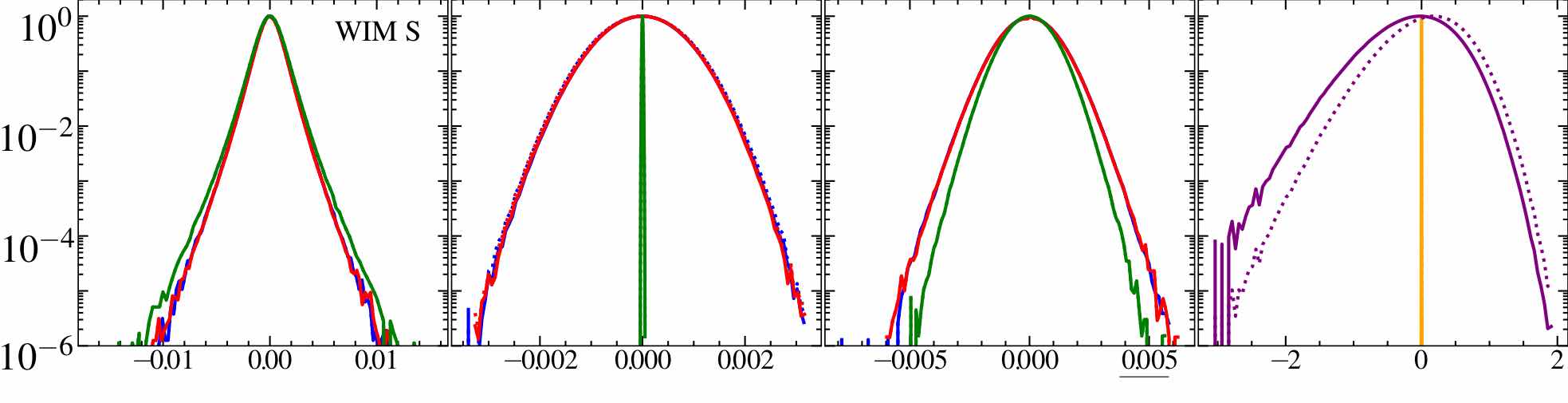}\vspace{-0.15cm} \\
\includegraphics[width=\columnwidth]{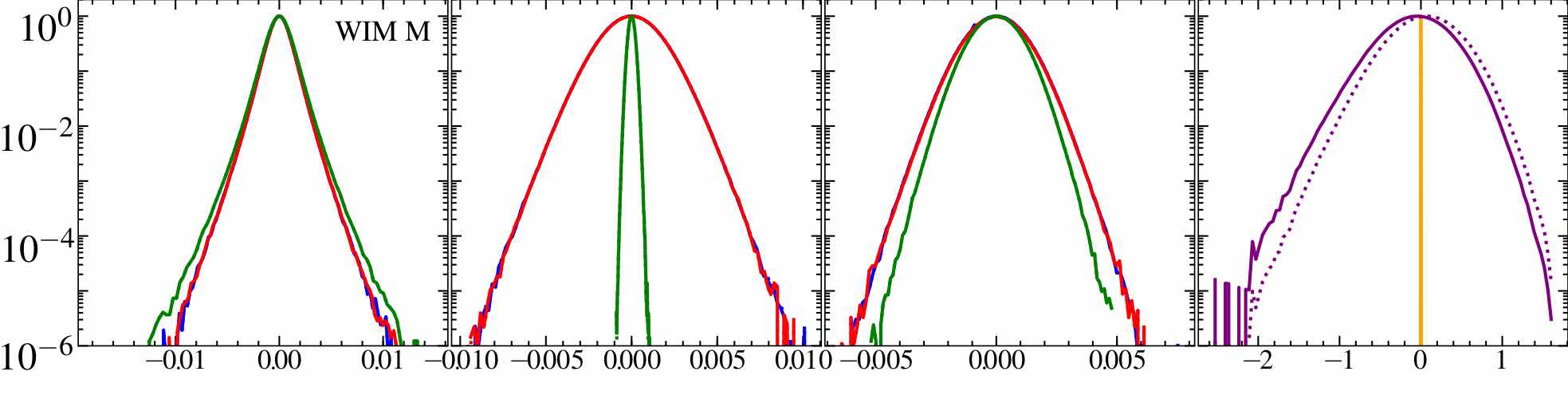}\vspace{-0.15cm} \\ 
\includegraphics[width=\columnwidth]{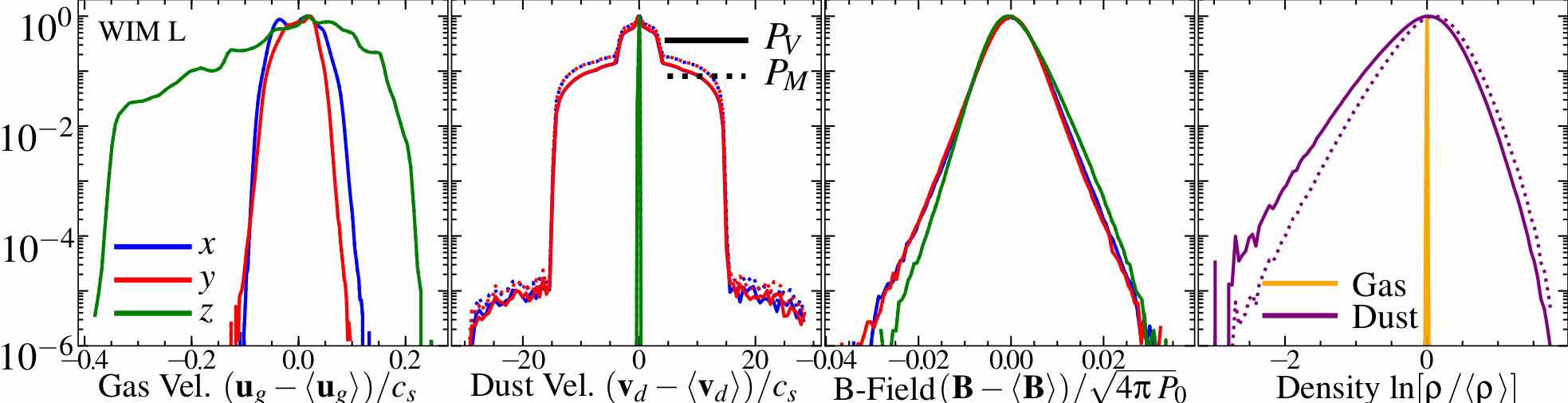} 
\vspace{-0.5cm}
\caption{PDFs for {\bf WIM} S/M/L from top-to-bottom, as \fref{fig:pdfs.agb}. Anisotropy is similar to {\bf Corona}, and gas is very weakly-compressible. $\dustden$ fluctuations are primarily to low-$\dustden$ (dust expelled from some regions), so e.g.\ the clumping factor $C\sim (1.2,\,1.02,\,1.1)$ (biased to high $\dustden$) is small. In box L, note the $\dustvel$ PDFs, symmetric perpendicular to $\B$, with progressively broader flat-topped PDFs ``super-posed'' to produce large tails.
\label{fig:pdfs.WIM}}
\end{center}
\end{figure}
%
%

%
%
\begin{figure}
\begin{center}
\includegraphics[width=\columnwidth]{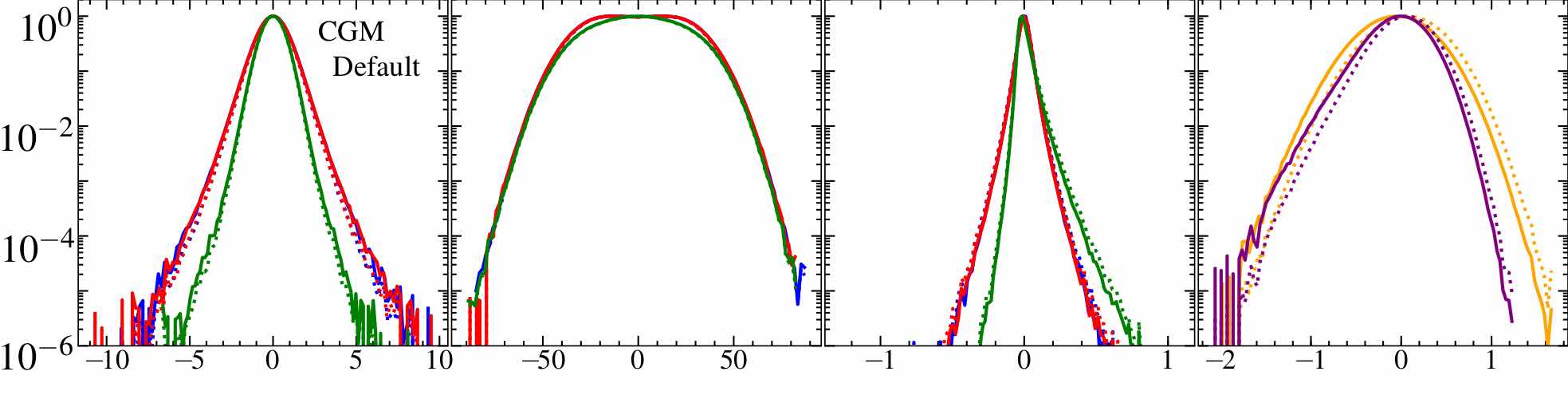}\vspace{-0.15cm} \\
\includegraphics[width=\columnwidth]{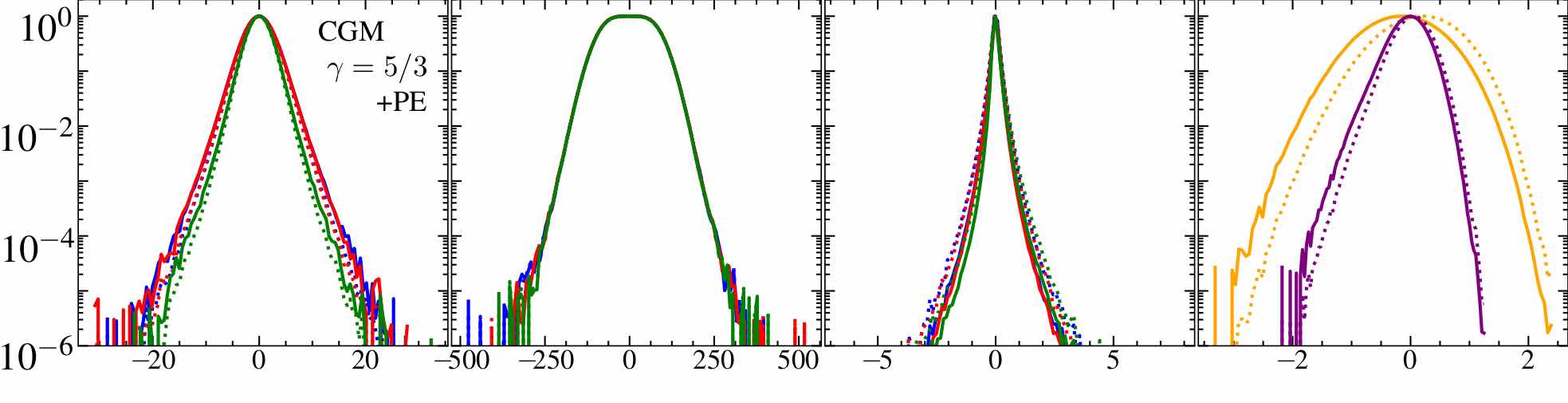}\vspace{-0.15cm} \\
\includegraphics[width=\columnwidth]{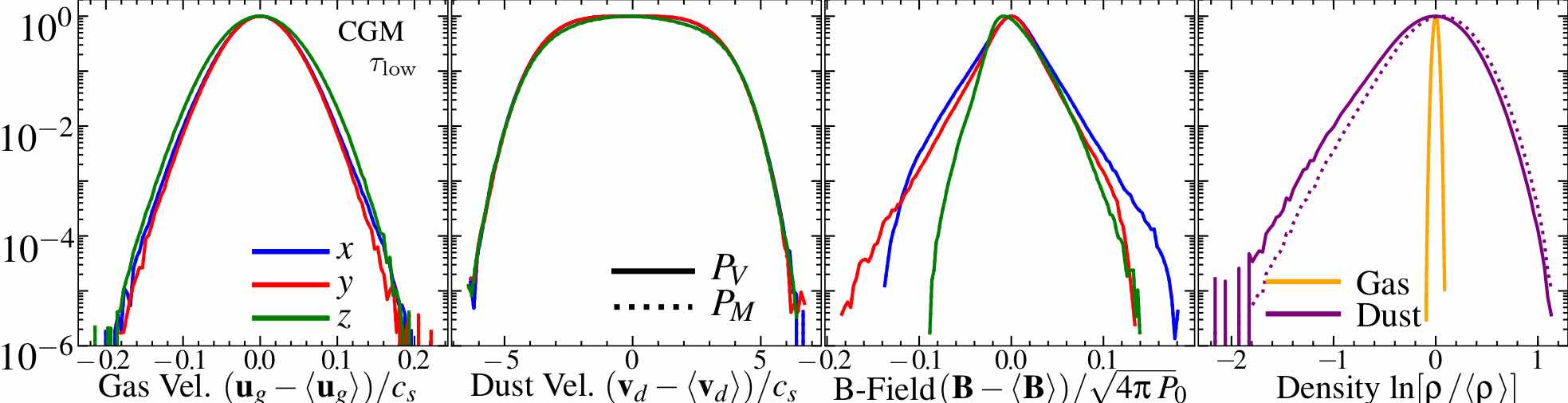} 
\vspace{-0.5cm}
\caption{PDFs for {\bf CGM}, as \fref{fig:pdfs.agb}, in our ``Default'' run ({\em top}), run with $\gamma=5/3$ and photo-electric grain charge scaling ({\em middle}), and low-charge (``$\tau_{\rm low}$'' i.e.\ $\tau\sim 5000$; {\em bottom}) run. Lower-$\tau$ (weaker Lorentz forces) produces weaker dust velocity dispersions, sourcing weaker gas motions. The photo-electric grain charging produces stronger dust acceleration. Fluctuations in all cases are close to isotropic. While $\gasvel$ and $\B$ PDFs are sub-Gaussian (exponential in $\B$), the PDFs for $\dustvel$ are super-Gaussian (flat-topped with rapid falloffs). 
\label{fig:pdfs.CGM}}
\end{center}
\end{figure}
%
%

%
%
\begin{figure}
\begin{center}
\includegraphics[width=\columnwidth]{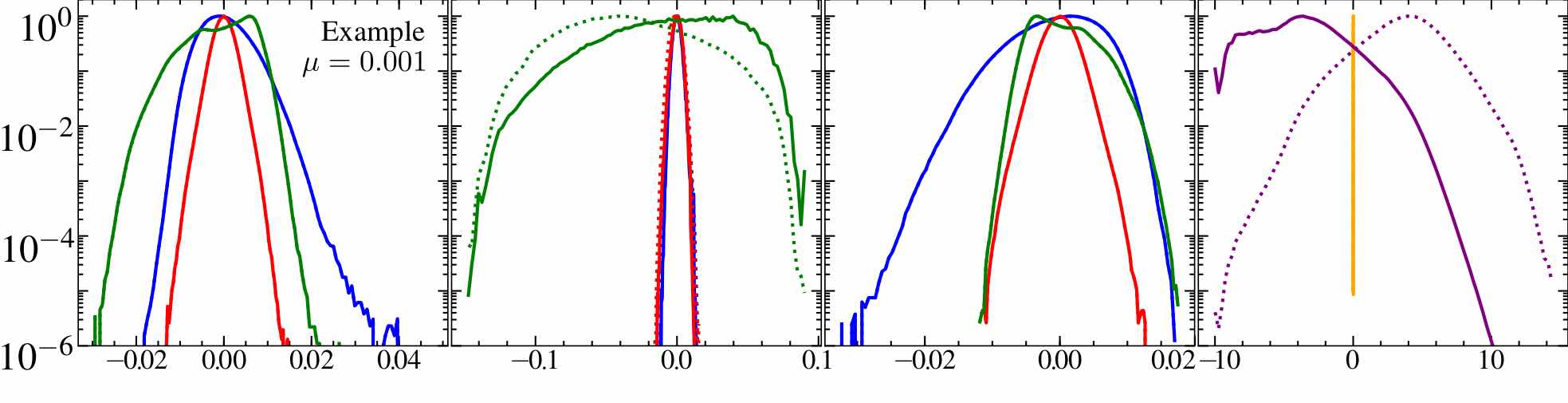}\vspace{-0.15cm} \\
\includegraphics[width=\columnwidth]{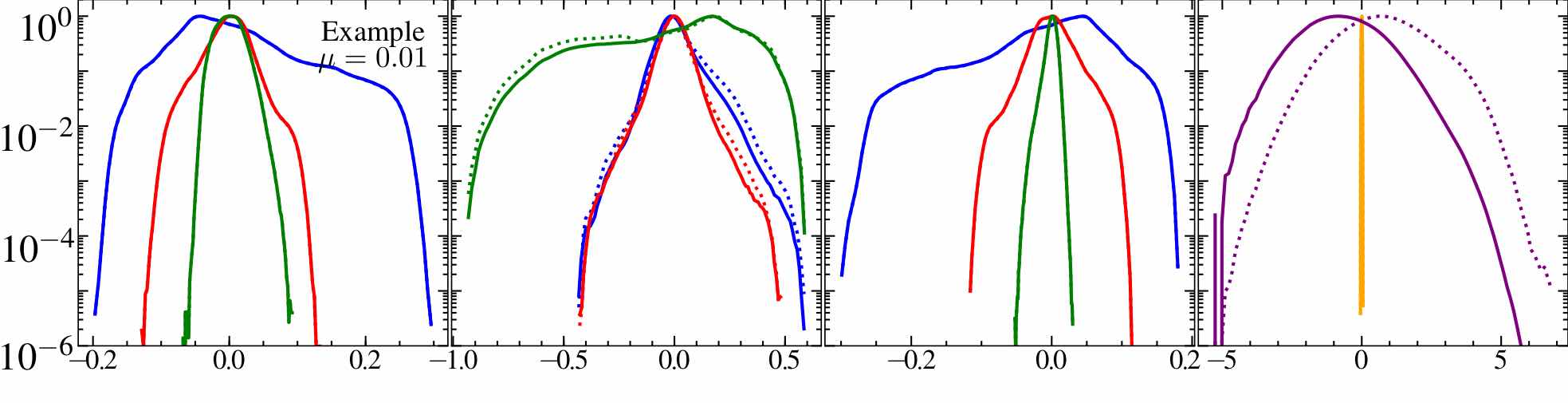}\vspace{-0.15cm} \\
\includegraphics[width=\columnwidth]{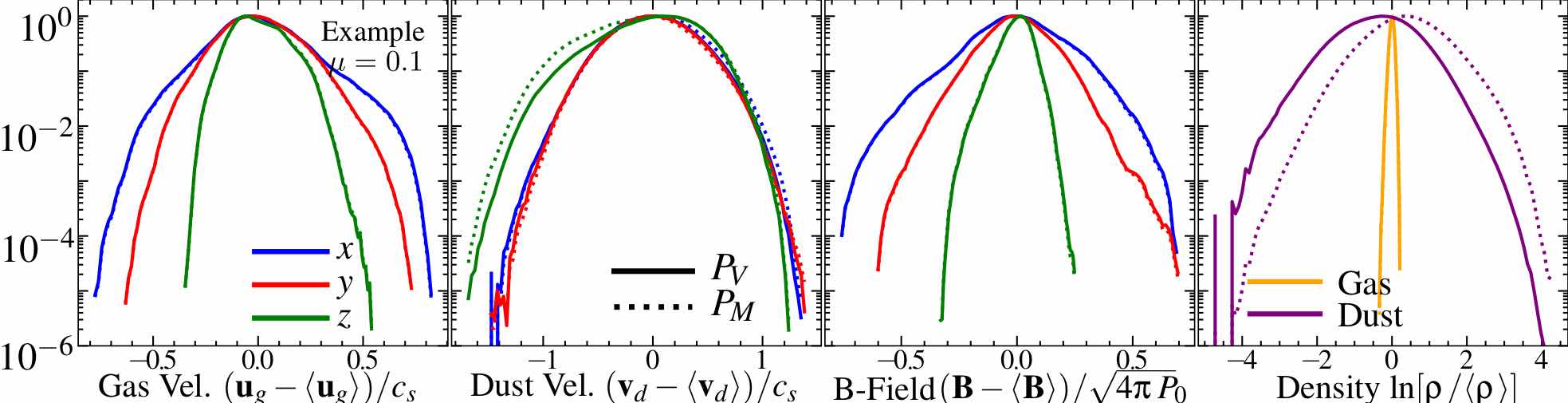} 
\vspace{-0.5cm}
\caption{PDFs for the runs in \fref{fig:example.dustgas}, where we vary the dust-to-gas ratio $\mu=0.001$ ({\em top}), $\mu=0.01$ (default; {\em middle}), $\mu=0.1$ ({\em bottom}) for run {\bf Example}. The PDFs for gas fluctuations and dust velocity become broader, more Gaussian, and more isotropic at higher-$\mu$, consistent with greater mixing and more uniform driving. But (per \fref{fig:example.dustgas}), the dust density fluctuations are {\em larger} at lower-$\mu$: 
\label{fig:pdfs.example.dustgas}}
\end{center}
\end{figure}
%
%

%
%
\begin{figure}
\begin{center}
\includegraphics[width=\columnwidth]{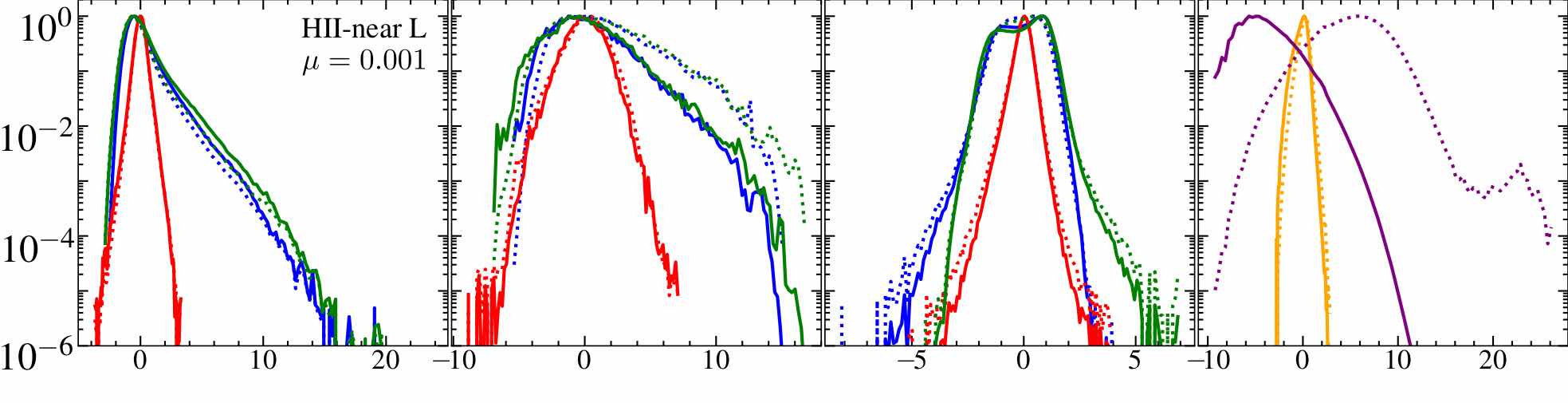}\vspace{-0.15cm} \\
\includegraphics[width=\columnwidth]{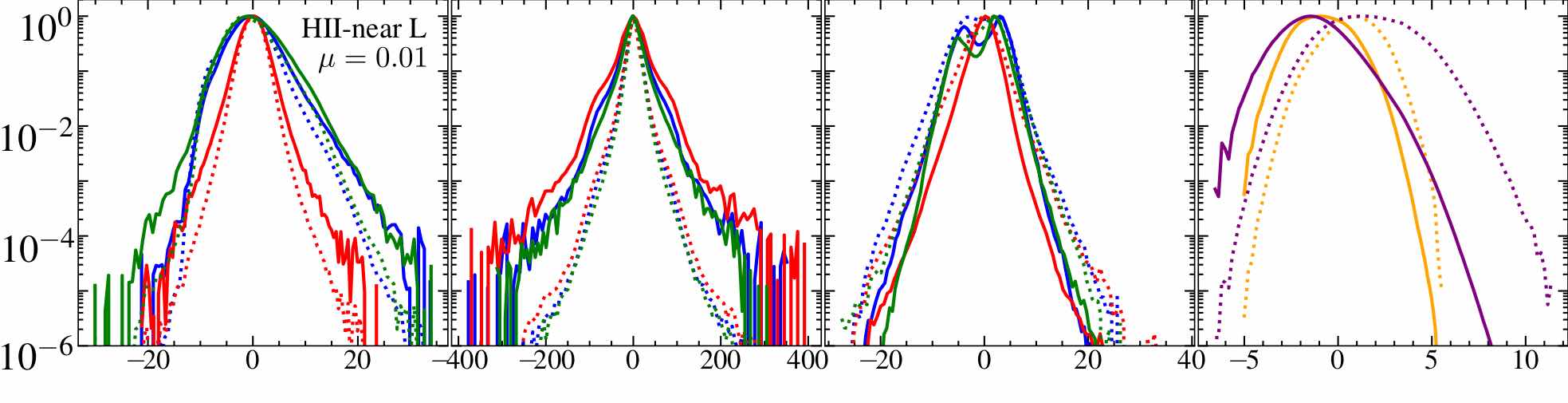}\vspace{-0.15cm} \\
\includegraphics[width=\columnwidth]{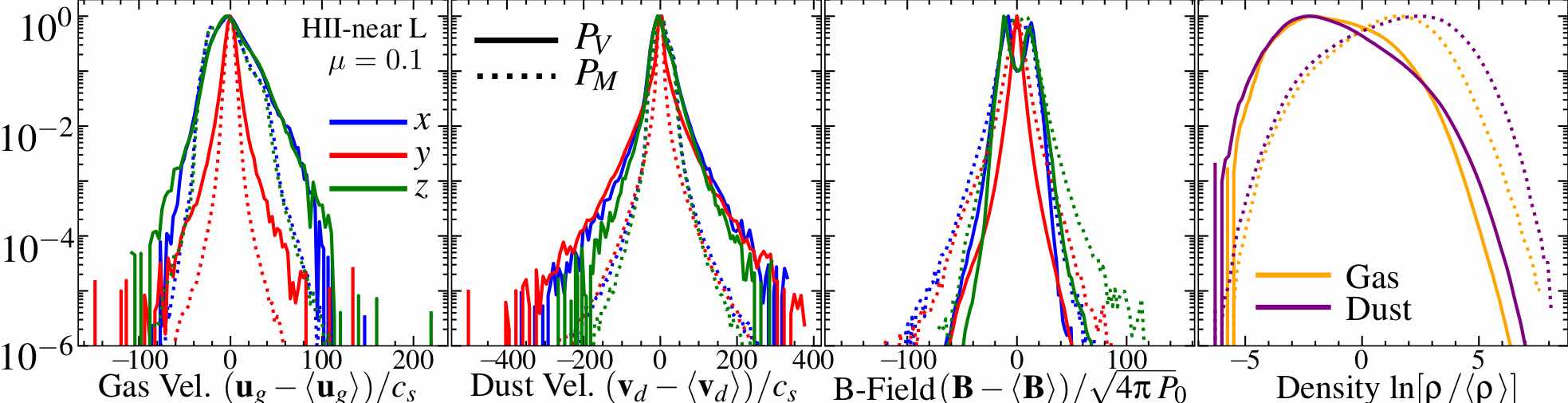} 
\vspace{-0.5cm}
\caption{As \fref{fig:pdfs.example.dustgas}, comparing different dust-to-gas ratios  $\mu=0.001$ ({\em top}), $\mu=0.01$ (default; {\em middle}), $\mu=0.1$ ({\em bottom}) for run {\bf HII-near} L (in \fref{fig:HIInear.dustgas}). Conclusions are similar to \fref{fig:pdfs.example.dustgas}. Note the extreme fluctuations in the low-$\mu$ case: most of the {\em volume} has $\dustden < 0.01\,\langle \dustden\rangle$ (i.e.\ is highly dust-depleted), while most of the dust {\em mass} resides in structures with  $\dustden \gtrsim 400\,\langle \dustden\rangle$ (highly dust-enriched), and $\gtrsim 0.1\%$ of the dust mass is at $\dustden \gtrsim 10^{9}\,\langle \dustden\rangle$ with the highest value here reaching $\sim 10^{12}\,\langle \dustden\rangle$.
\label{fig:pdfs.HIInear.dustgas}}
\end{center}
\end{figure}
%
%

\subsection{Distribution Functions}
\label{sec:results:PDFs}

Figures \ref{fig:pdfs.example}-\ref{fig:pdfs.CGM} examine the probability distribution functions (PDFs) of fluctuations for the seven simulations considered in this paper. We see that the large diversity of behavior in morphology, saturated amplitude, and anisotropy is mirrored in the diverse array of PDF shapes. The PDFs are often highly non-Gaussian; this comes (broadly speaking) in two ``forms.'' 

First, some PDFs exhibit substantial sub-structure, e.g.\ {\bf AGB}-M or {\bf HII-far} in $\delta\gasvel$ or $\delta \B$, which have multiple bumps (sharp inflection points/changes in curvature) or even second peaks. This is directly related to coherent, large-scale morphological structures in Figs.~\ref{fig:HIInear}-\ref{fig:WIM}. Each such ``peak'' corresponds to ``patches'' (sub-volumes of the parent box) which have either much larger or smaller dust density within them, and are have essentially de-coupled from one another, evolving non-linearly almost independently. 

Second, most of the ``smooth'' PDFs have non-Gaussian tails, which are most commonly exponential or ``stretched exponential'' in form: i.e.\ 
\begin{align} 
P(q)\propto \exp{(-|q/q_{0}|^{\gamma})}
\end{align}
for some $q$ and $\gamma$. For example,  the $\delta\B$ PDFs in almost all cases have this form, as do the $\delta \gasvel$ PDFs in {\bf Corona} and {\bf WIM} S/M and {\bf HII-far} S/M. PDFs with exponential or stretched-exponential tails are common in certain types of gas turbulence, velocity distributions of granular gases, and passive scalar concentrations in sub-sonic incompressible turbulence \citep{ruiz-chavarria:1996.passive.scalar.turb.poisson.tests,yakhot:1997.scalar.field.turb.pdfs,bennaim:2000.multiscaling.granular.pdfs,antal:2002.granular.vel.tails,ernst:2002.tail.boltzmann.scaling,kohlstedt:2005.granular.vel.pdfs,aranson:2006.granular.media.review,monchaux:2010.grain.concentration.experiments.voronoi,hopkins:2012.intermittent.turb.density.pdfs,colbrook:passive.scalar.scalings}. This generically arises from a competition between driving and dissipation. Consider the distribution of velocities in a statistically homogeneous system (so $P(\gasvel,\,{\bf x}) = P(\gasvel)$), where the velocities are driven by an uncorrelated stochastic process with the specific energy injection rate $\sim \dot{u}_{0}\,(v/v_{\ast})^{\xi}$ (the effective ``diffusion coefficient'' in velocity space), and damped\footnote{Note this damping corresponds to any process damping the fluctuations, not necessarily the dust drag.} with rate $t_{\rm damp}^{-1}\,(v/v_{\ast})^{\psi}$. In steady-state, if these driving and damping processes dominate, the PDF obeys 
\begin{align} 
\dot{u}_{0}\,(v/v_{\ast})^{\xi}\,\partial^{2} P(v)/\partial v^{2} \sim  -t_{\rm damp}^{-1}\,(v/v_{\ast})^{\psi}\,P(v)
\end{align} 
for each component, the solutions of which obey $P(v) \propto \exp{(-|v/v_{\ast}|^{\gamma})}$ at large $v$ with\footnote{For $\psi-\xi < -2$, the solutions become peaked functions $P(v)$ which asymptote to constant $P > 0$ as $v\rightarrow \infty$.} 
\begin{align}
\gamma &= 1 + (\psi-\xi)/2  \ \ \ \ , \ \ \ \ v_{\ast} \sim \sqrt{\dot{u}_{0}\,t_{0}} \ .
\end{align} So, for the simple case of white-noise (Brownian) driving and constant damping, or any case where driving and damping depend on $v$ in the same manner ($\psi\approx\xi$), the tails are exponential. The characteristic width of the PDF is simply given by $v_{\ast}^{2} \sim \dot{u}_{0}\,t_{0}$; i.e.\ the energy injected in a damping/dissipation time. 

While it is easy to qualitatively understand the range of PDF tails in this manner, such arguments fall far short of a predictive model. In other words, it is not possible (given the arguments here alone) to predict the PDFs and structure functions purely  from the various simulation parameters. For example, it is not {\em a priori} obvious what the relevant driving and damping rates should be in the saturated regime. If turbulence dominates both the non-linear forcing and the damping (e.g.\ eddies shearing apart growing modes, as we argue sets the saturation amplitude of some modes below) then both injection and damping times might scale with eddy turnover times, as in e.g.\ the standard theory of supersonic turbulent density fluctuations \citep{vazquez-semadeni:1994.turb.density.pdf,hopkins:frag.theory,hopkins:2012.intermittent.turb.density.pdfs,squire.hopkins:turb.density.pdf}. But driving could also arise directly from mode growth, or from non-linear parasitic modes, while damping could also stem from drag (acting on dust velocity fluctuations) or sound waves (for gas pressure fluctuations) -- the dominant terms do not have to be the same for each type of fluctuation.

Some of the PDFs exhibit strong asymmetries, with a much stronger ``tail'' in one direction (e.g.\ $\gasvel$ and $\dustvel$ in {\bf AGB} L/XL, $\dustden$ in {\bf HII-far} M/L or {\bf CGM}). In cases like {\bf AGB} or {\bf HII-far} where the tails extend to both larger $\gasvel$ in the direction of acceleration $\acc$ and larger $\dustden$, this relates to the fact that regions with non-linearly larger local $\dustden/\gasden$ experience faster local growth of the instabilities, and more efficient acceleration of the coupled dust-gas mixture (because the gas acceleration scales as $\sim \mu\,\acc$). Cases with a large tail in $\dustden$ towards smaller $\dustden$ ({\bf CGM}, {\bf WIM}) likely arise because dust is locally expelled from small pockets that are local vorticity maxima. This is essentially the generic mechanism studied in \citet{hopkins:2013.grain.clustering} and is well-known in terrestrial particulate ``preferential concentration'' studies \citep{squires:1991.grain.concentration.experiments,fessler:1994.grain.concentration.experiments,rouson:2001.grain.concentration.experiment,gualtieri:2009.anisotropic.grain.clustering.experiments,monchaux:2010.grain.concentration.experiments.voronoi}. Essentially, grains are centrifugally ``flung out'' of high-vorticity regions, and concentration is a side effect as grains collect in-between. Finally, there are examples that are highly non-Gaussian but do not neatly fit into any of the classes above (e.g.\ $\dustvel$ in {\bf CGM} or {\bf WIM} L). 

We also compare the volume-weighted ($P_{V}$, the probability that a given random volume has some value) and mass-weighted (dust weighted by dust mass, gas by gas mass; $P_{M}$) statistics. If the density variations are small, these must be similar. Indeed for the {gas} velocity and magnetic field ($\gasvel$ and $\B$) the two statistics rarely differ dramatically, as the gas density fluctuations are small. The few cases (e.g.\ {\bf HII-near} L) where some differences appear are those with the most dramatic gas density fluctuations, and they still do not differ qualitatively. Even for the dust, where $\dustden$ varies strongly, the differences are rarely dramatic in $\dustvel$ (an exception is {\bf HII-near} L, where dust in the high-$\dustden$ filaments is coherently moving relative to the low-$\dustden$ ``background,'' giving rise to larger dispersions when volume-weighted). 

In the density PDFs, volume and mass weighting makes more difference. Some of this is by definition: because $\rho = d{\rm Mass}/d{\rm Volume}$, 
\begin{align}
P_{M} \propto \rho\,P_{V}
\end{align}
if both are measured in differentially small regions. So $P_{M}(\rho)$ will always be biased to higher $\rho$ than $P_{V}(\rho)$. Let us define the mass and volume-weighted averages $\langle(\cdot)\rangle_{M}$ and $\langle(\cdot)\rangle_{V}$ respectively, the volume-weighted variance $S$ in the log-density field $\chi \equiv \ln{\rho}$, and the usual ``clumping factor'' $C$:
\begin{align}
S & \equiv S_{V} \equiv \langle \ln[\rho/\rho_{0}]^{2} \rangle_{V} - \langle \ln[\rho/\rho_{0}] \rangle_{V}^{2}  = (\delta \ln[\rho/\rho_{0}])^{2} \\ 
S_{M} & \equiv \langle \ln[\rho/\rho_{0}]^{2} \rangle_{M} - \langle \ln[\rho/\rho_{0}] \rangle_{M}^{2} \\
\rho_{0} & \equiv \langle \rho \rangle_{V} \\
C & \equiv \frac{\langle \rho^{2} \rangle_{V}}{\langle \rho \rangle_{V}^{2}}
\end{align}
where $\rho_{0} = \langle \rho \rangle_{V}$ follows from our definitions. 
{\em If} $P_{V}(\rho)$ is exactly log-normal, then mass-conservation implies: (1) the volume-weighted mean/median/mode of $\chi$ is $\langle \ln(\rho/\rho_{0})\rangle_{V} = -S_{V}/2$, (2) $P_{M}$ is also lognormal with $S_{M}=S_{V}$ and mass-weighted $\langle \ln(\rho/\rho_{0})\rangle_{M} = +S_{V}/2$, (3) $C=\exp{(S_{V})}$. 

One way that we can quantify the deviations from log-normal in these PDFs is via the extent to which $S_{M} \ne S_{V}$ or $C \ne \exp{(S_{V})}$ (this is commonly used as a diagnostic in  supersonic  turbulence studies). This can be read off from the values of $S_{V} = (\delta\ln{[\rho/\rho_{0}]})^{2}$ and $C$ in \tref{table:saturation}. In several cases (e.g.\ {\bf AGB}) the results are consistent with log-normality ($C \approx \exp{(S_{V})}$). But in some very strongly-clumped cases we have, for the dust, $S_{M} > S_{V}$ and $C \gg \exp{(S_{V})}$ (e.g.\ {\bf HII-near} and {\bf HII-far} M,L have $S_{V}\sim 2$ and $S_{M}\sim4-6$). Specifically, given the measured $\delta\ln{(\dustden/\langle \dustden\rangle)} = S_{V}^{1/2}$ (for the dust), and assuming log-normality ($C=\exp{(S_{V})}$) would predict $C\sim 5-10$ in these cases. In contrast the measured $C$ is $C\sim 100-300$ in  {\bf HII-near} L and {\bf HII-far} L, and $C\sim 1500$ in {\bf HII-far} M. Conversely, the PDFs with tails towards low $\dustden$ have smaller $C$ than would be predicted from $\delta\ln{(\dustden/\langle \dustden\rangle)}$; e.g.\ the {\bf CGM} and {\bf WIM} runs have $C\sim 1.0-1.1$. This owes directly to the large, asymmetric tails in $\dustden$ visible in Figs.~\ref{fig:pdfs.example}-\ref{fig:pdfs.CGM}.

Another, perhaps simpler, way of emphasizing this is to note that the most extreme cases (e.g.\ {\bf HII-far} M, or {\bf HII-near} L and {\bf Example} with lower $\mu=0.001$) have extreme tails, where $\gtrsim 0.1\%$ of the dust mass lies at densities $\gtrsim 10^{7}$ times  larger than the mean  (reaching as high as $\sim10^{9}-10^{12}$). This corresponds to values $\gtrsim 15$ standard deviations from the volume-weighted  median of the density PDF.

\section{Discussion}
\label{sec:discussion}

From the analysis in \sref{sec:results}, we can identify a number of important conclusions about both the linear and early non-linear phases, as well as the saturated and late non-linear phases.

\subsection{Linear and Early Growth Phases}

\begin{enumerate}

\item{In all cases, the instabilities grow and produce strongly non-linear properties in the dust (see Figs.~\ref{fig:HIInear}-\ref{fig:CGM}). This is not surprising, since all cases here are linearly unstable (\fref{fig:rates}). However, it does show that non-linear growth can occur even when the growth timescale is shorter than the dust ``stopping time'' or Larmor time (i.e.\ $\Im{(\omega\,\ts)} > 1$ or $\Im{(\omega\,\ts)} > \tau$, in \fref{fig:rates}). This implies that the instabilities grow faster than the timescale for the system to reach a ``new'' equilibria, which is not a trivial conclusion.}

\item{Generally, the instabilities exhibit the most rapid initial growth at the smallest scales. All wavelengths here are linearly unstable, with growth rates generally increasing at smaller wavelengths (\fref{fig:rates}), because we do not include explicit dissipation (e.g.\ viscosity).\footnote{In fact (see \paperone) some of  instabilities (e.g.\ the \Alf\ wave and gyro resonances) have growth rates that can continue to rise with decreasing wavelength even below the field-parallel viscous scale, potentially down to the ion gyro radii or even further.} Thus, within each parameter set, the smaller boxes evolve more rapidly in fixed physical units (e.g.\ $\initvalupper{\ts}$). Moreover {\em within} each box, in most cases we see small-scale modes saturate first, with the scale of structures increasing until box-scale modes saturate (compare e.g.\ early and later times in \fref{fig:HIInear}, \ref{fig:HIIfar}, \ref{fig:example.physics}).}

\item{The simulations broadly reproduce the expected growth rates from linear theory in the early linear growth phase, as depicted in \fref{fig:growth}. We demonstrate and discuss this in more detail for a few representative acoustic RDIs in \citet{moseley:2018.acoustic.rdi.sims} (Fig.~2 therein) and our {\bf Example} case in \citet{seligman:2018.mhd.rdi.sims} (Figs.~4-5). For modes that are unstable on all scales (e.g.\ the \Alf\ MHD-wave RDI), the early growth rate appears to match the linear theory prediction for wavenumbers $k \sim (1 - 3)/\resolution$ (where $\resolution = \Lbox/N_{\rm 1D}$ is the initial inter-element grid spacing; again see also \citealt{moseley:2018.acoustic.rdi.sims}, Fig.~2). This suggests that the growth rate can be recovered even if the full wavelengths are resolved by just $\sim 2-6$ inter-particle spacings.\footnote{For rigorous demonstration of this in idealized test problems, as well as formal convergence studies, see \citet{moseley:2018.acoustic.rdi.sims}, Appendices~A-B. This is also consistent with the results using different codes and numerical methods, e.g.\ \citet{johansen:2007.streaming.instab.sims}.} At later times, the box-scale modes take over as the strongest growing modes, until saturation.}

\item{The presence of magnetic fields and grain charge, along with associated Lorentz forces, play a critical role in the linear development of the instability. For several cases studied here, if we artificially remove the dust charge (Lorentz forces), or magnetic fields, the instabilities become stable or orders-of-magnitude more slowly-growing.  \fref{fig:example.uncharged} shows that no structure  develops at late times in the {\bf Example} run with zero grain charge. Moreover, increasing the grain charge ($q_{\rm grain}$ and $\tau$) produces faster-growing instabilities with more violent saturated dust behavior, in {both} the ``clumped'' and ``dispersed'' regimes (see  Figs.~\ref{fig:HIInear.physics.variations}, and \ref{fig:CGM}). This is despite the fact that Lorentz forces decrease the ``equilibrium'' dust drift velocities, and magnetic fields increase the pressure support of the gas, which would naively appear to be ``stabilizing'' effects. But these terms also introduce a variety of new dust and gas modes (e.g.\ \Alf\ and slow waves, dust gyro motion) which in turn dramatically increase the number of accessible ``resonances'' for the instabilities (as well as introducing new energy sources for the instabilities).}

\item{The ``resonances'' where linear growth rates are maximized are sufficiently vigorous that they can often be identified  well into the non-linear evolution. These maximal resonances occur where the ``natural frequency'' of advection $\driftvel \cdot {\bf k}$ or gyro motion $\tL^{-1}$ in the dust matches the ``natural frequency'' of an \Alf\ or magnetosonic wave in the gas. Even in the non-linear phases of evolution, these resonances manifest as  particular angles or wavelengths of the structures that form, as can be seen in Figs.~\ref{fig:example.dustgas}-\ref{fig:rates.example.dustgas}.}

\end{enumerate}

These results are all, to some extent, predicted from the linear theory in \paperone, but we  both verify the linear theory results and confirm that these conclusions persist even well into non-linear evolution.

\subsection{Saturation \&\ Late Non-Linear Phases: Generic Conclusions}

As noted above, the saturated states exhibit some qualitatively different behaviors, but there are some generic conclusions that apply to all of our runs. We discuss these first, before exploring the physics that is distinct in different saturated states. 

\begin{enumerate}

\item{All of the systems saturate in a turbulent quasi-steady state. This is evident in \fref{fig:growth}, where the velocity dispersion has clearly reached  saturation,  although in some cases certain fluctuations continue to grow very slowly. The finite velocity dispersion of the dust grains in the saturated state does not shut down fluctuations, even in the cases that reach a nearly isotropic grain velocity distribution function (e.g.\ \fref{fig:pdfs.CGM}). In other words, the turbulence reaches a saturated steady state, rather than the instability quenching itself.}

\item{The predicted structure here is qualitatively completely different from that formed in ``passive'' dust experiments, in which the  dust moves  as a tracer particle in externally driven turbulence (neglecting the forces from  dust on gas, i.e.\ the momentum-conserving ``back-reaction'' terms). In other words, ``passive'' dust cannot generate these instabilities or structures -- indeed, the RDIs do not exist with ``passive'' dust. Some of the most noticeable differences are found in the large-scale dust morphology; there is a much stronger prevalence of filaments and sheets here, compared to  no strong anisotropy present in ``passive dust'' simulations. Also in ``passive dust'' simulations,  the PDFs of dust density do not have the same shape or qualitative scalings as those presented here (compare \citealt{hopkins.2016:dust.gas.molecular.cloud.dynamics.sims,lee:dynamics.charged.dust.gmcs}, or the discussion in \citealt{moseley:2018.acoustic.rdi.sims}). In most cases we study, the dust density fluctuations are vastly larger in our simulations (with ``active'' dust which can drive the RDIs) as compared to an analogous ``passive-dust'' case. For example, with  $\tilde{\alpha}\sim 0.001$, similar to some of our most strongly-unstable and dust-clumped cases in {\bf HII-near}, {\bf HII-far}, and {\bf AGB} L, \citealt{lee:dynamics.charged.dust.gmcs}  typically found $\delta \ln{(\dustden/\gasden)} \sim 0.01$ for ``passive'' grains (orders-of-magnitude smaller than our result here). Perhaps most importantly,  the scaling of the strength of dust clustering with $\tilde{\alpha}$ ($\Lbox$) or $\tilde{\phi}$ ($\tau$) in ``passive dust'' studies is, in many cases, almost {\em opposite} those here (e.g.\ \citealt{lee:dynamics.charged.dust.gmcs} found fluctuations in dust-to-gas ratio with ``passive'' dust were only strong at $\tilde{\alpha} \gg 1$). }

\item{Details of the gas equation-of-state, the functional form of the drag law (Coulomb+Epstein or just Epstein), or the grain charge scaling (dependence on local temperature and density) do not qualitatively alter our conclusions, although they certainly have quantitative effects (\tref{table:saturation}, and Figs.~\ref{fig:example.physics},\,\ref{fig:CGM}, and \ref{fig:HIInear.physics.variations}). Larger dust charge generally produces more violent saturation (Figs.~\ref{fig:HIInear.physics.variations} and \ref{fig:CGM}). The non-linear behavior of the instabilities does not depend sensitively on particular alignments or anti-alignments between acceleration and magnetic field directions (except insofar as the resonant angles change; e.g.\ \fref{fig:HIInear.angle.variations}), and in fact, cases where the two are more strongly anti-aligned can even grow faster, despite weaker grain drift. Likewise, modest variation in parameters like the equilibrium grain drift velocity (relative dust-gas acceleration) or magnetic $\beta$ do not qualitatively alter the behavior or character of saturation (\fref{fig:example.physics} \&\ \ref{fig:WIM.lowVdust}). Lowering the dust-to-gas ratio produces slower initial growth and weaker {\em gas} turbulence, as expected. However, surprisingly, it can produce non-linear clustering in the dust that is as strong, or even stronger, than higher-$\mu$ cases (Figs.~\ref{fig:example.dustgas}, \ref{fig:HIInear.dustgas}, \ref{fig:pdfs.example.dustgas}, \ref{fig:pdfs.HIInear.dustgas}).}

\item{Most systems are driven towards approximate equipartition between gas velocity and magnetic field fluctuations. This agreement -- i.e.\ 
\begin{align} 
\langle \gasden \rangle\,\delta\gasvel^{2}/2 \sim \delta\B^{2}/8\pi 
\end{align}
is at the order-of-magnitude level, as shown in \fref{fig:correlations}.\footnote{There are a few notable exceptions with $\langle \gasden \rangle\,\delta\gasvel^{2}/2 \gg \delta\B^{2}/8\pi$. Two are {\bf WIM}-L  and {\bf Corona}-L:$\tau$=100, although the variant {\bf WIM}-L:LoV (``low drift velocity'') and ``default'' (higher-$\tau$) {\bf Corona}-L have $\langle \gasden \rangle\,\delta\gasvel^{2}/2 \approx \delta\B^{2}/8\pi$ to within $\sim 10\%$. In both of these exceptions, the gas moves nearly incompressibly and two-dimensionally, so the $\B$ fields are moved in the $xy$ plane but not compressed, generating negligible $\B$ fluctuations (see \fref{fig:WIM}). Our default (high-$\tau$) {\bf CGM} boxes also exhibit low $\delta\B$, though here it may be because the fluctuations are dominated by small-$k$, random gas motions which do not cause an effective coherent dynamo.} This result is independent of the initial $\beta$ (from $\initvallower{\beta}\sim0.001-1000$). In some cases this involves strong amplification of  $\B$ fields  (e.g.\ from $\beta\sim 1$ to $\sim 0.01$ in {\bf AGB}-XL). Because of this, while the instabilities can drive highly super-sonic turbulence in some cases, it is usually trans-\Alf{ic}.}

\item{All of the instabilities examined saturate with sustained gas turbulence. To rough order-of-magnitude, saturation often occurs when the eddy turnover timescale on the box scale becomes shorter than the box-scale linear growth timescale i.e.\ $|\delta \gasvel |/\Lbox \sim \Im{(\omega[\Lbox])}$. However, as discussed below, for both very small boxes and some magnetically-dominated boxes, other  criteria (e.g.\ equipartition between magnetic tension and driving by dust) may instead set the saturation amplitude (see \fref{fig:correlations}). In any of these cases, the strength of the saturated {\em gas} turbulence increases with dust-to-gas ratio and box size/wavelength (see \fref{fig:correlations}, \tref{table:saturation}). This can be understood physically, since the forcing of the dust onto the gas becomes stronger relative to pressure and magnetic forces. Provided some gas velocity fluctuations, the gas density fluctuations roughly follow the usual relation for pure isothermal MHD turbulence, 
\begin{align}
[\delta\ln{(\gasden/\langle\gasden\rangle)}]^{2}=\ln{[1+(b\,|\delta\gasvel|/\cs)^{2}]} \ , 
\end{align}
but with substantial variation in the ``compressibility'' $b\sim 0.2-1$ (see \fref{fig:correlations}). $\B$ can be related to $\gasvel$ as described above. Anisotropy in the gas properties can usually be understood as a direct reflection of the anisotropy in the fastest-growing linear modes at the box scale (see below).}

\item{The dust saturation is ubiquitously more complex than the gas saturation. In some cases, the dust exhibits extremely strong ``clumping'' or clustering, with a wide range of distinct morphologies and topologies (e.g.\ differently-connected sheets, filaments, or clumps). In the most extreme cases simulated here, the dust  over-densities reach magnitudes of $\sim 10^{9-12}$ times the mean, as seen in Figs.~\ref{fig:pdfs.example}-\ref{fig:pdfs.HIInear.dustgas}! In other cases, the dust is ``dispersed'' throughout the box, with nearly-isotropic, large, velocity dispersions. Qualitatively, the anisotropy of $\dustvel$ and relation between $\dustvel$ and $\gasvel$ reflect those of the fastest-growing linear modes at the box scale (see below). Generically, on ``intermediate'' and ``large'' scales $k\lesssim 1/(\mu\,\cs\,\ts)$, we expect and see $|\delta \dustvel|\gtrsim |\delta\gasvel|$, while on ``small'' scales $k\gtrsim 1/(\mu\,\cs\,\ts)$, $|\delta \dustvel|\lesssim |\delta\gasvel|$ (see \fref{fig:correlations}).\footnote{This follows from the linear RDI behavior and  can be understood from a local-balance-type argument from the equations for ``forcing'' the dust via gas. Dimensionally, a linear perturbation $\delta \dustvel$ should have $\omega\,\delta\dustvel \sim \delta\gasvel/t_{0}$ where $t_{0}\approx\ts$ if drag dominates, or $t_0\approx\tL$ if Lorentz forces dominate, so $|\delta\dustvel| \sim |\delta\gasvel|/(\omega\,t_{0})$. But generically for the MHD-wave RDI-type modes, $\omega \sim t_{0}^{-1}\,(\mu\,k\,\cs\,t_{0})^{\nu}$ with $\nu\sim 1/3-2/3$ depending on wavelength, so the scaling switches from $|\delta\dustvel| \gtrsim |\delta\gasvel|$ to $|\delta\dustvel| \lesssim |\delta\gasvel|$ around $k\sim 1/(\mu\,\cs\,t_{0})$.} On the very largest scales, the dust density fluctuations $\delta\dustden$ are comparable to gas density fluctuations $\delta\gasden$. But,  while $\delta\gasden$ decreases with scale, $\delta\dustden$ does not, because  there is no internal pressure resisting compressions. In fact, some of the most extreme dust-density fluctuations appear when the gas is nearly incompressible, and they actually become {\em stronger} at lower dust-to-gas ratios (Figs.~\ref{fig:pdfs.example.dustgas}-\ref{fig:pdfs.HIInear.dustgas}).}

\item{The statistics of both dust and gas fluctuations are often highly non-Gaussian (Figs.~\ref{fig:pdfs.example}-\ref{fig:pdfs.HIInear.dustgas}), with  exponential or ``stretched exponential'' tails and, in some cases, coherent sub-structure. This is generally associated with strong intermittency and stochastic driving in dissipative systems (\sref{sec:results:PDFs}). These strong deviations from Gaussianity mean, for example, that the mass-weighted dust density fluctuations can deviate substantially from volume-weighted fluctuations, and in some cases a significant fraction of the dust mass ($>0.1\%$) can reside at values $\gtrsim 15-20\,$ standard deviations from the median.}

\end{enumerate}

\subsection{Saturation: ``Clumped'' States}\label{sec:sub.clumped}

Although it is clear here that the saturated states are diverse and occupy a continuum of properties, we attempt  to classify them into two very broad ``regimes'', based on their morphology and resemblance to intuition from linear theory. First, we note that despite their obvious differences, boxes {\bf Example}, {\bf AGB}, {\bf HII-near} and {\bf HII-far} have several qualitative properties in common. These runs all have $\beta \gtrsim 1$, and $\tau \lesssim 100$, a value that is not {\em too} large. They all share a defining feature,  that the dust is strongly ``clumped'' and remains highly anisotropic even in saturation. Prominent clumps, plumes, filaments, and sheets appear, even when the gas is only weakly compressible.

In these, the ``medium'' and ``large'' boxes ({\bf Example}, {\bf AGB} M/L/XL, {\bf HII-near} M/L, {\bf HII-far} M/L) saturate with significant anisotropy or bias in $\delta\gasvel$ along the direction of the acceleration $\acc$ (as opposed to e.g.\ $\driftvel$ or $\B$). The components in the perpendicular direction are not negligible and the strength of the anisotropy varies, owing to mixing from the Lorentz forces. These runs also generally have 
\begin{align}
|\delta \B |/|\B| \sim |\delta \gasvel|/v_{A}
\end{align} 
(kinetic/magnetic energies similar) with anisotropies oriented in the same plane(s). Moreover 
\begin{align}
|\delta\dustvel|\sim |\delta\gasvel| \ ,
\end{align}
 with $\delta\dustvel$ typically slightly larger, but not by more than a factor of a few. In the largest boxes, $\delta\dustden/\dustden\sim \delta\gasden/\gasden$, (\fref{fig:correlations}), and the PDFs become increasingly Gaussian/lognormal (Figs.~\ref{fig:pdfs.agb} and \ref{fig:pdfs.HIInear}) especially at high-$\mu$ (\fref{fig:HIInear.dustgas}). In the intermediate-size-scale (M) boxes $\delta\dustden/\dustden > \delta\gasden/\gasden$. 

These behaviors can all, remarkably,  be predicted (at least qualitatively) by the {\em linear} properties of the fastest-growing modes at the box scale. These predictions are discussed in detail in \citet{hopkins:2017.acoustic.RDI} (Fig.~2) and \paperone\ (\S~4-5), and we briefly summarize them here. If the relevant modes at the box scale are the MHD-wave RDI modes or the aligned modes ($\hat{\bf k}=\driftvelhat$, also called ``pressure-free'' or quasi-sound/drift modes), then for modest magnetization ($\beta^{-1}$ and $\tau$ not too large) most of the insight can be gained from considering the much simpler pure hydro case (see \paperone\ for further discussion). At $k \lesssim \mu/\cs\,\ts$ ({\bf AGB} L/XL and {\bf HII-near} L, where $\Lbox/\cs\,\ts \gtrsim 1000$), the aligned ``pressure-free'' mode dominates, where internal pressure effects of the gas are weak compared to the bulk force from dust on gas. In this mode $\delta\dustden /\langle \dustden\rangle \approx \delta\gasden/\langle \gasden\rangle$ and $\delta \dustvel \approx \delta \gasvel$ (i.e.\ dust and gas fluctuate together; see \fref{fig:correlations}), with $\delta \gasvel \propto \hat{\bf k}$ (fluctuations are longitudinal) and maximum growth rates at $\hat{\bf k} = \driftvelhat$. Because of the weak pressure effects, $\B$ is driven passively by the velocity fluctuations so $|\delta{\B}|/|\B| \sim |\delta\gasvel|/v_{A}$ (with $\delta{\B}$ orthogonal to $\delta\gasvel$ in the $\B-\delta\gasvel$ plane).\footnote{For example, if we assume $\delta\gasvel\propto \acc \propto (\sin{\theta_{\B\acc}},\,0,\,\cos{\theta_{\B\acc}})$ in $(\hat{x},\,\hat{y},\,\hat{z})$, then since $\hat{\bf B} = \hat{z}$, for linear perturbations $\delta\B \propto (-\cos{\theta_{\B\acc}},\,0,\,\sin{\theta_{\B\acc}})$. So for our runs with $|\initvallower{\B}\cdot\acc|=\cos{\theta_{\B\acc}}=1/\sqrt{2}$, this gives $\delta\gasvel \propto (1/\sqrt{2},\,0,\,1/\sqrt{2})$ and $\delta\B \propto (-1/\sqrt{2},\,0,\,1/\sqrt{2})$ (i.e.\ the absolute magnitude of the anisotropy is the same in each direction for $\delta\gasvel$ and $\delta\B$). For our runs with $|\initvallower{\B}\cdot\acc|=0.05$, this gives $\delta\gasvel \propto (0.99,\,0,\,0.05)$ and $\delta\B \propto (-0.05,\,0,\,0.99)$ (so the dominant direction for $\delta\gasvel$ is $\hat{x}$, while that for $\delta\B$ is $\hat{z}$). These compare well to the results in \tref{table:saturation}.} Note that initially, $\driftvelhat$ is not aligned with $\acc$ (for non-zero $\tau$), so the modes produce the ``sheets'' of overdensity in dust perpendicular to $\driftvelhat$. However, because the pressure effects are weak in these modes, the non-linear forcing from $\acc$ tends to overwhelm competing forces like magnetic tension, and push the system to drift in the $\hat{\acc}$ direction (giving $\delta\gasvel \propto \acc$).

In the ``intermediate'' boxes ({\bf Example}, {\bf AGB}, {\bf HII-near}, and {\bf HII-far} M) the ``mid-$k$'' MHD-wave modes dominate at $\mu \lesssim k\,\cs\,\ts \lesssim \mu^{-1}$. Again the {linear} modes have $\delta \dustvel \sim \delta \gasvel$, $|\delta{\B}|/|\B| \sim |\delta\gasvel|/v_{A}$. At mid-$k$, the initially fastest-growing modes approximately satisfy $\hat{\bf k} \bot \B$  if $|\driftvel|\ll \cs$ (the \Alf\ or slow RDIs) or $\hat{\bf k} \bot \driftvelhat$ if $|\driftvel|\gg \cs$ (the fast RDI). This produces the perpendicular sheets and filaments extended along $\driftvelhat$, which are seen at early times. It also  explains the observed anisotropies, although these are weaker because the linear modes have a mix of components in each direction. Perhaps most notably, these boxes have $\delta\dustden /\langle \dustden\rangle \gg \delta\gasden/\langle \gasden\rangle$, which, as shown in \paperone, are likely related to the linear modes, which satisfy $\delta\dustden \sim \delta\gasden\,\Im{(\omega\,\ts)}$, i.e.\ $\delta\dustden /\langle \dustden\rangle \approx (\Im{(\omega\,\ts)}/\mu)\,\delta\gasden/\langle \gasden\rangle$. Because $\Im{(\omega\,\ts)} \sim (k\,\cs\,\ts)^{1/2}$ in this mode, the relative strength of $\delta\dustden / \delta \gasden$ {\em increases} at smaller scales and smaller $\mu$, consistent with our experiments (Figs.~\ref{fig:correlations} and \ref{fig:pdfs.example.dustgas}-\ref{fig:pdfs.HIInear.dustgas}). While this provides a reasonable qualitative explanation for the observed trends, we do caution that the {\em magnitude} of the saturated $\delta\dustden/\delta\gasden$ is often significantly larger than that predicted by linear theory. 

In the smaller boxes ({\bf AGB} S, {\bf HII-near} S, {\bf HII-far} S), the ``high-$k$'' MHD-wave modes dominate, with $k\gtrsim 1/(\mu\,\cs\,\ts)$. The fastest-growing mode directions are the same as the ``mid-$k$'' modes (\paperone), with a similar anisotropy (here $\delta\dustvel \propto \delta\gasvel \propto \hat{\bf k}$ to leading order, giving anisotropy in the $xy$ plane for $\delta\gasvel$, and in $z$ for $\delta\B$). Again $|\delta{\B}|/|\B| \sim |\delta\gasvel|/v_{A}$ in the linear mode and saturated turbulence (\tref{table:saturation}). As in the mid-$k$ modes, we have $\delta\dustden \sim \delta\gasden\,\Im{(\omega\,\ts)}$, but now with $\Im{(\omega\,\ts)} \sim (k\,\cs\,\ts)^{1/3} \gtrsim 1$, so the ratio $\delta\dustden/\delta\gasden$ continues to rise  with smaller $\mu$ or at smaller $k$ (making $\delta\dustden$ weakly-dependent on box size, with a small decrease to smaller $\Lbox$). This occurs even though  $\delta\gasden \rightarrow 0$ (see \fref{fig:correlations}). One notable difference from the mid-$k$ modes, however, is that the linear perturbations feature $|\delta\dustvel| \sim |\delta \gasvel| / \Im{(\omega\,\ts)} \ll |\delta \gasvel|$. This feature is also seen in the saturated turbulence (\fref{fig:correlations}).

We stress that despite their common elements, there are important differences across these clumped boxes, beyond just the magnitude of the effects. The morphology, topology, and even dimensionality (e.g.\ clumps, filaments, sheets) of the dust structures varies and depends on a complicated mix of both the global parameters (e.g.\ $|\driftvel|/\cs$, $\tau$, $\beta$, etc.), as well as scale, owing to the complex superposition of different modes. Box {\bf AGB}, with initial $\tau \ll 1$, is closest to the pure-hydrodynamic cases studied in \citet{moseley:2018.acoustic.rdi.sims}. As a result it saturates in primarily compressible, supersonic magnetosonic turbulence, with the saturation amplitudes for $\gasvel$ in boxes M/L/XL  well-predicted by the eddy turnover time argument (tested in detail therein), and $\delta\B$ following from $\delta\gasvel$. Box {\bf Example}, with higher $\tau \sim 30$, saturates in primarily incompressible MHD turbulence \citep[see][]{seligman:2018.mhd.rdi.sims}.  In this case, the saturation amplitude of $\delta\B$ (especially its variations with $\mu$) is more accurately predicted by assuming force balance between  forcing from dust and magnetic tension of the dominant (box-scale) modes, with $\delta\gasvel$ following from $\delta\B$, i.e.: 
\begin{align}
\frac{(\B\cdot\nabla)\B}{4\pi} \sim \frac{|\B |\,|\delta\B|}{4\pi\,\Lbox} \sim \frac{\dustden\,\driftvel}{\ts} \approx \dustden\,\acc
\end{align}
see \fref{fig:correlations} and \citealt{seligman:2018.mhd.rdi.sims}).

Although it is beyond the scope of this work to study cases where the external drift driving these instabilities initially is time-variable, it is worth noting that even if that drift were somehow ``turned off'' (which should allow the induced turbulence to decay), there is no obvious mechanism to disperse the dust-to-gas fluctuations formed. Also, in environments with some {\em externally} driven turbulence, it would be interesting to explore whether the net effect of this turbulence is to enhance the dust-to-gas ratio fluctuations (as occurs in the {\em absence} of RDIs; see \citealt{hopkins.2016:dust.gas.molecular.cloud.dynamics.sims,lee:dynamics.charged.dust.gmcs}) or to limit the saturation of the RDI-induced clumping.

\subsection{Saturation: ``Dispersed'' or ``Granular'' States}\label{sec:sub.dispersed}
We refer to the second  regime as ``Dispersed'' or ``Granular'', because the dust is generally more dispersive in these runs (boxes {\bf CGM}, {\bf WIM}, {\bf Corona}). It appears that the transition between the two regimes occurs as $\tau$  becomes very large, specifically $\tau\gtrsim 100$. In this regime, the saturated states of the instabilities begins to differ from the description above, and the dust  has more isotropic velocity dispersion and notably smaller density fluctuations  (especially at the high-$\dustden$ end, which is suppressed relative to low-$\dustden$; see Figs.~\ref{fig:pdfs.Corona}-\ref{fig:pdfs.CGM}). 

Many of the saturated properties are consistent with the dominant linear modes, as observed in the previous regime. Unlike the ``clustered'' boxes, which are dominated by a combination of the low-$k$ ``pressure free'' (and quasi-sound/drift) modes and mid/high-$k$ MHD-wave modes, at sufficiently high-$\tau$ the instabilities become increasingly confined along $\B$. Boxes {\bf CGM}, {\bf WIM}, {\bf Corona} are dominated by a combination of the strong $\B$-aligned ``cosmic-ray-like'' instabilities (see \paperone\ for details), together with the related gyro RDIs in {\bf WIM} S and {\bf Corona} S/M (\fref{fig:rates}). In linear theory, the fastest-growing modes (in both cases) have 
\begin{align} \hat{\bf k} \approx \hat{\B} \ ,
\end{align} 
with $\B$ field fluctuations transverse ($\delta\B$ preferentially in the $xy$ plane, similar to an \Alf\ wave), with\footnote{As discussed in \fref{fig:WIM}, {\bf WIM} L (the default run, with larger drift velocity) is the one notable exception with $|\delta \B|/|\B| \ll |\delta\gasvel|/v_{A}$.} 
\begin{align}
|\delta \B|/|\B| \sim |\delta\gasvel|/v_{A}\ .
\end{align} 
Like the ``clumped'' case the gas fluctuations $\delta\gasden$ can be related to $\delta\gasvel$ with the usual MHD turbulence scalings. However, the anisotropy is often weak, because (1) the overall turbulence is isotropized and (2) the linear modes have components in all directions. Also like the ``clumped'' case, the intermediate/large-scale boxes ({\bf CGM} M/L, {\bf Corona} M/L, {\bf WIM} L) have a saturation amplitude of the {\em gas} that is reasonably well explained by equating eddy turnover and growth timescales ($\delta \gasvel \sim \Im{(\omega[\Lbox]\,\Lbox)}$),\footnote{The notable apparent exceptions in \fref{fig:correlations} are the {\bf CGM} low-$\tau_{\rm low}$ runs. However, \fref{fig:CGM} show this initially grows vigorously at high-$k$ ($\sim 100/\Lbox$) and in fact it reaches a large $|\delta\gasvel|\sim \cs$, before the ``sheets'' break up and disperse the dust suppressing growth of larger-scale modes, and $\delta\gasvel$ actually decays somewhat before reaching its equilibrium value. If we use the higher $k$ at which the largest rapidly growing modes are present, and the larger $\gasvel$, before the isotropized dust orbits lead to less coherent turbulent motions in the gas, then these runs are plausibly consistent with the ``eddy turnover time'' saturation scaling.} while the small-scale\footnote{As discussed in \paperone, when $\tau \gg 1$, the dividing line between ``small'' and ``intermediate'' scales is not simply $k \gtrsim 1/(\mu\,\cs\,\ts)$, but can become a rather complicated expression of $\tau$, $\beta$, etc. Here they can be effectively defined by the presence in \fref{fig:rates} of modes with the ``high-$k$'' asymptotic scalings.} boxes ({\bf Corona} S, {\bf WIM} S/M) have a saturation amplitude  $\delta\gasvel$ that is better explained by the same ``high-$k$'' scaling as the ``clumped'' cases above (see final paragraph of \sref{sec:sub.clumped}).

In the linear modes of these high-$\tau$ cases, the $\dustvel$ perturbation is approximately a gyro orbit, i.e.\ preferentially equal power in the $xy$ direction. The scaling of $\delta\dustvel/\delta\gasvel$ is similar to the ``mid-$k$'' and ``high-$k$'' MHD-wave cases discussed above (i.e.\ substantially smaller in the high-$k$ limit) but enhanced by a factor between $\sim \mu^{-1/2}$ and $\mu^{-1}$ (depending on wavelength in the out-of-resonance gyro or cosmic-ray like mode; see \paperone, \S~4). This is directly evident in \fref{fig:correlations}, which shows that $|\delta\dustvel|/|\delta\gasvel|\gtrsim 1$ in many of these strongly magnetized cases. Finally, $\delta\dustden$ also scales qualitatively like the mid/high-$k$ MHD-wave modes,  in that it is weakly dependent on $\Lbox$ or $k$ (while $\delta\gasden$ decreases at lower $\Lbox$). However, in both the aligned cosmic-ray-like and gyro modes, the Lorentz motion is (to leading order) incompressible, with the dust density fluctuations suppressed by a factor $\sim \mu^{1/2}$ and gas density fluctuations by a factor $\sim\mu$ (see \paperone, \S~6.4). This suggests that $\delta\dustden \sim \exp{(\delta\ln{\dustden})}-1$ should be an order-of-magnitude lower compared to similar ``clumped'' runs. This intuition provides a surprisingly good fit to the difference between ``clumped'' and ``dispersed'' runs in \fref{fig:correlations}. 

This regime is analogously very broad, and there is no single, transcendent behavior that defines it. In {\bf Corona} M/L and {\bf CGM}, the aligned modes initially produce rather thick ``sheets'' in the $xy$ plane perpendicular to $\B$. These form as  dust particles move  slowly relative to each other {\em along} the field lines, and collapse into thin sheets, with an increase in $\delta\dustden$. However, once the sheet becomes thin, the acceleration on dust in direction $\hat{\acc} \ne \hat{\B}$ pushes the dust with a component transverse to $\B$ only at one point along $\B$. This excites gyro motion of dust about $\B$, but also drags the field and ``bends'' $\B$ locally (as opposed to simply pushing the entire field line uniformly, as in the initial state), generating a magnetic tension and exciting \Alf\ waves. That, in turn, can re-orient the gyro motion (bending or ``dispersing'' the sheet). How ``isotropized'' the dust is -- and, correspondingly, how uniformly the dust is spread -- depends on how easily the field can be bent. Thus in {\bf CGM}, with high-$\beta$, the fields and corresponding $\dustvel$ can be fully isotropized; in contrast, in {\bf Corona}-M, the low-$\beta$ and small scales mean the energy in the dust  cannot fully re-orient the fields, and the dust motion remains primarily in the $xy$ plane. The maximum dust velocity dispersion is set by equating the ``pumping'' of the gyro motion (acceleration $\acc$) with damping by drag, which just gives 
\begin{align}
|\dustvel| \sim |\acc|\,\ts \sim |\driftvel| \ ,
\end{align}
 with isotropic $\driftvel$. 

In contrast, in {\bf Corona} S, and {\bf WIM} S/M, the dust is collected in dense ``lines,'' or, more precisely, closed {\em vertical} sheets or ``tubes'', oriented along the $\B$ direction. The dominant modes are gyro modes (not the aligned modes), so the ``horizontal sheets'' discussed above do not form.\footnote{We have checked that there is no apparent correlation between the presence of these tubes and whether or not the box is large enough that the dust ``wrapping'' time ($\Lbox/\driftvel$) is shorter or longer than the mode growth time, which might artificially contribute to such structures. We have also checked for any dependence on whether the dust gyro radii are resolved or unresolved in the {\em gas} cells (because of our super-particle approach,  dust orbits are always resolved).} This produces the ``granular'' appearance of $xy$ slices. Without the ``sheets'' to bend the fields strongly at individual points, the field lines move coherently in aligned ``columns'' or ``tubes.'' Therefore, the turbulence is essentially two-dimensional, and because the scales are small the gas is weakly-compressible. These features are also evident in box {\bf WIM}-L, although this also has some aligned modes, giving it a mix of properties.

It is also worth noting that in some boxes (e.g.\ {\bf Corona}-S), the modes do not appear to reach the grid scale, even after the simulation has been evolved for longer than the box-scale mode growth time. This tends to occur when there are strongly-growing, dominant gyro modes on somewhat smaller scales which can retain their dominance, even at late times. There is some hint that these may continue growing, along with box-scale modes, more slowly (e.g.\ linearly in time). Unfortunately it is computationally prohibitive to run these boxes to arbitrarily long times.

We emphasize that there does not appear to be a sharp ``threshold'' where behavior changes between ``clumped'' and ``dispersed'' modes. Obviously, some ``clumped'' runs feature gyro modes and closer-to-isotropic grain velocity dispersions, while some ``disperse'' runs here (like {\bf Corona}-M) retain coherent dense grain structures well into their non-linear evolution. Rather, there is a spectrum of different behavior in different regimes, and different parameters involve a different mixture of these.

\section{Other Physics (Not Included Here)}
\label{sec:other.physics}

The simulations here are intentionally idealized, designed to study the physics and non-linear behaviors of the instabilities identified in \paperone. Of course, in different specific physical applications, there are an infinite variety of additional physics which could also be important, some of which may modify the instabilities themselves. Some of these cases will be studied in future work modeling observables in specific physical systems (e.g.\ AGB outflows and dust extinction/density estimation; Steinwandel et al., in prep.), or analytic studies focused on the linear instabilities in more complex systems (e.g.\ three-fluid radiation-MHD RDIs, Squire et al., in prep.). Other examples (viscosity, non-ideal MHD, stratification, differential rotation) are discussed in detail in \citet{hopkins:2017.acoustic.RDI}, \citet{squire:rdi.ppd}, and \citet{hopkins:2018.mhd.rdi}. But we briefly review some here for context. 

Small-scale transport processes (e.g.\ viscosity, conductivity) are discussed in \paperone. For the physical examples which motivate our parameter study in \tref{table:sims}, we specifically chose the parameters so that our smallest resolvable scales in the smallest (``S'') boxes correspond to the scale where viscous effects could begin to become significant ($k^{2} \gtrsim |\omega|/\nu$, where $\nu$ is the kinematic viscosity), so that all resolved scales in all our boxes correspond to the ``inertial range'' where viscous effects are weak (see \S~9 in \paperone, where these scales are estimated). But we also note, as shown explicitly in \paperone\ (e.g.\ Fig.~7 there and \citealt{squire:rdi.ppd}), that most of the instabilities here are undamped below the viscous scale (down to at least the ion gyro-radii), like e.g.\ the cosmic ray resonant and non-resonant instabilities \citep{wentzel:1968.mhd.wave.cr.coupling,bell.2004.cosmic.rays}. We also choose motivating parameters where non-ideal MHD effects should be negligible (these terms generally require an ionization fraction $f_{\rm ion}\ll 10^{-8}$ to be significant for the modes studied here; see \S~8.3 in \paperone), but note that even in overwhelmingly neutral gas (e.g.\ proto-planetary disks), ambipolar diffusion and the Hall effect do not actually damp many of the instabilities studied here (but instead modify them into additional branches; see e.g.\ \citealt{squire:rdi.ppd}, Figs.~7-8). Other small-scale effects such as the current carried by dust grains (\S~8.3 in \paperone), finite inter-grain separation effects, or  degeneracy pressure are completely negligible in any regime motivating our study here, or any regime where ideal MHD could actually apply (e.g.\ the dust current is sub-dominant by a factor $\sim 10^{10}\,f_{\rm ion}\,(T/10^{4}\,{\rm K})$ in the induction equation). 

On the largest spatial scales, global effects such as stratification, gradients in the external forces driving drift, rotation and shear make our local periodic-box approximation invalid. We have therefore specifically chosen parameters such that the largest scales in our largest boxes (``L/XL'') in \tref{table:sims} correspond to these scales. All of these global effects introduce new RDIs such as the \BV, vertical settling, and epicyclic RDIs (see e.g.\ Appendix~C in \citealt{hopkins:2017.acoustic.RDI} or \S~5 \&\ 6 in \citealt{squire:rdi.ppd}). 

There are also various effects which directly involve dust-gas interactions and operate on relatively long timescales such as gas cooling; dust accretion, mantle formation, and chemical enrichment (generally relevant in cold or very dense gas like AGB outflows); or sputtering of dust in hot gas (relevant in gas with $T\gtrsim 10^{6}\,{\rm K}\,(\mu/0.01)$). These can modify the equation-of-state of the gas, dust-to-gas ratios or dust size/charge, but generally operate on timescales much longer than (or at most comparable to) the RDI growth times. It is therefore unlikely that these strongly modify the RDIs directly: instead a medium with e.g.\ efficient cooling or sputtering would simply shift the parameters of interest here, having effectively lower temperature or dust-to-gas ratios, respectively. But it is likely that these processes are themselves strongly modified by the RDIs, as they depend sensitively on clumping factors of the dust and gas and their (grain-size-dependent) cross-correlation, a subject worthy of future study.

One effect which could modify the instabilities here is externally-driven turbulence. Although we clearly show the RDIs persist in the presence of self-excited turbulence, it is possible that if some other process drives gas turbulence much more strongly (on a given scale) it could significantly modify the RDIs. This could potentially set the saturation amplitude by e.g.\ shearing apart different modes if the eddy turnover times become much shorter than RDI growth times (see \citealt{squire:rdi.ppd}, \S~8), although it is also well-known from both theory and laboratory experiments that even test-particle grains ($\mu=0$, so no RDIs are present) are actually strongly clustered on small-scales in externally-driven turbulence \citep[see e.g.][]{squires:1991.grain.concentration.experiments,fessler:1994.grain.concentration.experiments,bracco:1999.keplerian.largescale.grain.density.sims,cuzzi:2001.grain.concentration.chondrules,rouson:2001.grain.concentration.experiment,monchaux:2010.grain.concentration.experiments.voronoi,pan:2011.grain.clustering.midstokes.sims,hopkins.2016:dust.gas.molecular.cloud.dynamics.sims}. This is most likely to be important in physical systems motivating boxes like {\bf WIM} or {\bf CGM:}$\tau_{\rm low}$, where strength of the saturated gas turbulence from the RDIs alone (\tref{table:saturation}) is significantly weaker than the level of turbulence observed.

A variety of physical effects will also become important specifically when the local dust-to-gas ratio $\mu$ becomes very large. The most obvious is grain-grain collisions, as the ratio of stopping to collision time scales as $t_{\rm stop}/t_{\rm collision} \sim \mu\,(\delta v_{\rm dd} / c_{s})$ (where $\delta v_{\rm dd}$ is the rms relative grain-grain velocity difference on infinitesimally small spatial scales). Obviously this will become important at the very high $\mu$ reached by some of our simulations on small scales, but how it modifies the distribution of $\mu$ itself and subsequent RDIs depends on the outcome of said collisions (which can produce grain growth/coagulation, elastic or inelastic scattering/bouncing, or grain shattering, depending sensitively on the local conditions and grain chemistry). In some cases, the self-gravity of a cluster of grains can become important before collisions (though this is more likely important in proto-planetary disks, it could occur under exceptional ISM conditions, see e.g.\ \citealt{hopkins:totally.metal.stars}). And if radiation pressure drives the dust drift, then the dust can become self-shielding if the dust-to-gas ratio within a patch of size $\sim \lambda$ exceeds $\mu(\lambda) \gg \tilde{\alpha}\,(\lambda/L_{\rm box})^{-1}\,(1+\lambda_{\rm rad}/\grainsize)$ (where $Q\sim (1+\lambda_{\rm rad}/\grainsize)^{-1}$ crudely approximates the grain absorption efficiency for incident radiation with wavelength $\lambda_{\rm rad}$). In that regime the the radiation can be multiply-scattered, leading to more complicated three-fluid (radiation, dust, gas) RDIs. The very high local $\mu$ reached here in the absence of these effects should be considered motivation for studying these different regimes.

\section{Summary \&\ Conclusions}
\label{sec:conclusions}

We have presented the first simulation parameter study of the non-linear regime of the \citet{squire.hopkins:RDI} ``resonant drag instabilities'' of charged dust grains in magnetized gas. Because the parameter space of the instabilities is large and this is a first study, we focused on several sets of initial conditions that are broadly representative of different astrophysical regimes. In all cases studied, the instabilities produce highly non-linear structures and fluctuations, often including strong turbulence and magnetic field amplification. Strong anisotropy and non-linear features appear in the dust,  including orbit-crossing, density fluctuations, and complicated velocity distribution functions. This  necessitates numerical simulations using dust particle methods which can follow the full non-linear dust velocity distribution function, as opposed to just the fluid limit.

Even within our small survey, the simulations exhibit a diverse array of behaviors. In \paperone\ we demonstrated using linear perturbation theory that a homogeneous MHD gas coupled to dust via Lorentz forces and drag exhibits around $10$ different ``instability families''. These families include the \Alf, slow, and fast magnetosonic MHD-wave RDIs; the three corresponding gyro RDIs; the ``pressure-free'' and related ``quasi-drift'' and ``quasi-sound'' modes; and the ``cosmic ray streaming''-type modes. These families all have different linear growth rates and mode structures, but often overlap and occur within the same system. Without introducing additional physics or constraints (e.g.\ Braginskii viscosity, which suppresses the growth rate of the magnetosonic modes, but not the \Alf\ modes), it is generally impossible to construct a simulation that isolates a single instability family.  Our range of initial conditions were chosen both  to be representative of different astrophysical regimes and also to exhibit different dominant, fastest-growing instability families. The resultant non-linear evolution yields a remarkable diversity of outcomes.

We show that over the course of the simulations, the instabilities become violently non-linear on all scales. Their non-linear outcomes can result in dust being highly concentrated (the ``clumped'' regime), or dispersed with large isotropic velocity dispersions (the ``disperse'' regime). In the ``clumped'' case, there are a wide range of morphologies, topologies, and dimensionalities of the clumped structures, depending on the parameters of the system, with dust in e.g.\ multiply-connected sheets, filaments, or point-like clumps. In the most extreme cases, the dust can reach enormous overdensities in these idealized tests ($>0.1\%$ of the dust mass at $\gg 10^{9}$ times the mean dust density, with volume-averaged ``clumping factors'' $>10^{4}$). The dust clumping does not depend systematically on the spatial scale or the compressibility of the gas: clumping can be stronger on small scales in nearly incompressible gas than on large scales in highly compressible cases. Surprisingly, the clumping is {\em stronger} at lower dust-to-gas ratios, although the growth times of the instabilities are longer. 

In contrast, in the ``disperse'' cases, the dust can be accelerated to highly super-sonic isotropic velocity dispersions (even undergoing first-order Fermi acceleration with stretched-exponential velocity ``tails'' that would reach relativistic velocities in some physical systems) and dispersed nearly-uniformly over the box (clumping factors as small as $\sim 1.02$).  These cases  are akin to well-studied cosmic ray instabilities that self-excite diffusive behavior (i.e.\ self-generating dust diffusion). The growth times of these instabilities can be shorter than either the dust drag/stopping or gyro times, and are often extremely short relative to other timescales in the gas (e.g.\ the sound-crossing or dynamical times, for physical systems of interest). This will have a huge range of important physical ramifications for essentially all regimes where dust is present. 

The dust drives anisotropic turbulence in the gas, whose properties may be {\em qualitatively} understood via  heuristic quasi-linear theory. The gas turbulence is stronger and more compressible if the  box scale is larger, or if the dust-to-gas ratio is larger. Moreover, the gas velocity and density fluctuations are correlated in approximately the same manner observed for pure-gas MHD turbulence. In the saturated state, the instabilities tend to produce equipartition between gas velocity and magnetic field fluctuations. The latter means that systems can have the magnetic fields strongly amplified by the instabilities (e.g.\ decreasing $\beta$ by factors of $\sim 10^{4}$, in the most extreme cases here).


The gas turbulence can be qualitatively magnetosonic or \Alf{ic}, but in addition to the dust properties, it has many characteristics that are unique. PDFs of various fluctuations ($\gasvel$, $\B$, $\gasden$, $\dustvel$, $\dustden$) are typically highly non-Gaussian, with exponential or stretched-exponential tails indicative of strong intermittency and stochastic driving in a highly-dissipative system. These large tails mean that sometimes the mass and volume-weighted statistics can yield quite different results, and the tails can be many orders-of-magnitude more populated than Gaussian (e.g.\ many of our systems have $>0.1\%$ of their dust or gas mass at $\gtrsim 10-20$ ``standard deviations'' in some property). 

The parameter space here is highly multi-dimensional, so it is difficult to make conclusions that can be robustly applied to all regimes. However it appears that choices such as the detailed equation-of-state, form of the dust charge scaling with ambient gas properties, exact magnetic $\beta$, initial direction of field orientiation (relative to the relative dust-gas acceleration), or exact dust-to-gas ratio do not dramatically ({qualitatively}) change the character of the solutions. As anticipated from linear theory in \paperone, the most important parameters that determine the qualitative behavior appear to be the physical scale and ratio of magnetic (Lorentz) to drag (aerodynamic+Coulomb) forces on dust.

Future work will be necessary to investigate how these instabilities will manifest in physical systems. We stress that the models here are intentionally idealized: we follow a single grain species subject to a constant external differential dust/gas acceleration in periodic, initially homogeneous gas boxes, with an ideal equation-of-state. Therefore our model names (e.g.\ {\bf AGB}) should not be taken literally -- these are {\em not} intended to be realistic physical simulations of those systems. Rather, the names are chosen reflect an example of a system where the key dimensionless parameters {\em for these instabilities} ($\tau$, $\beta$, $|\driftvel|/\cs$) are similar to the box simulated. Nor is it obvious, yet, how the simulations here directly translate to observational consequences: even if the behavior in fully ``realistic'' systems is similar to that predicted here, this will clearly have consequences for dust extinction curves and emission, cooling and dust chemistry, and many other areas, but the magnitude or even the sign of these effects depends sensitively on the exact observations considered, as well as physical conditions (e.g.\ chemical conditions, optical depth, additional radiative transfer effects) beyond those modeled here. The physical systems  simply provide helpful motivation for our survey and in future work we will explore more realistic scenarios. 

This investigation has begun to elucidate the broad non-linear behaviors of the resonant drag instabilities in magnetized gas. However, it raises more questions than it answers, some of which include:

\begin{enumerate}

\item{What are the effects of a broad spectrum of grain sizes and charges? Here we intentionally simulate a single species of grains, so that  the growth rates and dominant modes can be clearly defined and studied. However in almost all astrophysical situations there will be a wide range of grain sizes.  In some circumstances the largest grains (which tend to dominate the dust mass) will dominate the dynamics, while in others there is an intricate mix of which grains dominate which terms in the relevant equations (for a more detailed discussion, see \paperone). Further simulations will be required to understand when different grain sizes are effectively independent, and when they will have strong non-linear interactions via the gas \citep[see e.g.][for examples in the streaming RDI]{bai:2010.streaming.instability}.}

\item{Is there a meaningful way to define convergence, and incorporate the effects of all relevant size scales in a single simulation? The dynamic range over which the instabilities are present with interesting growth rates is enormous, and far larger than can be simulated at present. Because we find that the {\em gas} turbulence has most of its power at the largest (driving) scales, there is hope that, like MHD turbulence, certain bulk properties (e.g.\ the bulk power in turbulence, the dissipation rate, and most of its effects) can be ``converged'' even if the Kolmogorov scale is unresolved. It is unclear if a meaningful convergence criteria can even be defined for the dust, since some properties, like the dust density fluctuations, are {\em not} uniquely dominated by the driving-scale modes.}

\item{What is the effect of stratification or time-variability on the instabilities? As the size scale of the simulations increases, the physical system would become stratified and non-uniform in space or time.  We have shown that stratification and other large-scale terms such as shear, rotation, and differential acceleration (e.g.\ Coriolis forces) can all introduce {\em additional} instabilities, some of which have faster growth rates on large scales than those here \citep[see][]{squire:rdi.ppd}. So, this could introduce unique and important phenomena. On large time-scales, the ``forcing'' terms driving dust drift (hence the instabilities) could fluctuate: if this occurs rapidly compared to growth timescale it may produce unique phenomena as well. It would be particularly interesting to study the analog of ``decaying turbulence'' when external acceleration/forcing is suddenly ``turned off'' once the simulations have reached saturation.}

\item{What is the nature of the non-linear gas turbulence and its intermittency?
Any of these simulations provide ample opportunity to study the nature of the turbulence. We have only explored simple, ``zeroth order'' diagnostics, but more detailed studies of the Eulerian and Lagrangian power spectra and structure functions will provide substantial insight into the structure of the turbulence. The highly non-Gaussian behavior we see suggests that the turbulence may be quite different from a simple Kolmogorov-type picture (although of course pure-MHD turbulence is already substantially non-Gaussian in some measures).}

\item{Can we develop a predictive theory for the dust clumping and turbulence? Here we are able to develop some qualitative insights into when dust should exhibit clustering. Further, from quasi-linear theory, we can heuristically understand why the dust density fluctuations become stronger at lower dust-to-gas ratio, and are weakly-dependent on the spatial scale or compressibility of the gas. However, this is far from a predictive or quantitative theory. Most past work on the non-linear saturation of particle clustering in turbulence assumed ``passive'' grains (i.e.\ neglected the forces of dust on gas). However, in the ``passive'' case  these instabilities do not exist, and the predicted saturated dust clustering (neglecting back-reaction terms) can be very different from what we find here. It is clearly important to develop new theoretical models which can explain the strength and nature of the observed dust structures.}

\item{What are the important regimes of parameter space yet to explore? Our parameter survey is far from complete. We have chosen a few examples which are interesting and plausibly motivated, but there are many variations possible within similar physical systems. Many physical systems with dust have fast-growing RDIs, but  with parameters quite different to those studied here (e.g.\ AGN and their outflows, dense GMCs, supernova remnants, the ISM of primordial neutral galaxies, planetary atmospheres, proto-stellar disks; see \paperone\ for further discussion).}

\item{How does the introduction of additional grain physics modify the predictions here? As noted above, in the non-linear regime, some of our simulations reach dust concentrations $\gg 10^{9}$ times the mean density. Obviously, other physics will become important long before these densities are reached, including e.g.\ dust self-shielding (if radiation drives the differential acceleration), self-gravity of the grains, grain collisions, and dust current. Some of these could  suppress  clustering, but others could make it stronger. The high densities and low {\em local} relative grain-grain velocities in these regimes may make them ideal for coagulation. But much of the volume being dust-poor might suppress dust growth via accretion of ions from the gas. In the disperse regime, some of our simulations produce isotropic dust velocities many times the sound speed, which may introduce sputtering and shattering in grain collisions.}

\end{enumerate}

Finally, we stress that we have explored only the ideal MHD case of the \citet{squire.hopkins:RDI} instabilities. A host of other instabilities exist which appear when other physics are present (e.g.\ external or self gravity, stratification, centrifugal or Coriolis forces, non-ideal MHD, kinetic MHD, strong coupling of multiply-scattered radiation, etc.). In future work, we hope to explore these cases in more detail, together with some of the questions above.

\acknowledgments{We thank Alexander Kaurov and Ulrich Steinwandel for a number of enlightening discussions and useful comments. Support for PFH was provided by an Alfred P. Sloan Research Fellowship, NSF Collaborative Research Grant \#1715847 and CAREER grant \#1455342, and NASA grants NNX15AT06G, JPL 1589742, 17-ATP17-0214. Support for J.S. was provided by   Marsden Fund grant UOO1727 and a Rutherford Discovery Fellowship,  managed through the Royal Society Te Ap\=arangi. Numerical calculations were run on the Caltech compute cluster ``Wheeler,'' allocations from XSEDE TG-AST130039 and PRAC NSF.1713353 supported by the NSF, and NASA HEC SMD-16-7592.}

\bibliography{ms_extracted}

\end{document}